\documentclass[reprint,superscriptaddress,amsmath,amssymb,twocolumn,prl]{revtex4-2}

\usepackage{amsmath}
\usepackage{mathtools}
\usepackage{amssymb}
\usepackage{graphicx}
\usepackage{bm}
\usepackage[colorlinks, linkcolor=blue, citecolor=blue, urlcolor=blue, breaklinks=true]{hyperref}
\usepackage{color,dsfont,multirow,booktabs}
\usepackage{braket}
\usepackage{tabularx}
\usepackage{lipsum,booktabs}
\usepackage{subfiles}
\usepackage{subcaption}
\usepackage{caption}
\usepackage[dvipsnames]{xcolor}
\usepackage{ragged2e}
\usepackage{orcidlink}
\usepackage[T1]{fontenc}
\usepackage[normalem]{ulem}

\newcommand{\commentold}[1]{}
\DeclareMathSymbol{:}{\mathpunct}{operators}{"3A}

\newcommand{\red}[1]{\textcolor{Black}{#1}}

\newcommand{\figpanel}[2]{\hyperref[#1]{\ref*{#1}(#2)}}

\begin{document}
\date{\today}

\title{Charging Quantum Batteries with Chiral Squeezing}

\author{Borhan Ahmadi\orcidlink{0000-0002-2787-9321}}
\email{borhan.ahmadi@ug.edu.pl}
\address{International Centre for Theory of Quantum Technologies, University of Gdańsk, ul. prof. Marii Janion 4, 80-309 Gdańsk, Poland}
\author{André H. A. Malavazi\orcidlink{0000-0002-0280-0621}}
\address{International Centre for Theory of Quantum Technologies, University of Gdańsk, ul. prof. Marii Janion 4, 80-309 Gdańsk, Poland}
\author{Janine Splettstoesser \orcidlink{0000-0003-1078-9490}}
\affiliation{Department of Microtechnology and Nanoscience (MC2), Chalmers University of Technology, 412 96 Gothenburg, Sweden}
\author{Paweł Horodecki}
\address{International Centre for Theory of Quantum Technologies, University of Gdańsk, ul. prof. Marii Janion 4, 80-309 Gdańsk, Poland}
\author{Lei Du \orcidlink{0000-0003-0641-440X}}
\email{lei.du@chalmers.se}
\affiliation{Department of Microtechnology and Nanoscience (MC2), Chalmers University of Technology, 412 96 Gothenburg, Sweden}

\begin{abstract}

We propose a quantum-battery charger based on a driven bosonic Kitaev chain (BKC), where chiral squeezing converts passive input fluctuations into ordered, non-passive battery states. While a coherent input pulse exhibits phase-sensitive chiral transport, the charging dynamics is dominated by bidirectionally propagating fluctuations that are amplified and squeezed into orthogonal quadratures at opposite chain ends. In contrast to conventional phase-preserving amplifiers, our scheme stores largely extractable energy and achieves a work-like signal-to-noise ratio (SNR) near unity, even in the presence of thermal noise and moderate symmetry-preserving disorder.

\end{abstract}

\maketitle
\textit{Introduction}---We address a key challenge in quantum batteries, namely how to inject energy while converting unavoidable input fluctuations into useful, extractable work without sacrificing operational performance. Quantum batteries (QBs) are devices designed to store and retain energy in their internal--both classical and genuinely quantum--degrees of freedom~\cite{PhysRevE.87.042123,ferraro2026opportunities,RevModPhys.96.031001,2jtp-jpkn,Campbell_2026,ahmadi2026LandauZener}.
Over the past decade, substantial progress has been achieved, including the design of charging protocols~\cite{PhysRevLett.134.130401,PhysRevA.105.062203,PhysRevLett.131.240401,PhysRevLett.122.210601,PhysRevLett.132.090401,PhysRevLett.120.117702,sp5l-c6m8,43vv-521p,9vv8-s8r1,43n6-rnj3}, the development of strategies to mitigate energy loss~\cite{PhysRevE.105.064119,PhysRevA.102.060201,bv4w-jr6q,PhysRevResearch.2.013095,Liu2019,stable1,d9k1-75d4,PhysRevE.105.054115}, and the exploration of reservoir-engineering~\cite{ahmadi2025harnessing,p93y-jflt,Li_2026,PhysRevLett.132.210402,67wh-1fxv,Lin_2026,ahmadi2025reservoir} and topological effects~\cite{PhysRevLett.134.180401} as resources for performance enhancement. From an experimental perspective, a variety of QB architectures have been proposed across different physical implementations~\cite{Hu_2022,PhysRevA.107.023725,PhysRevA.109.062432,Cruz_2022,PhysRevA.106.042601,d9k1-75d4}.
Most QB charging schemes  optimize  how much energy can be stored or how fast~\cite{binder2015quantacell,PhysRevLett.118.150601,PhysRevLett.127.100601,Dou2021,Rodriguez_2024,deMoraes_2024,10.3389/fphy.2022.1097564,PhysRevApplied.23.024010,PhysRevLett.133.243602}. However, energy fluctuations constitute an equally important aspect that must be carefully accounted for~\cite{mohan2026fundamental,6kwv-z6fx,Perarnau-Llobet_2019,Friis2018precisionwork,PhysRevA.107.022215,Lobejko2022workfluctuations,PhysRevResearch.2.023095,e25111528,PhysRevE.109.014131,Crescente_2020,e24060820,sarkar2025fluctuation,PhysRevLett.125.040601,Henning2026QB}. A key open question is therefore how to exploit input fluctuations for charging while keeping the extractable energy large compared with the energy fluctuations of the battery state. This issue is highly nontrivial because conventional energy amplification is typically accompanied by substantial added noise, making it difficult to combine a large charging output with a high signal-to-noise ratio (SNR)~\cite{LimitAmplifier1982,IntroAmplificationRMP2010}. It is thus desirable to identify charging platforms in which both the coherent input and input fluctuations are processed such that amplification leads to high extractable work.

\begin{figure}
    \centering
    \includegraphics[width=0.95\linewidth]{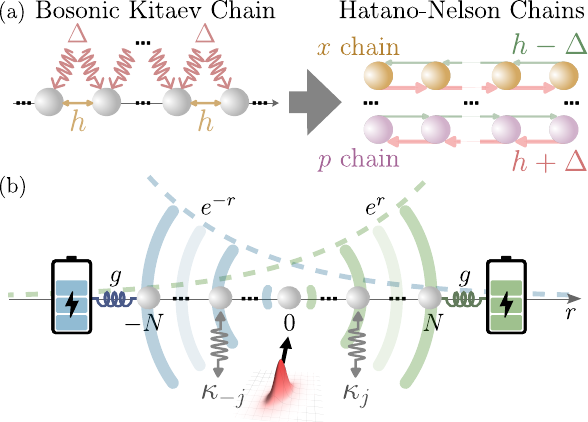}
    \caption{(a)~A bosonic Kitaev chain (BKC) with nearest-neighbor hopping $h$ and parametric pairing $\Delta$, which is equivalent to two decoupled Hatano-Nelson chains for the $x$ and $p$ quadratures with opposite chiralities. (b)~Schematic of the chiral charger: a driven-dissipative BKC of length $2N+1$, with local losses $\kappa_j$ and chiral squeezing weights $e^{\pm r}$, is excited at the central site $j=0$ by a short coherent pulse. The two end sites ($j=\pm N$) are coupled to the batteries through coupling $g$.\justifying}\label{Sketch}
\end{figure}

In this Letter, we show that a driven bosonic Kitaev chain (BKC) provides such a platform that can charge batteries, see Fig.~\ref{Sketch}. 
The BKC is a paradigmatic one-dimensional (1D) bosonic lattice in which nearest-neighbor hopping is supplemented by parametric-pump terms that create and annihilate excitations in pairs~\cite{PhysRevX.8.041031}. The pairing breaks excitation-number conservation and gives rise to chiral transport. This unusual structure enables a series of extraordinary phenomena absent in conventional passive lattices, including phase-sensitive and topologically protected directional amplification~\cite{PhysRevX.8.041031,Wanjura2020NC,NonuniformLossBKC2025}, exceptionally enhanced quantum sensing~\cite{McDonald2020NC}, and realization of bosonic Majorana zero modes~\cite{MajoranaBosons2021}. With the required ingredients now accessible in various platforms, such as superconducting circuits~\cite{BKCrealizationCircuit2024} and optomechanical systems~\cite{BKCrealizationOM2024}, the BKC has become a controllable and robust platform for engineering energy transport in driven bosonic networks~\cite{Du2025prr}.

While previous studies have focused on BKC dynamics driven by coherent inputs, here we investigate how continuously injected reservoir fluctuations can participate in the charging process alongside a coherent pulse. We find that a coherent input pulse with a well-defined initial phase undergoes phase-sensitive chiral transport, whereas the fluctuation sector propagates bidirectionally and is amplified and squeezed into orthogonal quadratures toward opposite ends of the chain. Due to its finite duration and the dissipation accumulated across the chain, the coherent pulse contributes negligibly to the stored battery energy, so that the overall charging dynamics is dominated by the fluctuation sector. More surprisingly, the active BKC dynamics converts the originally passive fluctuations into ordered, non-passive battery energies. Our protocol achieves a nearly unity work-like SNR and remains robust against thermal noise and symmetry-preserving disorder. In this Letter, we set the Planck constant to $\hbar=1$.

\textit{Model}---We model the BKC as a uniform 1D bosonic chain of length $2N+1$, with nearest-neighbor hopping $h$ and parametric-pump induced two-particle pairing $\Delta$, as shown in Fig.~\figpanel{Sketch}{a}. \red{Let $\omega_{\rm C}$ denote the bare resonance frequency of the BKC sites. We take the parametric pump to be resonant at $2\omega_{\rm C}$ and work in a frame rotating at $\omega_{\rm C}$. The resulting rotating-frame chain Hamiltonian reads} 
\begin{equation}\label{HB_realspace}
\hat H_\mathrm{C}=\frac{1}{2}\sum_{j=-N}^{N-1}
\bigl(i\,h\,\hat a^{\dagger}_{j+1}\hat a_{j}
+ i\,\Delta\,\hat a^{\dagger}_{j+1}\hat a^{\dagger}_{j}
+\mathrm{H.c.}\bigr),
\end{equation}
where $\hat{a}_{j}$ ($\hat{a}_{j}^{\dag}$) annihilates (creates) a bosonic excitation on the $j$th chain site, satisfying $[\hat{a}_{j},\hat{a}_{k}^{\dagger}]=\delta_{jk}$. By introducing the position and momentum quadratures, $\hat{x}_{j}=(\hat{a}_{j}+\hat{a}^{\dag}_{j})/\sqrt{2}$ and $\hat{p}_{j}=i(\hat{a}^{\dag}_{j}-\hat{a}_{j})/\sqrt{2}$, the Hamiltonian can be rewritten as
\begin{equation}
\hat H_\mathrm{C}=\frac{1}{2}\sum_{j}
\Bigl[-(h-\Delta)\,\hat x_{j+1}\hat p_j
+ (h+\Delta)\,\hat p_{j+1}\hat x_j\Bigr].
\label{eq:HB_quadratures}
\end{equation}
In the absence of dissipation and external drive, the Heisenberg equations of motion derived from Eq.~(\ref{eq:HB_quadratures}) read
$\dot{\hat{x}}_{j}=\frac{h+\Delta}{2}\hat{x}_{j-1}-\frac{h-\Delta}{2}\hat{x}_{j+1}$ and 
$\dot{\hat{p}}_{j}=\frac{h-\Delta}{2}\hat{p}_{j-1}-\frac{h+\Delta}{2}\hat{p}_{j+1}$. 
These equations show that the BKC decomposes into two decoupled Hatano-Nelson chains~\cite{HNchain1996} for the $x$ and $p$ quadratures, with opposite nonreciprocal couplings, as illustrated in Fig.~\figpanel{Sketch}{a}, where the pink (thick) and green (thin) arrows are swapped for the $x$- and $p$-dynamics. As a result, an injected wavepacket with a \emph{well-defined} initial phase exhibits a phase-dependent chirality: the $x$ component couples preferentially to the right neighbor, whereas the $p$ component couples preferentially to the left neighbor.

We emphasize that $\hat{H}_\mathrm{C}$ is an effective rotating-frame Hamiltonian that contains parametric pump terms. These terms, which are proportional to $\Delta$, represent an external \emph{active} resource that underlies the chirality and amplification of the dynamics. We therefore account for the corresponding energetic cost through the pump work defined in the End Matter. 


For open boundaries, a lossless BKC is dynamically stable when $|h|>|\Delta|$. In contrast, for $|h|<|\Delta|$, the dynamics becomes unstable and no well-defined wave transport exists~(see Note~1 of the Supplemental Materials (SM)~\cite{SuppMat}). In what follows we restrict ourselves to the stable regime, where one can introduce the squeezing parameter $r=\frac{1}{2}\ln{\left(\frac{h+\Delta}{h-\Delta}\right)}$ and 
an effective hopping amplitude $\tilde h=\sqrt{h^{2}-\Delta^{2}}=h/\cosh r$.
The chiral transport is reflected in the opposite squeezing for the $x$ and $p$ components: they propagate toward opposite ends of the chain with amplitudes weighted by $e^{\pm jr}$; see Eq.~\eqref{xp-chiral} below. 

As shown in Fig.~\figpanel{Sketch}{b}, in this work we consider a BKC of length $2N+1$, and couple its two ends to storage modes (``batteries''). The batteries are modeled as two identical harmonic modes with Hamiltonian $\hat{H}_{\rm bat}=\sum_{\alpha}\hat{H}_{\mathrm{bat},\alpha}=\sum_{\alpha}\Omega\, \hat{b}_{\alpha}^\dagger \hat{b}_{\alpha}$ ($\alpha\in\{\rm L, \rm R\}$), where $\Omega=\omega-\omega_{\rm C}$ is the detuning between the battery frequency $\omega$ and the BKC bare frequency $\omega_{\rm C}$. \red{Hereafter, we focus on the resonant case $\Omega\equiv0$, such that $\hat{H}_{\rm bat}$ vanishes in the rotating frame.} The BKC-battery interaction is described by $\hat{H}_{\rm int} = g_\mathrm{L}(t)\left(\hat{x}_{-N} \hat{x}_\mathrm{L} + \hat{p}_{-N}\hat{p}_\mathrm{L}\right)+g_\mathrm{R}(t)\left(\hat{x}_{N} \hat{x}_\mathrm{R} + \hat{p}_{N}\hat{p}_\mathrm{R}\right)$, with the battery quadratures defined as $\hat{x}_{\alpha}=(\hat{b}_{\alpha}+\hat{b}_{\alpha}^{\dag})/\sqrt{2}$ and $\hat{p}_{\alpha}=i(\hat{b}_{\alpha}^{\dag}-\hat{b}_{\alpha})/\sqrt{2}$.
The total Hamiltonian in the rotating frame is given by $\hat{H}_{\rm{tot}}=\hat{H}_\mathrm{C}+\hat{H}_{\rm int} + \hat H_{\rm drive}$, where $\hat H_{\rm drive}=\sqrt{\kappa_{0}}(p_{\rm in}\hat{x}_{0}-x_{\rm in}\hat{p}_{0})$ describes a coherent drive applied to the central site ($j=0$), with effective driving amplitudes $\sqrt{\kappa_0}x_{\rm in}$ and $\sqrt{\kappa_0}p_{\rm in}$ for the two quadrature components. \red{The coherent drive is assumed to be resonant with the BKC sites, so that its carrier-frequency oscillation is removed in the rotating frame.}

\textit{Equations of motion}---Assuming that every chain site is subject to (uniform) losses, modeled by Markovian reservoirs, the quantum Langevin equations for the chain quadratures are given by
\begin{equation}\label{xp-chiral}
\begin{aligned}
    \dot{\hat{x}}_{j}&=\frac{\tilde{h}}{2}\Big(e^{r}\,\hat x_{j-1}-e^{-r}\,\hat x_{j+1}\Big)-\frac{\kappa_{j}}{2}\,\hat x_{j}-\sqrt{\kappa_{0}}\, x_{{\rm in}}(t)\,\delta_{j,0}\\&\quad+\sqrt{\kappa_{j}}\,\hat{\xi}_{j}^{(x)}(t)+g_\mathrm{R}(t)\,\hat p_\mathrm{R}\,\delta_{j,+N}+g_\mathrm{L}(t)\,\hat p_\mathrm{L}\,\delta_{j,-N},\\\dot{\hat{p}}_{j}&=\frac{\tilde{h}}{2}\Big(e^{-r}\,\hat p_{j-1}-e^{r}\,\hat p_{j+1}\Big)-\frac{\kappa_{j}}{2}\,\hat p_{j}-\sqrt{\kappa_{0}}\, p_{{\rm in}}(t)\,\delta_{j,0}\\&\quad+\sqrt{\kappa_{j}}\,\hat{\xi}_{j}^{(p)}(t)-g_\mathrm{R}(t)\,\hat x_\mathrm{R}\,\delta_{j,+N}-g_\mathrm{L}(t)\,\hat x_\mathrm{L}\,\delta_{j,-N}.
\end{aligned}
\end{equation}
Here $\kappa_j$ is the decay rate of the $j$th chain site and $\hat{\xi}_j^{(\beta)}(t)$ ($\beta=x,p$) is the corresponding quadrature noise operator. The noise is delta-correlated, $\langle\hat\xi_j^{(\beta)}(t)\hat\xi_k^{(\beta')}(t')\rangle=(n_{\rm th} + 1/2)\delta_{jk}\delta_{\beta\beta'}\delta(t-t')$ with $n_{\mathrm{th}}$ being the thermal occupation, and has by definition zero mean, $\langle\hat\xi^{(\beta)}_j(t)\rangle=0$. The dynamics of the right battery is described by 
\begin{equation}\label{xp-battery}
    \begin{aligned}
    \dot{\hat{x}}_\mathrm{R} &= - \frac{\gamma}{2} \hat x_\mathrm{R} + g_\mathrm{R}(t) \hat p_N + \sqrt{\gamma}\hat{\zeta}^{(x)}_\mathrm{R}(t),\\
    \dot{\hat{p}}_\mathrm{R} &= - \frac{\gamma}{2}\,\hat p_\mathrm{R} - g_\mathrm{R}(t) \hat x_N + \sqrt{\gamma}\hat{\zeta}^{(p)}_\mathrm{R}(t),
    \end{aligned}
\end{equation}
with similar equations (R $\to$ L and $N\to-N$) for the left battery. The batteries are subject to dissipation at rate $\gamma$, with associated noise operators $\hat{\zeta}_\mathrm{R/L}^{(x/p)}$, which also have zero mean and obey similar delta-type correlations. For simplicity, we hereafter assume $g_\mathrm{R}(t) \equiv g_\mathrm{L}(t) =g(t)$, meaning that the batteries are detached/attached simultaneously with the same coupling strength.

In our protocol, we coherently drive only the center site $j=0$ with amplitude $\mathcal{A}$, phase $\theta$, and a Gaussian envelope $f(t)=\exp\!\left[-\frac{(t-t_0)^2}{2\sigma^2}\right]$, so that
%
    \(x_{\rm in}(t)=\mathcal{A}\cos\theta\,f(t),\quad
    p_{\rm in}(t)=\mathcal{A}\sin\theta\,f(t) \), 
%
where $t_0$ is the pulse-center time and $\sigma$ sets the pulse width. Because the quantum Langevin equations for the $x$ and $p$ quadratures decouple and exhibit opposite chiralities, the $x$ component of the injected wavepacket propagates predominantly to the right, whereas the $p$ component propagates predominantly to the left. Consequently, a coherent input field with a well-defined initial phase exhibits phase-sensitive chiral transport~\cite{PhysRevX.8.041031}. For fields \emph{without} a definite phase, the transport remains bidirectional. As we show below, however, the right-moving and left-moving components of an incoherent signal are amplified and squeezed into orthogonal phase-space directions. This chiral squeezing mechanism allows the active BKC dynamics to convert otherwise passive fluctuations into ordered, non-passive energies.
\begin{figure}
    \centering
    \includegraphics[width=1\linewidth]{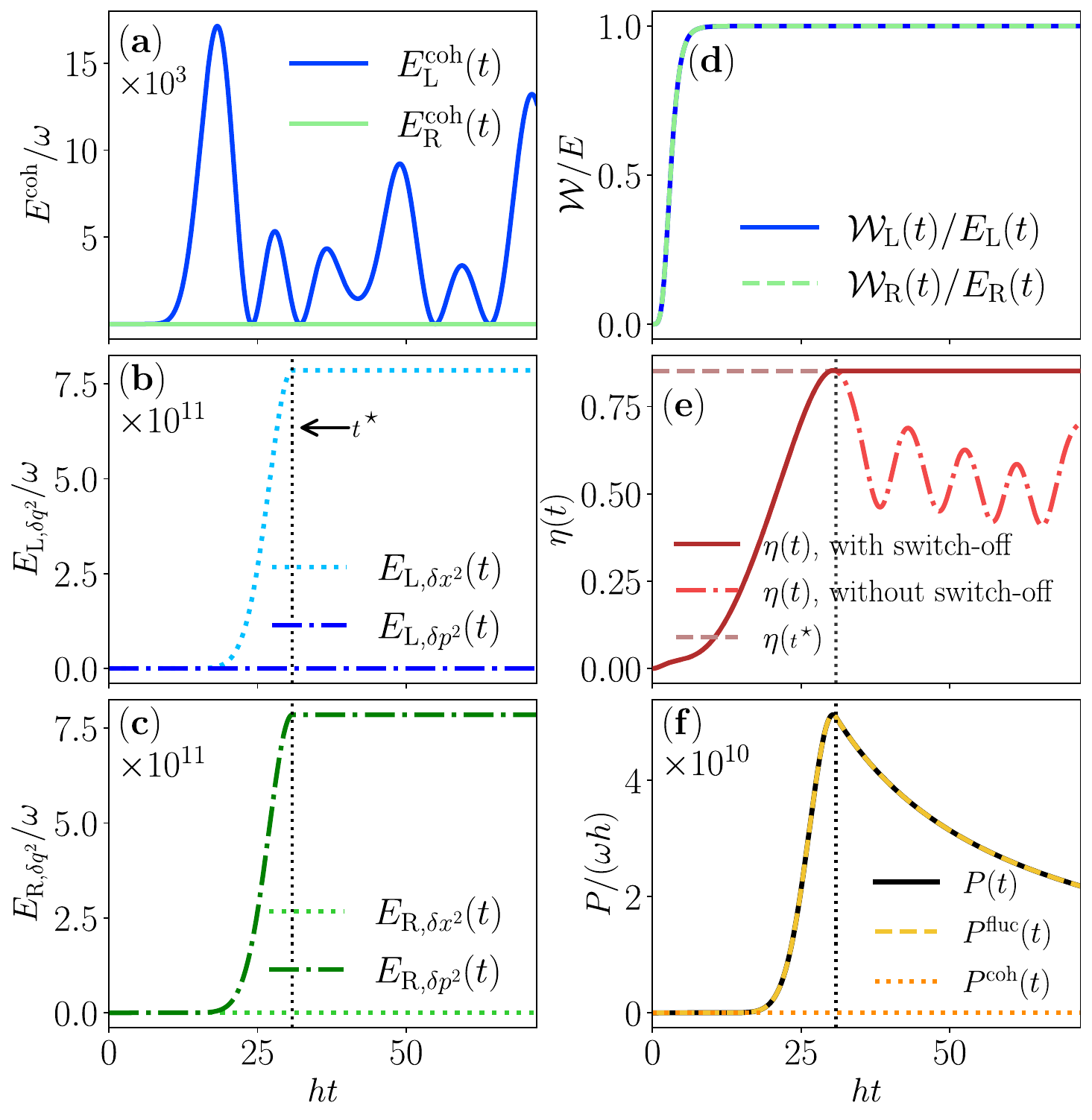}
    \caption{Charging dynamics. (a)~Coherent contributions to the left- and right-battery stored energies, $E_\mathrm{L}^{\rm coh}$ and $E_\mathrm{R}^{\rm coh}$. (b)~Quadrature-resolved fluctuation contributions to the left-battery stored energy, $E_{\mathrm{L},\delta x^2}$ and $E_{\mathrm{L},\delta p^2}$. (c)~Same as (b) for the right battery, $E_{\mathrm{R},\delta x^2}$ and $E_{\mathrm{R},\delta p^2}$. (d)~Ergotropy-energy ratios $\mathcal{W}_\mathrm{L}(t)/E_\mathrm{L}(t)$ and $\mathcal{W}_\mathrm{R}(t)/E_\mathrm{R}(t)$. \red{(e)~Overall charging efficiency $\eta$, with and without switch-off at $t=t^\star$. (f) Average charging power $P(t)$ and its coherent and fluctuation contributions.} The vertical dotted lines indicate the switch-off time \(t=t^\star\). Parameters: \(N=5\), \(\Delta/h=11/12\), \(g/h=5/12\), \(\kappa_j/h=8.3\times10^{-4}\) ($j=-N,\cdots,N$), \(n_{\rm th}=0\), \(\gamma=0\), \(\mathcal{A}/\sqrt{h}=2.887\), \(\theta=\pi/2\), \(h\sigma=1.2\), and \(ht_0=1.2\). \justifying}
    \label{5_Combined_abcd}
\end{figure}

\textit{Chiral charging}---We first use the stored energy and the ergotropy as the central figures of merit of our charging protocol~\cite{alicki1979quantum,ahmadi2023work}. \red{Since $\hat{b}_{\alpha}^{\dag}\hat{b}_{\alpha}$ is invariant under the rotating-frame transformation, the stored energy can be evaluated directly from the rotating-frame moments as} 
\(
\hat E_\alpha \equiv \omega\hat b_\alpha^\dagger \hat b_\alpha
=\frac{\omega}{2}\left(\hat x_\alpha^2+\hat p_\alpha^2-1\right)
\)
and
\(
E_{\alpha}(\hat\rho)\equiv{\rm Tr}\{\hat\rho \hat E_{\alpha}\}
\) (\(\alpha=\)L,R).
The ergotropy quantifies the maximum amount of work that can be extracted from a quantum state via a unitary process, and is defined as~\cite{Allahverdyan_2004}
\begin{equation}
    \mathcal{W}_{\alpha}(\hat\rho) \equiv E_{\alpha}(\hat\rho) - E_{\alpha}(\hat\rho_{\rm pa}),
\end{equation}
where $\hat\rho_{\rm pa}$ is the passive state associated with $\hat\rho$, obtained by assigning the eigenvalues of $\hat\rho$ (in descending order) to the eigenstates of $\hat E_{\alpha}$ (in ascending order). 

Figure~\ref{5_Combined_abcd} shows representative charging dynamics. To illustrate the chiral features, we first plot in Fig.~\figpanel{5_Combined_abcd}{a} the coherent contributions to the stored energies, $E_\mathrm{L}^{\rm coh}$ and $E_\mathrm{R}^{\rm coh}$, originating solely from the injected Gaussian pulse with initial phase $\theta=\pi/2$. These curves are obtained by solving the mean-value parts of Eqs.~(\ref{xp-chiral}) and (\ref{xp-battery}) (see Note~2 of the SM~\cite{SuppMat}). As expected from the phase-sensitive chirality, the coherent wavepacket is transported predominantly toward the left and is eventually loaded into the left battery (with oscillations reflecting the dynamics of the battery-chain coupling), whereas its contribution to the right battery remains negligible.  However, this directional coherent contribution remains small and shows no appreciable amplification in the batteries. For the finite-duration pulse considered here, the dissipation accumulated across the chain dominates over the amplification.


In Figs.~\figpanel{5_Combined_abcd}{b} and~\figpanel{5_Combined_abcd}{c}, we show the stored energies originating from the fluctuations, which are obtained from the second moments of Eqs.~\eqref{xp-chiral} and \eqref{xp-battery}. We refer to them as the \emph{fluctuation contributions} to the stored energies. For these plots, we switch off the setup at $t=t^\star$, i.e., the time at which the charging efficiency reaches its maximum; see Fig.~\figpanel{5_Combined_abcd}{e}. 
The behaviors of the coherent and fluctuation contributions are qualitatively different. 
\red{Unlike the finite coherent pulse, the fluctuation sector is continuously driven by the Markovian input-noise fields at all BKC sites. Thus, the same dissipation channels that attenuate the propagating mean field also inject vacuum (or thermal) fluctuations into the covariance dynamics. The parametric pump then reshapes and amplifies these fluctuations into ordered, non-passive energy, allowing the fluctuation contribution to become orders of magnitude larger than the coherent one. 
Since the coherent pulse enters only the first-moment dynamics, removing it leaves the covariance dynamics unchanged. In our protocol, the coherent pulse primarily serves to illustrate the phase-sensitive chiral transport established in previous studies of the BKC.}

In contrast to the injected Gaussian pulse, the fluctuations are amplified in both directions. However, the components \(E_{\alpha,\delta x^2}\) and \(E_{\alpha,\delta p^2}\), which quantify the energies stored in the two orthogonal battery quadratures~\cite{SuppMat}, are strongly direction dependent. The fluctuation energy in the left battery is carried almost entirely by the $x$ sector, while that in the right battery is dominated by the $p$ sector. More interestingly, the complementary quadratures of the two batteries are squeezed below the vacuum level: specifically, we have $\langle\delta p_\mathrm{L}^{2}(t^\star)\rangle<1/2$ and $\langle\delta x_\mathrm{R}^{2}(t^\star)\rangle<1/2$ (see Note~2 of the SM~\cite{SuppMat}). Consistent with this strongly anisotropic fluctuation structure, the stored energies are almost fully extractable. As shown in Fig.~\figpanel{5_Combined_abcd}{d}, the ratios $\mathcal{W}_\mathrm{L(R)}(t)/E_\mathrm{L(R)}(t)$ approach unity after a short transient. This transient reflects the time required for the chiral dynamics to convert initially passive fluctuation energy into strongly squeezed, non-passive battery states~\cite{SuppMat}.


\red{We next quantify the overall charging performance (including both the coherent and fluctuation contributions), through the charging efficiency \(\eta(t)\) and average charging power $P(t)$. The former is defined as the ratio between the total energy stored in the two batteries and the active energy cost of the protocol, while the latter quantifies the total energy obtained by the batteries per unit charging time; details are given in the End Matter. 
As shown in Fig.~\figpanel{5_Combined_abcd}{e}, when the chain is kept on, \(\eta(t)\) first increases and reaches a maximum at \(t=t^\star\) (see Note~3 of the SM~\cite{SuppMat}), after which it decreases due to the continued chain-battery energy exchange and the accumulation of energy cost. This maximum therefore identifies an operational switch-off time. We thus implement a finite-time charging protocol by switching off the setup at \(t=t^\star\). As shown in Fig.~\figpanel{5_Combined_abcd}{f}, the average charging power likewise increases monotonically until the setup is switched off at $t^\star$ and is dominated by the fluctuation contribution. Moreover, we find that
$P(t^\star)$ exhibits a strongly \emph{superextensive} growth with the half-chain length $N$, substantially exceeding the extensive scaling $P\propto N$ over
the investigated range (see Note~2 of the SM~\cite{SuppMat}).}

\emph{High signal-to-noise ratio}---To quantify how the extractable work compares with the energy fluctuations of the charged battery state, we introduce a work-like SNR
\begin{equation}\label{SNRW}
\mathrm{SNR}^{\rm work}(t^\star) =
\frac{\mathcal{W}(t^\star)}{\sigma_E(t^\star)}\,,
\end{equation}
where \(\mathcal{W}(t^\star)=\mathcal{W}_\mathrm{L}(t^\star)+\mathcal{W}_\mathrm{R}(t^\star)\) and \(\sigma_E(t^\star)=\sqrt{\mathrm{Var}[\hat{E}_{\rm bat}](t^\star)}\) is the standard deviation of the battery energy in the same state, with $\hat{E}_{\rm bat}=\hat{E}_\mathrm{L}+\hat{E}_\mathrm{R}$.
As we show below, our BKC-based charger exhibits another key advantage: unlike conventional phase-preserving amplifiers, such as a 1D tight-binding chain with on-site gain, it enables a nearly unity $\mathrm{SNR}^{\rm work}(t^\star)$.

\begin{figure}
    \centering
    \includegraphics[width=0.95\linewidth]{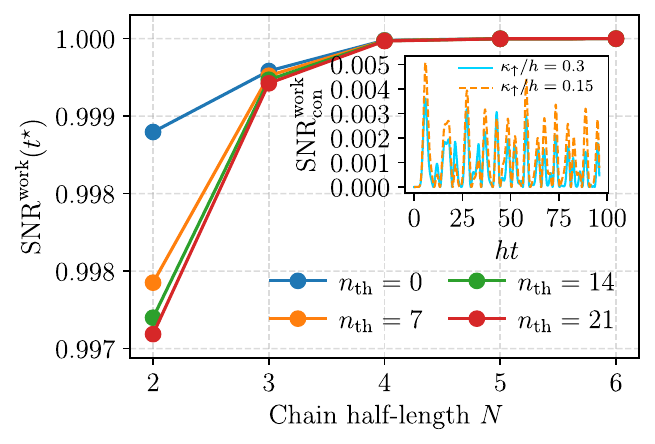}
    \caption{Work-like signal-to-noise ratio ${\rm SNR}^{\rm work}(t^\star)$, evaluated at $t=t^\star$, as a function of the chain half-length $N$ for several thermal occupancies $n_{\rm th}$. Inset: signal-to-noise ratio ${\rm SNR}^{\rm work}_{\rm con}(t)$ for the conventional phase-preserving amplifier (see Note~4 of the SM~\cite{SuppMat}), \red{with $n_{\rm th}=0$ and two representative on-site gain rates}. All other parameters are the same as in Fig.~\ref{5_Combined_abcd}. \justifying}
    \label{SNR}
\end{figure}

Figure~\ref{SNR} shows $\mathrm{SNR}^{\rm work}(t^\star)$ as a function of the BKC half-length \(N\) for several values of the thermal occupancy \(n_{\rm th}\) (applied uniformly to all chain sites). We find that \(\mathrm{SNR}^{\rm work}(t^\star)\) rapidly approaches unity as \(N\) increases. Even for moderate system sizes, \(\mathcal W(t^\star)\) becomes comparable to \(\sigma_E(t^\star)\), yielding an asymptotically size-independent SNR. Moreover, the curves for different \(n_{\rm th}\) nearly coincide for large $N$. Only a slight difference can be observed for the smallest \(N\), while the dependence on \(n_{\rm th}\) becomes negligible for \(N\gtrsim 4\).
This indicates that, at the optimal time, the injected thermal noise does not qualitatively degrade the work-to-fluctuation performance.

The result in Fig.~\ref{SNR} does not mean that usable work can be extracted from a thermal bath alone. Indeed, a thermal (Gibbs) state is passive and has vanishing ergotropy. In our protocol, extractable work is generated by \emph{actively} converting the incoming fluctuations through the externally pumped dynamics. In other words, the BKC acts as an active Gaussian processor that can convert vacuum or thermal fluctuations into non-passive battery states. Moreover, $\sigma_{E}$ should not be viewed as a work resource by itself, since passive states can exhibit substantial energy fluctuations while having zero ergotropy. In an active Gaussian charging process such as ours, instead, the same pump-driven dynamics that generates non-passivity (and hence ergotropy) can also amplify energy fluctuations, so that $\mathcal{W}(t^\star)$ and $\sigma_E(t^\star)$ co-vary~\footnote{The amplified fluctuations not only enhance the total stored energy, but also broaden the intrinsic energy distribution of the resulting battery state.}. In the End Matter, we examine the scaling of these two quantities separately, showing that the near-unity \(\mathrm{SNR}^{\rm work}(t^\star)\) results from the correlated (nearly linear) scaling of the extractable work and the battery-energy fluctuation.

This favorable work-to-fluctuation behavior clearly distinguishes our protocol from conventional phase-preserving gain media, where amplification is inevitably accompanied by added quantum noise. In a tight-binding chain with uniform on-site gain~\cite{SuppMat}, for example, the added fluctuations are predominantly isotropic and contribute mainly as passive, thermal-like energy. This energy contribution therefore increases the battery-energy fluctuation while providing little ergotropy. Increasing the gain does not overcome this limitation, because the coherent contribution and the added fluctuations acquire the same leading amplification factor (see Note~4 of the SM~\cite{SuppMat}). Accordingly, the inset of Fig.~\ref{SNR} shows SNRs of only $\sim 10^{-3}$ \red{for two representative on-site gain rates $\kappa_{\uparrow}$}. By contrast, the chiral squeezing mechanism of our BKC scheme (encoded in the drift matrix) converts amplified fluctuations into non-passive energy, avoiding this passive-noise penalty.

\begin{figure}
    \centering
    \includegraphics[width=1\linewidth]{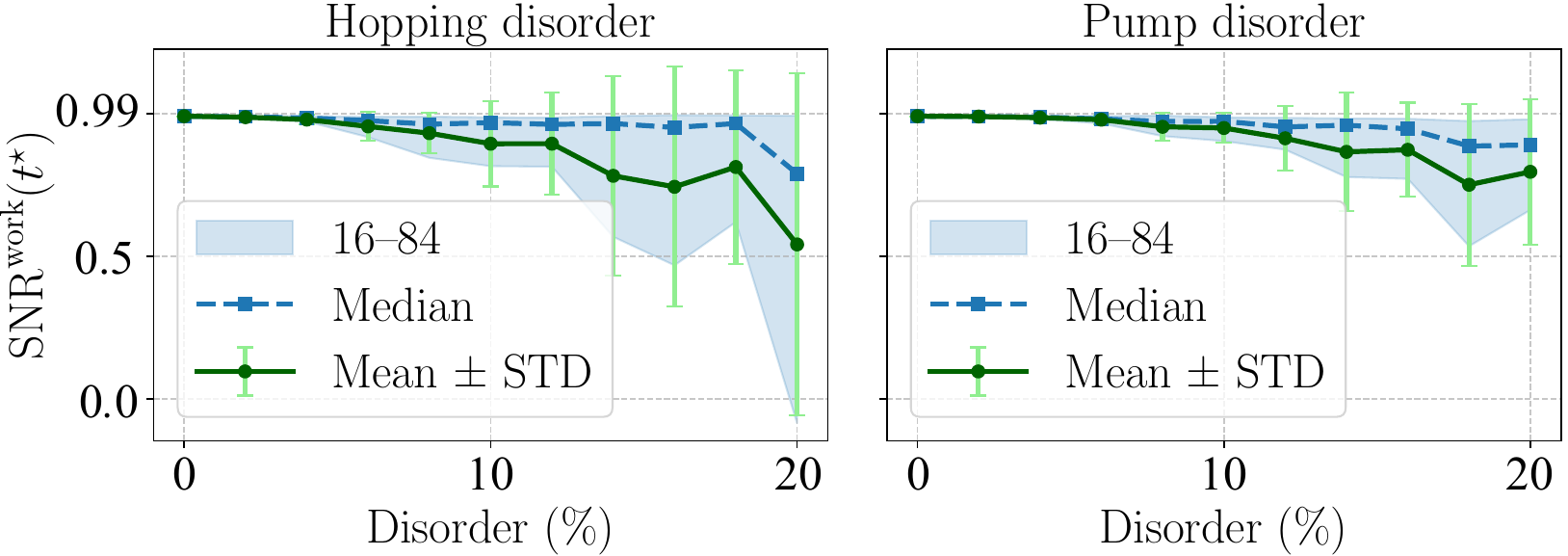}
\caption{${\rm SNR}^{\rm work}(t^\star)$ versus static hopping disorder (left) and pump disorder (right), for a fixed chain half-length $N=5$. For each disorder strength, $50$ random realizations are sampled. The green solid curves with error bars show the mean $\pm$ standard deviation, the blue dashed curves denote the median, and the shaded regions indicate the 16th–84th percentile range. All other parameters are the same as in Fig.~\ref{5_Combined_abcd}. \justifying}\label{Disorder_SNR}
\end{figure}

\emph{Topological protection}---We now examine the robustness of the work-like SNR against specific imperfections. Figure~\ref{Disorder_SNR} quantifies this robustness by showing $\mathrm{SNR}^{\rm work}(t^\star)$ in the presence of static disorder in both the hopping amplitudes (left panel) and the parametric pump strength (right panel). For each disorder strength, we sampled 50 random realizations and summarize the corresponding distribution by plotting the median together with the 16–84 percentile band (which contains the central $68\%$ of the realizations)~\footnote{Note that the percentile band reflects the empirical spread across realizations and should not be interpreted as an uncertainty of the mean.}. For comparison, we also show the mean, $\overline{\mathrm{SNR}}^{\rm work}$, together with error bars (one standard deviation), i.e., $\overline{\mathrm{SNR}}^{\rm work} \pm \rm STD$.  

We find that $\mathrm{SNR}^{\rm work}$ remains close to its disorder-free value over a broad range of disorder strengths. This robustness can be understood from the fact that both types of disorder preserve the BKC structure and do not mix the $x$- and $p$-quadrature sectors. The chiral dynamics remains protected by the \emph{point-gap topology}~\cite{GongPRX2018,KunstRMP2021} of the $x$- and $p$-sector HNCs: their complex spectra remain separated from the reference point (here set by the BKC on-site energy), so that the winding numbers cannot change under weak disorder~\cite{WanjuraDisorder2021}. 
The slightly stronger degradation under hopping disorder is mainly because it directly perturbs the field transport and enhances reflections, whereas the pump disorder primarily modifies the local squeezing and amplification.

The hopping and pump disorders considered above are examples of \emph{symmetry-preserving} disorder. For symmetry-breaking disorder, however, the protocol becomes more fragile. A representative example is random on-site detuning disorder, incorporated through perturbations $\sum_{j}\delta\omega_{j}\hat{a}_{j}^{\dag}\hat{a}_{j}$ in Eq.~(\ref{HB_realspace}). Physically, such random on-site energies mix the $x$ and $p$ quadratures and thereby degrade the transport chirality. For sufficiently strong detunings, this quadrature mixing can also trigger an instability, leading to uncontrolled amplification during the charging process~\cite{PhysRevX.8.041031,Du2025prr}. In practice, however, these effects can be kept under control by operating with shorter chains and/or weaker parametric pumps. A detailed analysis of the impact of on-site detuning disorder is provided in Note~5 of the SM~\cite{SuppMat}.

\emph{Conclusions}---In summary, we have proposed a quantum-battery charger based on a driven BKC. While a coherent input pulse with a well-defined phase exhibits phase-sensitive chiral transport, the input fluctuations propagate bidirectionally and are reshaped into orthogonal quadratures at opposite ends of the chain. We showed that these amplified and squeezed fluctuations dominate the total stored energies and ergotropies of the batteries, thereby enabling fluctuation-driven quantum charging beyond the coherent channel alone. Moreover, the resulting protocol achieves a nearly unity work-like SNR at the optimal charging time, in sharp contrast to conventional phase-preserving gain media, and remains robust against thermal noise and moderate symmetry-preserving disorder. Our proposal is \red{conceptually distinct from previous chiral charging schemes (see End Matter)} and could be implemented in several experimentally relevant platforms, including parametric superconducting resonators~\cite{BKCrealizationCircuit2024} and optomechanical arrays~\cite{BKCrealizationOM2024}.


Looking ahead, our results suggest a design principle for quantum batteries: input fluctuations need not only be a source of noise, but can be reshaped into useful non-passive energy. This idea may extend beyond the specific BKC implementation to other nonreciprocal or topological bosonic networks. A natural next step is to explore whether going beyond Gaussian dynamics, for example through weak nonlinearities or conditional operations, can further enhance or control this fluctuation-to-work conversion and its robustness.


\acknowledgments{\emph{Acknowledgments}---BA and PH acknowledge support from IRA Programme (project no. FENG.02.01-IP.05-0006/23) financed by the FENG program 2021-2027, Priority FENG.02, Measure FENG.02.01., with the support of the FNP. A.H.A.M. acknowledges support from the National Science Centre, Poland (SONATINA-9, grant agreement no. UMO-2025/56/C/ST2/00368 entitled “Modularne kwantowe urządzenia termiczne: integrowanie termicznych funkcjonalności”). JS and LD acknowledge financial support by the Knut och Alice Wallenberg stiftelse through project grant no. 2022.0090, as well as through an individual Wallenberg Academy fellowship grant and from the European Research Council (ERC) under the European Union’s Horizon Europe research and innovation program (101088169/NanoRecycle).}

\emph{Data and code availability}---
The numerical code used to generate the data and figures in this work is available at \url{https://github.com/Borhan19/Chiral-QB/releases/tag/v1.0-submission}. 
The repository contains the simulation scripts, plotting routines, and parameter files required to reproduce the results presented in the main text and Supplemental Material.

\bibliography{References}

\section*{End Matter}\label{EndMatter}
\appendix

\subsection{Charging efficiency and power}
Here we provide the energetic bookkeeping used to define the charging efficiency in Figs.~\figpanel{5_Combined_abcd}{e} and \figpanel{5_Combined_abcd}{f}. For each battery $\alpha=\{\rm L,\rm R\}$, the stored energy is
\begin{equation}
E_\alpha(t)=\frac{1}{2}\omega\mu_{\alpha}^{T}\mu_{\alpha}
+\frac{1}{2}\omega\left\{{\rm Tr}[V_\alpha(t)]-1\right\},
\label{EachEnergy}
\end{equation}
where $\omega=\omega_{\rm C}$ is the battery frequency (in the laboratory frame), \(\mu_\alpha=\left(\langle \hat{x}_{\alpha}\rangle, \langle\hat{p}_{\alpha}\rangle\right)^T\) and \(V_\alpha\) are the first-moment vector and covariance matrix of the corresponding battery mode. The total stored energy is \(E_{\rm bat}(t)=E_\mathrm{L}(t)+E_\mathrm{R}(t)\). The (active) energy cost of our charging protocol has two contributions, 
\[
E_{\rm cost}(t)=E_{\rm pulse}(t)+E_{\rm pump}(t).
\]
The first term is the work supplied by the coherent Gaussian pulse applied to the central site (for details see Note ~2 of the SM~\cite{SuppMat}),
\begin{eqnarray}
E_{\rm pulse}(t)&=&\int_0^t dt'\,P_{\rm pulse}(t'), \\
P_{\rm pulse}(t)&=&-\omega_{\rm C}\sqrt{\kappa_0}\,[x_{\rm in}(t)\langle x_0(t)\rangle+p_{\rm in}(t)\langle p_0(t)\rangle].
\end{eqnarray}
The second term accounts for the energetic cost of the parametric pump, which is obtained as~\cite{SuppMat} 
\begin{equation}
E_{\rm pump}(t)=\int_0^t dt'\,P_{\rm pump}(t'),
\label{Egain}
\end{equation}
with
\begin{equation}
\begin{split}
P_{\rm pump}(t)&=\omega_{\rm C}\tilde h(t)\sinh r(t)
\sum_{j=-N}^{N-1}
\left[
\langle x_j(t)x_{j+1}(t)\rangle \right.\\
&\left.\quad\,- \langle p_j(t)p_{j+1}(t)\rangle
\right].
\end{split}
\label{Ptrans}
\end{equation}
Indeed, this contribution vanishes in the absence of the parametric pump, namely for a squeezing parameter \(r=0\).

The overall charging efficiency is then defined as
\begin{equation}
\eta(t)=\frac{E_{\rm bat}(t)}{E_{\rm cost}(t)}
=\frac{E_\mathrm{L}(t)+E_\mathrm{R}(t)}{E_{\rm pulse}(t)+E_{\rm pump}(t)} .
\label{eta_overall}
\end{equation}
Equivalently, one may introduce the individual charging efficiencies
\(\eta_\alpha(t)=E_\alpha(t)/E_{\rm cost}(t)\) for \(\alpha=\)L,R, so that
\(\eta(t)=\eta_\mathrm{L}(t)+\eta_\mathrm{R}(t)\). As shown in Fig.~\figpanel{5_Combined_abcd}{e}, when the setup is kept on continuously, the efficiency initially increases since the amplified, ordered fluctuations are transported to the chain ends and loaded into the batteries. It subsequently decreases after \(t=t^\star\) because of the continued chain-battery energy exchange together with the accumulation of energetic cost. We therefore choose \(t^\star\), the time at which \(\eta(t)\) is maximal, as the operational switch-off time. The finite-time protocol is implemented by setting \(h=\Delta=g=0\) at \(t=t^\star\).

\color{Black}
The average charging power shown in Fig.~\figpanel{5_Combined_abcd}{f} is defined as the energy gained by the batteries per unit charging time,
\begin{equation}
P(t)=\frac{E_{\rm bat}(t)-E_{\rm bat}(0)}{t}.
\label{average_power}
\end{equation}
Correspondingly, decomposing the stored energy into coherent and fluctuation contributions, \(E_{\rm bat}=E_{\rm bat}^{\rm coh}+E_{\rm bat}^{\rm fluc}\), gives
\(
P(t)=P^{\rm coh}(t)+P^{\rm fluc}(t)
\)
with \(P^{\rm coh(fluc)}(t)=\left[E_{\rm bat}^{\rm coh(fluc)}(t)-E_{\rm bat}^{\rm coh(fluc)}(0)\right]/t\).
\color{Black}

\subsection{Ergotropy Versus Intrinsic Energy Fluctuation}
\begin{figure}[pth]
    \centering
    \includegraphics[width=0.9\linewidth]{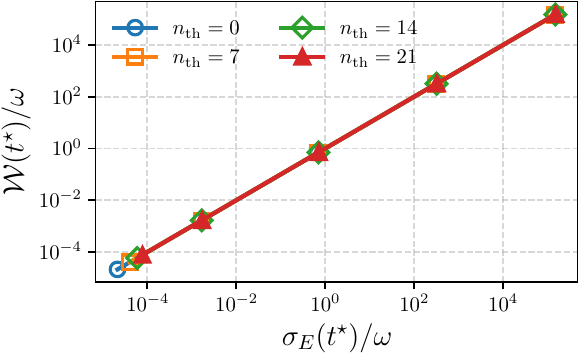}
    \caption{Ergotropy \(\mathcal{W}(t^\star)\) as a function of the battery-energy fluctuation \(\sigma_E(t^\star)\), evaluated at the optimal charging time $t^\star$. Each curve is obtained by varying the chain half-length $N$ from $2$ to $6$ at fixed bath thermal occupancy $n_{\rm th}$, as in Fig.~\ref{SNR}. The data lie almost on the same straight line, showing that \(\mathcal{W}(t^\star)\) and \(\sigma_E(t^\star)\) grow in a correlated manner as $N$ increases. All other parameters are the same as in Fig.~\ref{5_Combined_abcd}. 
\justifying}
\label{W_vs_sigmaE_Efficiency}
\end{figure}

To better understand the origin of the nearly unity work-like SNR of our protocol, in Fig.~\ref{W_vs_sigmaE_Efficiency} we replot the data of Fig.~\ref{SNR} in terms of the ergotropy \(\mathcal{W}(t^\star)\) versus the battery-energy fluctuation \(\sigma_{E}(t^\star)\), both evaluated at the optimal charging time $t^\star$. The different data points along each curve are obtained by varying the chain half-length $N$ from $2$ to $6$ (as in Fig.~\ref{SNR}), while each curve corresponds to a fixed bath thermal occupancy $n_{\rm th}$, taken to be uniform across the chain.

For all considered values of \(n_{\rm th}\), the data points lie almost on the same straight line. This nearly linear co-scaling shows that increasing the chain length enhances the extractable work and the intrinsic energy fluctuation in a correlated manner. For the smallest chain sizes, particularly \(N=2\), the data retain a visible dependence on the bath thermal occupancy \(n_{\rm th}\). As \(N\) increases, however, the curves rapidly converge, indicating that the influence of the thermal occupancy becomes progressively  negligible for the larger chains considered here. As a result, their ratio \({\rm SNR}^{\rm work}(t^\star)=\mathcal{W}(t^\star)/\sigma_E(t^\star)\) remains close to unity and becomes insensitive to \(n_{\rm th}\), consistent with the saturation behavior in Fig.~\ref{SNR}.

\color{Black}
\subsection{Relation to other nonreciprocal and chiral charging schemes}

We finally distinguish the present mechanism from previous quantum-battery proposals based on nonreciprocal or chiral energy transfer. In the reservoir-engineered nonreciprocal charging scheme~\cite{PhysRevLett.132.210402}, directionality is created by balancing the coherent and dissipative charger--battery couplings so as to suppress energy backflow. In the recently proposed chiral waveguide implementation~\cite{Chiral2026JHAn}, chirality instead originates from spin-momentum locking, which produces asymmetric couplings to oppositely propagating waveguide modes and thereby enables nonreciprocal energy transfer. Moreover, a recent giant-atom scheme~\cite{43n6-rnj3} exploits self-interference between spatially separated coupling points to suppress dissipative loss while preserving coherent energy exchange, with related configurations also enabling chiral long-range transfer. These mechanisms primarily control the directionality or loss of coherently supplied energy, without introducing parametric amplification.

In our BKC scheme, by contrast, chirality arises from the \emph{parametric pairing} and is intrinsically \emph{quadrature dependent}: the $x$ and $p$ sectors realize effective Hatano--Nelson chains with opposite directional amplification. Importantly, the active pairing process also gives rise to chiral amplification and squeezing, thereby converting passive fluctuations into ordered, non-passive energy. Therefore, the key resource here is the active chiral squeezing and fluctuation-to-work conversion, rather than the directional redistribution or protection of coherent energy alone.


\color{Black}

\clearpage
\onecolumngrid  


\setcounter{section}{0}
\setcounter{subsection}{0}
\setcounter{equation}{0}
\setcounter{figure}{0}
\setcounter{table}{0}

\renewcommand{\theequation}{S\arabic{equation}}
\renewcommand{\thefigure}{\arabic{figure}}
\renewcommand{\thetable}{S\arabic{table}}

\newcommand{\suppnote}[1]{%
    \refstepcounter{section}%
    \setcounter{subsection}{0}%
    \section*{Supplementary Note \arabic{section}. #1}%
    \addcontentsline{toc}{section}{Supplementary Note \arabic{section}. #1}%
}

\newcommand{\suppsubsection}[1]{%
    \refstepcounter{subsection}%
    \subsection*{\Alph{subsection}. #1}%
    \addcontentsline{toc}{subsection}{\Alph{subsection}. #1}%
}
\setcounter{section}{0}
\renewcommand{\thesection}{Supplementary Note \arabic{section}}

\setcounter{equation}{0}
\renewcommand{\theequation}{\arabic{equation}}

\newcommand{\R}{\mathrm{R}}
\renewcommand{\L}{\mathrm{L}}

\setcounter{figure}{0}
\setcounter{table}{0}
\renewcommand{\thefigure}{\arabic{figure}}
\renewcommand{\thetable}{\arabic{table}}

\section*{Supplementary Materials for "Charging Quantum Batteries with Chiral Squeezing"}

In these Supplementary Materials, we provide additional details that support the results presented in the main text. We omit operator hats when no confusion can arise, but retain them when distinguishing an observable from its expectation value. We also assume $\hbar=1$ throughout the Supplementary Materials.


\suppnote{Bosonic Kitaev chain and a compact description of its driven-dissipative dynamics}
\label{app:gaussian_formalism}

In this Supplementary Note, we briefly review the physical features of the bosonic Kitaev chain (BKC), discuss its stability, and present a compact Gaussian description of a driven BKC coupled to two storage modes (batteries) at its two edges. We also include (\textit{i}) a coherent pulse injected at the central site of the BKC and (\textit{ii}) Markovian dissipation and bath noise acting on each chain site and on the batteries. This formulation is convenient because it yields closed equations for the first moments and the full covariance matrix, and it makes explicit how the bath fluctuations (vacuum or thermal) are transported through the chain and can feed into the batteries via the chain--battery coupling.

\suppsubsection{Quadratures, state vector, moments, and covariance}
\label{app:quadratures}

In our proposed scheme, the dynamics remains Gaussian for a Gaussian input field, since the total Hamiltonian is quadratic as shown below \cite{Serafini2017QuantumCV}. Consequently, the full evolution is completely characterized within a Gaussian description by the first and second moments. For each bosonic mode (including the chain sites and batteries), we work with dimensionless quadratures
\begin{equation}
    x \equiv \frac{o+o^\dagger}{\sqrt{2}},\qquad
    p \equiv \frac{o-o^\dagger}{i\sqrt{2}}
\end{equation}
with $o$ ($o^\dag$) the corresponding annihilation (creation) operator,
so that $[x,p]=i$ and the vacuum covariance is $V_{\rm vac}=\frac12\mathbb{I}_2$.
The chain consists of $M=2N+1$ sites labeled $j=-N,\dots,+N$, with quadratures $\{x_j,p_j\}$. The right (left) battery is coupled to the edge site $j=+N$ ($j=-N$) and has quadratures $\{x_\R,p_\R\}$ ($\{x_\L,p_\L\}$). We collect all quadratures into a single real vector
\begin{equation}\label{Full_R_SM}
    R \equiv
    \big(x_{-N},\dots,x_{+N},\;
    p_{-N},\dots,p_{+N},\;
    x_\R,p_\R,\;
    x_\L,p_\L\big)^{\mathsf T}
    \in \mathbb{R}^{2M+4}.
\end{equation}
We also define the first moments (mean values) and the covariance matrix as
\begin{equation}\label{CovarianceMatrix_SM}
    \mu(t)\equiv \langle R(t)\rangle,
    \qquad
    V(t)\equiv \frac12\Big\langle \delta R(t)\,\delta R^{\mathsf T}(t)
    +\big(\delta R(t)\,\delta R^{\mathsf T}(t)\big)^{\mathsf T}\Big\rangle,
    \qquad
    \delta R(t)\equiv R(t)-\mu(t).
\end{equation}
Below, we introduce a compact Gaussian description of the driven BKC dynamics in terms of $\mu (t)$ and $V(t)$. 

\suppsubsection{Bosonic Kitaev chain and its dynamical stability}

The chirality and stability of the BKC are characterized by the nearest-neighbor hopping amplitude $h(t)$ and the parametric pump amplitude $\Delta(t)$; see the chain Hamiltonian $H_{\rm C}$ given in Eq.~(1) of the main text. Specifically, we define effective parameters
\begin{equation}\label{defhr}
\tilde h(t) \equiv \sqrt{h^2(t)-\Delta^2(t)},\qquad
r(t)\equiv \frac12\ln\!\left[\frac{h(t)+\Delta(t)}{h(t)-\Delta(t)}\right],
\qquad
e_\pm(t)\equiv e^{\pm r(t)}.
\end{equation}
For fixed $h$ and $\Delta$ (and in the absence of gain-saturation nonlinearities), the BKC is dynamically stable only when the spectrum of its Bogoliubov dynamical matrix is purely real.
For an \emph{open and lossless} chain, this occurs provided $|h|>|\Delta|$~\cite{PhysRevX.8.041031,Du2025prr}, which is equivalently the condition that $\tilde h=\sqrt{h^{2}-\Delta^{2}}$ defined in Eq.~\eqref{defhr} is real.
In this stable regime, one can make the stability and transport properties explicit by performing a position-dependent squeezing transformation of quadratures~\cite{PhysRevX.8.041031},
\begin{equation}
x_j \to e^{rj} \, \tilde x_j, 
\qquad
p_j \to e^{-rj} \, \tilde p_j,
\label{eq:localsqueeze}
\end{equation}
with $r$ given in Eq.~\eqref{defhr}.
For a lossless, uniform chain, the transformation in Eq.~\eqref{eq:localsqueeze} maps the BKC dynamics onto an excitation-conserving nearest-neighbor tight-binding model with effective hopping amplitude \(\tilde h/2\), as in the standard description of phase-dependent chiral transport in the bosonic Kitaev chain~\cite{PhysRevX.8.041031,Du2025prr}. The chiral amplification observed in the original quadratures then comes from the inverse position-dependent squeezing factors \(e^{\pm rj}\).
The Heisenberg equations of motion of the chain quadratures read
\begin{equation}
\dot{\tilde x}_{j}=\frac{\tilde{h}}{2}\left(\tilde x_{j-1}-\tilde x_{j+1}\right), \qquad \dot{\tilde p}_{j}=\frac{\tilde{h}}{2}\left(\tilde p_{j-1}-\tilde p_{j+1}\right).
\end{equation}
Consequently, the transformed model shows a dispersion relation $\omega_{k}=\tilde{h}\sin{k}$ and wave packets propagate with a well-defined group velocity $v_{k}=\tilde{h}\cos{k}$.
The apparent chiral amplification in the original quadratures ($\hat{x}_{j}$ and $\hat{p}_{j}$) arises from the spatially varying squeezing factors $e^{\pm rj}$.

In contrast, when $|h|<|\Delta|$, $\tilde h=\sqrt{h^2-\Delta^2}$ becomes \emph{purely imaginary} and the dynamics is parametrically unstable, leading to exponential energy growth (i.e., the system cannot support steady transport with controlled amplification)~\cite{Du2025prr}. In this regime, a well-defined group velocity is generally absent.
At the threshold $|h|=|\Delta|$, one has $\tilde h\to 0$, which marks the onset of the instability.
We therefore restrict our protocol to the stable regime, $|h(t)|>|\Delta(t)|$, throughout the charging process, which allows us to engineer phase-sensitive charging processes with controlled amplification.

\suppsubsection{Heisenberg--Langevin equations}
\label{app:hle}

Besides the lattice parameters $h(t)$ and $\Delta(t)$, we also allow time-dependent edge couplings $g_{\R}(t)$ and $g_{\L}(t)$, which are used in the protocol to switch off the charging process at a chosen time. Each chain site $a_j$ experiences linear loss at rate $\kappa_j$, and the central site $a_0$ is driven by a coherent input pulse with coupling rate $\kappa_0$. In our protocol, we always assume that the batteries are resonant with the chain (i.e., $H_{\rm bat}=0$ in the rotating framework; see the main text), and the BKC-battery interaction is given by 
\begin{equation}
H_{\rm int}(t) = g_\L(t)\left(x_{-N} x_\L + p_{-N}p_\L\right)+g_\R(t)\left(x_{N} x_\R + p_{N}p_\R\right).
\label{eq:Hint_SM}
\end{equation}

We derive the equations of motion using the standard Heisenberg--Langevin formalism associated with a Lindblad master equation. For any system operator \(O\), one may write \cite{gardiner2004quantum}
\begin{equation}
\dot O
= i[H_{\rm tot}(t),O]
+ \sum_\ell\!\left(L_\ell^\dagger O L_\ell - \frac{1}{2}\{L_\ell^\dagger L_\ell, O\}\right)
+ \sum_\ell\!\left([L_\ell^\dagger,O]\;\xi_\ell(t) + \xi_\ell^\dagger(t)\,[O,L_\ell]\right),
\label{eq:HL_general_SM}
\end{equation}
where \(H_{\rm tot}(t)\) is the total Hamiltonian of the system, \(L_\ell\) are the jump operators (defined below), and \(\xi_\ell(t)\) are the corresponding Markovian input fields.

In our model, dissipation of the \(j\)-th chain site is described by the jump operator \(L_j=\sqrt{\kappa_j}\,a_j\). Similarly, the battery dissipation is
described by \(L_{\R(\L)} = \sqrt{\gamma_{\R(\L)}}\, b_{\R(\L)}\), with $\gamma_\L$ and $\gamma_\R$ the decay rates of the left and right batteries, respectively. The coherent Gaussian pulse applied to the central chain site is included via a classical drive Hamiltonian,
\begin{equation}
H_{\rm D}(t)=\sqrt{\kappa_0}\left[p_{\rm in}(t)\,x_0 - x_{\rm in}(t)\,p_0\right],
\label{eq:Hdrive_SM}
\end{equation}
where  
\begin{equation*}
x_{\rm in}(t)=\mathcal{A}\cos\theta\, f(t), \qquad
p_{\rm in}(t)=\mathcal{A}\sin\theta\, f(t),
\end{equation*}
with \( f(t)=\exp\left[-\frac{(t-t_0)^2}{2\sigma^2}\right] \). Here, $\mathcal{A}$ is the input quadrature amplitude, $\theta$ is the
input phase, $t_0$ is the pulse-center time, and $\sigma$ denotes the
pulse width. Equivalently, the coherent input-field amplitude is
$\alpha_{\rm in}(t)
=[x_{\rm in}(t)+ip_{\rm in}(t)]/\sqrt{2}
=\mathcal{A}e^{i\theta}f(t)/\sqrt{2}$.
In this way, the commutator term \(i[H_{\rm D}(t),O]\) generates the $c$-number drive contributions
\(-\sqrt{\kappa_0}\,x_{\rm in}(t)\delta_{j,0}\) in \(\dot x_j\) and \(-\sqrt{\kappa_0}\,p_{\rm in}(t)\delta_{j,0}\) in \(\dot p_j\), as shown below. 

In addition, we include Markovian bath noise on each dissipation channel via the terms $\sqrt{\kappa_j}\,\xi_j(t)$ and $\sqrt{\gamma_{\R(\L)}}\,\zeta_{\R(\L)}(t)$. For convenience, we work in quadratures and introduce the chain quadrature noises
\begin{equation}
\xi^{(x)}_j(t)\equiv \frac{\xi_j(t)+\xi_j^\dagger(t)}{\sqrt{2}},\qquad
\xi^{(p)}_j(t)\equiv \frac{\xi_j(t)-\xi_j^\dagger(t)}{i\sqrt{2}},
\label{eq:xi_quads_SM}
\end{equation}
as well as the battery quadrature noises [when \(\gamma_{\R(\L)}\neq 0\)]
\begin{equation}
\zeta^{(x)}_{\R(\L)}(t)\equiv \frac{\zeta_{\R(\L)}(t)+\zeta_{\R(\L)}^\dagger(t)}{\sqrt{2}},\qquad
\zeta^{(p)}_{\R(\L)}(t)\equiv \frac{\zeta_{\R(\L)}(t)-\zeta_{\R(\L)}^\dagger(t)}{i\sqrt{2}}.
\label{eq:zeta_quads_SM}
\end{equation}
With these conventions and the total Hamiltonian $H_{\rm tot}(t)=H_{\rm C}(t)+H_{\rm int}(t)+H_{\rm D}(t)$, the Heisenberg--Langevin equations for the chain read
\begin{equation}\label{xp_chain}
\begin{aligned}
    \dot x_j &=
    \frac{\tilde h(t)}{2}\Big[e_{+}(t)\,x_{j-1}-e_{-}(t)\,x_{j+1}\Big]
    -\frac{\kappa_j}{2}\,x_j
    -\sqrt{\kappa_0}\,x_{\rm in}(t)\,\delta_{j,0}
    +\sqrt{\kappa_j}\,\xi^{(x)}_j(t)
    + g_\R(t)\,p_\R\,\delta_{j,+N}
    + g_\L(t)\,p_\L\,\delta_{j,-N},\\
    \dot p_j &=
    \frac{\tilde h(t)}{2}\Big[e_{-}(t)\,p_{j-1}-e_{+}(t)\,p_{j+1}\Big]
    -\frac{\kappa_j}{2}\,p_j
    -\sqrt{\kappa_0}\,p_{\rm in}(t)\,\delta_{j,0}
    +\sqrt{\kappa_j}\,\xi^{(p)}_j(t)
    - g_\R(t)\,x_\R\,\delta_{j,+N}
    - g_\L(t)\,x_\L\,\delta_{j,-N},
\end{aligned}
\end{equation}
while the batteries follow
\begin{equation}\label{storLR}
\begin{aligned}
    \begin{pmatrix}\dot x_\R\\ \dot p_\R\end{pmatrix}
    &=
    \begin{pmatrix}
    -\gamma_\R/2 & 0\\
    0 & -\gamma_\R/2
    \end{pmatrix}
    \begin{pmatrix}x_\R\\ p_\R\end{pmatrix}
    +
    g_\R(t)
    \begin{pmatrix}
    p_{+N}\\
    -x_{+N}
    \end{pmatrix}
    +\sqrt{\gamma_\R}
    \begin{pmatrix}
    \zeta^{(x)}_\R(t)\\
    \zeta^{(p)}_\R(t)
    \end{pmatrix},
    \\[1mm]
    \begin{pmatrix}\dot x_\L\\ \dot p_\L\end{pmatrix}
    &=
    \begin{pmatrix}
    -\gamma_\L/2 & 0\\
    0 & -\gamma_\L/2
    \end{pmatrix}
    \begin{pmatrix}x_\L\\ p_\L\end{pmatrix}
    +
    g_\L(t)
    \begin{pmatrix}
    p_{-N}\\
    -x_{-N}
    \end{pmatrix}
    +\sqrt{\gamma_\L}
    \begin{pmatrix}
    \zeta^{(x)}_\L(t)\\
    \zeta^{(p)}_\L(t)
    \end{pmatrix}.
\end{aligned}
\end{equation}
Equations~(\ref{xp_chain}) and (\ref{storLR}) reduce to Eqs.~(3) and (4) in the main text when $\gamma_\R=\gamma_\L=\gamma$. In our work, the battery dissipation and noise terms are optional; in many simulations we set $\gamma_{\R/\L}=0$ so that the batteries are lossless and receive noise only through their coupling to the chain.

\suppsubsection{Markovian noise statistics}

For a Markovian bath, the input noises have zero mean and are delta-correlated in time. Thus, we have \cite{gardiner2004quantum}
\begin{equation}\label{noisexi}
    \left\langle \xi^{(x)}_j(t) \right\rangle=0,\qquad
    \left\langle \xi^{(p)}_j(t) \right\rangle=0,
\end{equation}
and the (symmetrized) correlations
\begin{equation}
    \left\langle \xi^{(x)}_j(t)\,\xi^{(x)}_k(t') \right\rangle
    = \delta_{jk}\left(n_{\rm th}+\frac12\right)\delta(t-t'),
    \qquad
    \left\langle \xi^{(p)}_j(t)\,\xi^{(p)}_k(t') \right\rangle
    = \delta_{jk}\left(n_{\rm th}+\frac12\right)\delta(t-t'),
\label{app:markov_noise_corr}
\end{equation}
where $n_{\rm th}$ is the thermal occupancy of the chain bath. For a thermal bath, there are no cross-correlations between the $x$- and $p$-quadrature noises. The same structure applies to the battery noises $\zeta_{\R}^{(x/p)}$ and $\zeta_{\L}^{(x/p)}$.

\suppsubsection{Compact linear form}
\label{app:compact_linear}

Equations~\eqref{xp_chain} and \eqref{storLR} can be written in the compact linear form
\begin{equation}
\dot R(t)=A(t)\,R(t)+C(t)+G\,\Xi(t),
\label{app:linear_R}
\end{equation}
where $R(t)\in\mathbb{R}^{2M+4}$ ($M=2N+1$ as introduced above) collects all system quadratures (chain and batteries), $C(t)$ encodes the coherent input pulse at the central site, $G$ is the noise-injection coefficient matrix, and $\Xi(t)$ collects all independent Markovian noise quadratures. The number of noise channels is $N_{\rm noise}\equiv \dim\Xi$ and depends on which baths are included (chain baths only, or chain plus battery baths, etc.). For instance, if only the chain baths are included and each chain site carries two independent noise quadratures $\{\xi^{(x)}_j,\xi^{(p)}_j\}_{j=-N}^{N}$, then $N_{\rm noise}=2M$ and one may choose the ordering
\begin{equation}
\Xi(t)\equiv\big(\xi^{(x)}_{-N},\ldots,\xi^{(x)}_{+N},\;\xi^{(p)}_{-N},\ldots,\xi^{(p)}_{+N}\big)^{\mathsf T}\in\mathbb{R}^{2M}.
\end{equation}
In this case, the noise-injection matrix $G\in\mathbb{R}^{(2M+4)\times(2M)}$ takes the explicit block-diagonal form
\begin{equation}
G=
\begin{pmatrix}
\mathrm{diag}\big(\sqrt{\kappa_j}\big) & 0\\
0 & \mathrm{diag}\big(\sqrt{\kappa_j}\big)\\
0 & 0\\
0 & 0
\end{pmatrix},
\label{app:G_example}
\end{equation}
where the diagonal matrices act on the chain $x$- and $p$-blocks, and the last four rows (corresponding to the battery quadratures) vanish when no baths are introduced to the batteries. If battery baths are included, additional noise components are appended to $\Xi(t)$ and the corresponding columns are added to $G$, with nonzero entries in the battery rows.

The drift matrix $A(t)\in\mathbb{R}^{(2M+4)\times(2M+4)}$ has the block structure
\begin{equation}
A(t)=
\begin{pmatrix}
A_{x_{\rm C}, x_{\rm C}}(t) & 0 & A_{x_{\rm C}, \R}(t) & A_{x_{\rm C}, \L}(t) \\
0 & A_{p_{\rm C}, p_{\rm C}}(t) & A_{p_{\rm C}, \R}(t) & A_{p_{\rm C}, \L}(t) \\
A_{\R, x_{\rm C}}(t) & A_{\R, p_{\rm C}}(t) & A_{\R,\R}(t) & 0 \\
A_{\L, x_{\rm C}}(t) & A_{\L, p_{\rm C}}(t) & 0 & A_{\L,\L}(t)
\end{pmatrix}.
\label{app:A_block}
\end{equation}
Here the block labels refer to the chain quadrature vectors \(x_{\rm C}=(x_{-N},\ldots,x_{+N})^{\mathrm T}\) and \(p_{\rm C}=(p_{-N},\ldots,p_{+N})^{\mathrm T}\), and to the right- and left-battery quadrature vectors \(\R=(x_\R,p_\R)^{\mathrm T}\) and \(\L=(x_\L,p_\L)^{\mathrm T}\), respectively. The diagonal blocks \(A_{x_{\rm C}, x_{\rm C}}\) and \(A_{p_{\rm C}, p_{\rm C}}\) describe the intrinsic BKC dynamics of the two quadrature sectors, \(A_{\R,\R}\) and \(A_{\L,\L}\) describe the local battery damping, and the off-diagonal blocks encode the chain--battery couplings. Specifically, the nonzero elements of the chain blocks are 
\begin{equation}\label{Axp}
\begin{aligned}
    (A_{x_{\rm C},x_{\rm C}})_{j,j}=-\frac{\kappa_j}{2},\qquad\quad
    (A_{x_{\rm C},x_{\rm C}})_{j,j-1}=\frac{\tilde h(t)}{2}e_{+}(t),\qquad\quad
    (A_{x_{\rm C},x_{\rm C}})_{j,j+1}=-\frac{\tilde h(t)}{2}e_{-}(t),
    \\
    (A_{p_{\rm C},p_{\rm C}})_{j,j}=-\frac{\kappa_j}{2},\qquad\quad
    (A_{p_{\rm C},p_{\rm C}})_{j,j-1}=\frac{\tilde h(t)}{2}e_{-}(t),\qquad\quad
    (A_{p_{\rm C},p_{\rm C}})_{j,j+1}=-\frac{\tilde h(t)}{2}e_{+}(t).
\end{aligned}
\end{equation}
The local battery blocks are
\begin{equation}
    A_{\R,\R}=
\begin{pmatrix}
    -\gamma_\R/2 & 0 \\
    0 & -\gamma_\R/2
\end{pmatrix},
\qquad
A_{\L,\L}=
\begin{pmatrix}
    -\gamma_\L/2 & 0 \\
    0 & -\gamma_\L/2
\end{pmatrix}.
\end{equation}
The off-diagonal blocks are sparse and have nonzero entries only at the edge sites $j=\pm N$, as dictated in Eqs.~(\ref{xp_chain}) and (\ref{storLR}).

The drive vector $C(t)$ is nonzero only in the entries corresponding to the central-site quadratures $x_0$ and $p_0$:
\begin{equation}
    C_{x_0}(t)=-\sqrt{\kappa_0}\,x_{\rm in}(t),\qquad
    C_{p_0}(t)=-\sqrt{\kappa_0}\,p_{\rm in}(t),
    \qquad
    C_{x_j}(t)=C_{p_j}(t)=0\ \text{for all other } j.
\end{equation}

\suppsubsection{Gaussian moment and covariance equations}
\label{gaussian_equations}

Taking the expectation value of Eq.~(\ref{app:linear_R}) and using $\langle \Xi(t)\rangle=0$ gives the closed first-moment equation
\begin{equation}
    \dot \mu(t)=A(t)\,\mu(t)+C(t),
\label{app:mu_eq}
\end{equation}
where $\mu(t)\equiv \langle R(t)\rangle$. Instead, the covariance matrix obeys the Lyapunov equation
\begin{equation}
    \dot V(t)=A(t)\,V(t)+V(t)\,A^{\mathsf T}(t)+D(t),
\label{app:V_eq}
\end{equation}
where the diffusion matrix $D(t)$ is determined by the noise correlations \cite{Serafini2017QuantumCV}. For independent Markovian baths, as in Eq.~(\ref{app:markov_noise_corr}), $D(t)$ is block diagonal in the quadrature basis. In particular, each chain site contributes
\begin{equation}\label{Djj}
D_{x_j, x_j}(t)=D_{p_j, p_j}(t)=\kappa_j\Big(n_{\rm th}+\frac12\Big),
\end{equation}
and each battery mode contributes (when $\gamma_{\R/\L}\neq 0$)
\begin{equation}\label{DRL}
D_\R=\gamma_\R\Big(n_\R+\frac12\Big)\mathbb{I}_2,\qquad
D_\L=\gamma_\L\Big(n_\L+\frac12\Big)\mathbb{I}_2.
\end{equation}
where $n_{\R}$ and $n_{\L}$ are the thermal occupancies of the right- and left-battery baths, respectively. This makes explicit how bath fluctuations injected into the chain (vacuum or thermal) can propagate to the battery modes through the chain-battery coupling blocks of $A(t)$, or equivalently, via the off-diagonal chain-battery covariance blocks of $V(t)$ in Eq.~\eqref{app:V_eq}.
Once $\mu(t)$ and $V(t)$ are known, all second moments follow from
\begin{equation}\label{Rab}
\frac{1}{2}\langle R_a R_b + R_b R_a \rangle = \mu_a \mu_b + V_{ab}.
\end{equation}
This identity is used throughout to compute energy and power flows in the present Gaussian model.

When the drift matrix $A(t)$ is \emph{phase sensitive}, as in the present BKC case, the injected vacuum fluctuations are generally reshaped into an ellipse, showing a squeezing-like process. Even if the input noise is isotropic (for vacuum noise $D \propto I$), the evolution generated by $A(t)$ can render the covariance matrix $V(t)$ anisotropic. That is apparent from the Lyapunov equation
\begin{equation}\nonumber
\dot{V} = AV + VA^\top + D,
\end{equation}
where the drift term $AV + VA^\top$ can transform an initially ``round'' covariance into a squeezed one whenever $A$ acts differently on the $x$- and $p$-quadratures.

Mathematically, the formal solutions to Eqs.~\eqref{app:mu_eq} and \eqref{app:V_eq} explicitly show the distinction between the coherent and fluctuation inputs:
\begin{equation}
    \mu(t) = \Phi(t,0)\mu(0) + \int_0^t \Phi(t,s)\, C(s)\, ds,
\end{equation}
and
\begin{equation}
    V(t) = \Phi(t,0)V(0)\Phi^\top(t,0) + \int_0^t \Phi(t,s)\, D(s)\, \Phi^\top(t,s)\, ds,
\end{equation}
where $\Phi(t,s)$ is the drift propagator, defined by
\begin{equation}\label{driftpropagator}
    \partial_t \Phi(t,s) = A(t)\,\Phi(t,s), \qquad \Phi(s,s) = I.
\end{equation}
Thus, the coherent sector is driven by a localized, finite-duration source, $C$, while the covariance sector is continuously fed by the diffusion matrix $D$, which collects fluctuations from all dissipation channels (associated with all chain sites). After being injected, these fluctuations are transferred and reshaped by $\Phi(t,s)$ in a direction-dependent manner. This is why the ordered fluctuation energy can become much larger than the coherent energy.

The batteries are coupled to the chain edges through beam-splitter interactions $H_{\rm int}$, which, in the quadrature representation, couples the edge and battery quadratures. As a result, the statistical properties of the edge modes, including squeezing and correlations, are partially transferred (``imprinted'') onto the battery covariance matrix. Consequently, even if the batteries start in vacuum, they can eventually have squeezed fluctuations due to the (phase-sensitive) chiral dynamics of the chain. This mechanism directly affects the subsequent work-extraction properties of the batteries, as discussed below.

\suppnote{Figures of merit of the charging process}\label{FOM_SM}


For each battery mode, we quantify the stored energy above vacuum in terms of its Gaussian first and second moments. The parametric pump (proportional to $\Delta$) makes the chain behave like a distributed parametric amplifier. This amplifier reshapes and amplifies the injected fluctuations in a chiral manner, and the resulting correlations propagate to the boundaries and enter the battery modes through the chain-battery coupling.

To make this mechanism explicit, we start from the full quadrature vector $R$ of the whole system, as given in Eq.~\eqref{Full_R_SM}.
We recall the covariance matrix $V(t)$ in Eq.~\eqref{CovarianceMatrix_SM} and express it in block form as
\begin{equation}
    V=
    \begin{pmatrix}
    V_{{\rm C},{\rm C}} & V_{{\rm C},\R} & V_{{\rm C},\L}\\
    V_{\R,{\rm C}} & V_{\R,\R} & V_{\R,\L}\\
    V_{\L,{\rm C}} & V_{\L,\R} & V_{\L,\L}
    \end{pmatrix},
\label{V_block_partition}
\end{equation}
where the subscript ``$\rm C$'' denotes the chain and ``R (L)'' corresponds to the right (left) battery. The covariance matrix evolves according to the Lyapunov equation~\eqref{app:V_eq} and, as an example, the right-battery block obeys the equation
\begin{equation}
    \dot V_{\R,\R}=A_{\R,\R}V_{\R,\R}+V_{\R,\R}A_{\R,\R}^{\mathsf T}+A_{\R,{\rm C}}V_{{\rm C},\R}+V_{\R,{\rm C}}A_{\R,{\rm C}}^{\mathsf T}+D_{\R}.
\label{VRR_exact}
\end{equation}
The left-battery block follows a similar equation with R $\to$ L.
Equation~\eqref{VRR_exact} shows that the chain fluctuations injected through the dissipative channels (encoded in $D$) generate nonzero chain-battery cross-covariances (e.g., $V_{\rm C,\R}$), which then feed into the battery covariance blocks (e.g., $V_{\R,\R}$) even when the batteries are lossless. 

\suppsubsection{A single-mode example}

Having established how noise enters the batteries, we now characterize the energy stored in a single battery mode in terms of its $2\times 2$ covariance block. Since the following derivation applies to an arbitrary single-mode Gaussian state, we use the label $B$ to denote a \emph{generic bosonic mode}. For a single bosonic mode with resonance frequency $\omega_{B}$ and quadrature vector
\begin{equation}
    R_B\equiv (X,P)^{\mathsf T},\qquad [X,P]=i,
\end{equation}
we define the first moments $\mu_B(t)\equiv \langle R_B(t)\rangle$ and the corresponding covariance matrix
\begin{equation}
    V_B \equiv \frac{1}{2}\Big\langle \delta R_B\,\delta R_B^{\mathsf T}
    +(\delta R_B\,\delta R_B^{\mathsf T})^{\mathsf T}\Big\rangle,\qquad \delta R_B\equiv R_B-\mu_B.
\end{equation}
The energy operator is given by 
\begin{equation}
    \hat{E}_{B}\equiv \omega_{B}\hat{b}^\dag\hat{b}=\frac{1}{2}\omega_{B}\left(\hat{X}^2+\hat{P}^2-1\right)
\label{E_def}
\end{equation}
and we denote its mean value by $E_{B}=\langle\hat{E}_B\rangle$. The vacuum covariance is given by $V_{B, \rm vac}=\frac12\mathbb{I}_2$.

Using $\langle X^2\rangle = \langle X\rangle^2 + \mathrm{Var}(X)$ (and similarly for $P$), together with $\mathrm{Var}(X)=V_{X,X}$ and $\mathrm{Var}(P)=V_{P,P}$, we obtain
\begin{equation}
    E_B = \underbrace{\frac12\omega_{B}\mu_B^{\mathsf T}\mu_B}_{E_{B}^{\rm coh}} + \underbrace{\frac12\omega_{B}\big(\mathrm{Tr}[V_B]-1\big)}_{E_{B}^{\rm fluc}}.
\label{E_coh_fluc}
\end{equation}
Here $E_{B}^{\rm coh}$ is the coherent energy stored in the mean field, while $E_{B}^{\rm fluc}$ is the fluctuation energy stored in the second moments.

Let $\Omega_{\rm sp}\equiv \begin{pmatrix}0&1\\-1&0\end{pmatrix}$ denote the single-mode symplectic form.  For any physical one-mode covariance matrix $V_B>0$ satisfying the Robertson--Schr\"odinger uncertainty relation $V_B+\frac{i}{2}\Omega_{\rm sp}\ge 0$, Williamson's theorem (for one mode) implies the existence of a symplectic matrix $S_B\in \mathrm{Sp}(2,\mathbb{R})$ (a rotation followed by a squeeze transformation) such that
\begin{equation}
    V_B = S_B\,(\nu_B\,\mathbb{I}_2)\,S_B^{\mathsf T},
\label{williamson_1mode}
\end{equation}
with the symplectic eigenvalue
\begin{equation}
    \nu_B=\sqrt{\det[V_B]}\ge \frac12.
\label{nu_det}
\end{equation}
Indeed, taking determinants of Eq.~\eqref{williamson_1mode} and using $\det S_B=1$ gives $\det[V_B]=\nu_B^2$.

\subsection*{Thermal-like (incoherent) contribution}

A thermal state with occupancy $n_{\rm th}$ has an isotropic covariance, i.e., $V_{\rm th}=(n_{\rm th}+\frac12)\mathbb{I}_2$, and therefore a symplectic eigenvalue $\nu_{\rm th}=n_{\rm th}+\frac12$. Motivated by the Williamson's normal form in Eq.~\eqref{williamson_1mode}, we define the effective thermal occupancy of a general single-mode Gaussian state by
\begin{equation}
    \nu_B \equiv n_{B,\rm th}^{\rm (eff)}+\frac12
    \qquad\Longleftrightarrow\qquad n_{B,\rm th}^{\rm (eff)} = \nu_B-\frac12.
\label{nth_eff}
\end{equation}
The corresponding thermal-like (incoherent) energy above vacuum is then obtained as
\begin{equation}
    E_{B}^{\rm th}\;\equiv\;\omega_{B}\,n_{B,\rm th}^{\rm (eff)}=\omega_{B}\left(\nu_B-\frac12\right).
\label{Eth_def}
\end{equation}

\subsection*{Squeezed/ordered contribution}

Besides the thermal-like contribution $E_{B}^{\rm th}$, the fluctuation energy also contains an \emph{ordered}, squeezing-like component, i.e., 
\begin{equation}
    E_{B}^{\rm sq}\;\equiv\;E_{B}^{\rm fluc}-E_{B}^{\rm th}.
\label{Esq_def1}
\end{equation}
Combining Eqs.~\eqref{E_coh_fluc} and \eqref{Eth_def}, we obtain
\begin{equation}
    E_{B}^{\rm sq} = \frac12\omega_{B}\big(\mathrm{Tr}[V_B]-1\big)-\omega_{B}\Big(\nu_B-\frac12\Big)
    = \frac12\omega_{B}\mathrm{Tr}[V_B]-\omega_{B}\nu_B.
\label{Esq_def2}
\end{equation}

\textit{Non-negativity of $E_{B}^{\rm sq}$}---Since $V_B$ is a real, symmetric, positive-definite $2\times 2$ matrix, its eigenvalues $\lambda_{1,2}>0$ obey the inequality
\begin{equation}
    \frac{\lambda_1+\lambda_2}{2}\ge \sqrt{\lambda_1\lambda_2}.
\end{equation}
Using $\mathrm{Tr}[V_B]=\lambda_1+\lambda_2$ and $\det[V_B]=\lambda_1\lambda_2=\nu_B^2$ [cf.\ Eq.~\eqref{nu_det}], we obtain
\begin{equation}
    \frac12\,\mathrm{Tr}[V_B] \ge \sqrt{\det[V_B]}=\nu_B
    \qquad\Longrightarrow\qquad
    E_{B}^{\rm sq}=\frac12\omega_{B}\mathrm{Tr}[V_B]-\omega_{B}\nu_B\ge 0.
\label{Esq_nonneg}
\end{equation}
Equality holds if and only if $\lambda_1=\lambda_2$, i.e., if and only if $V_B \propto \mathbb{I}_2$ (no squeezing or anisotropy). 

In this way, the total mean energy can be decomposed as
\begin{equation}
    E_B = \underbrace{\frac12\omega_{B}\mu_B^{\mathsf T}\mu_B}_{E_{B}^{\rm coh}}
    + \underbrace{\omega_{B}\Big(\frac12\,\mathrm{Tr}[V_B]-\nu_B\Big)}_{E_{B}^{\rm sq}}
    + \underbrace{\omega_{B}\Big(\nu_B-\frac12\Big)}_{E_{B}^{\rm th}}.
\label{E_total_split}
\end{equation}
In particular, $E_{B}^{\rm th}$ quantifies the passive, thermal-like contribution, while $E_{B}^{\rm sq}$ quantifies energy stored in anisotropic (squeezed) fluctuations.

\subsection*{Ergotropy}

For a single-mode Gaussian battery, the ergotropy can be expressed explicitly in terms of the above decomposition. By definition, the ergotropy is the difference between the mean energy $E_B$ of the battery state $\rho_B$ and the mean energy $E_{B}^{\rm pa}$ of the passive state $\rho_{B}^{\rm pa}$ (associated with $\rho_B$),
\begin{equation}
\mathcal W_B \equiv E_B - E_{B}^{\rm pa}.
\label{eq:WB_def}
\end{equation}
For a single-mode Gaussian state, the passive state is the centered thermal state with the same symplectic eigenvalue \(\nu_B=\sqrt{\det[V_B]}\), and therefore
\begin{equation}
E_{B}^{\rm pa}=\omega_{B}\left(\nu_B-\frac12\right)=E_{B}^{\rm th}.
\label{eq:EB_passive}
\end{equation}
Using Eq.~\eqref{E_total_split}, we immediately obtain
\begin{equation}
\mathcal W_B
=
E_B-\omega_{B}\left(\nu_B-\frac12\right)
=
\frac12\omega_{B}\mu_B^{\mathsf T}\mu_B+\frac12\omega_{B}\mathrm{Tr}[V_B]-\omega_{B}\nu_B
=
E_{B}^{\rm coh}+E_{B}^{\rm sq}.
\label{eq:WB_gaussian}
\end{equation}
Thus, \(E_{B}^{\rm th}\) is the passive, non-extractable part of the stored energy, whereas both the coherent contribution \(E_{B}^{\rm coh}\) and the squeezing contribution \(E_{B}^{\rm sq}\) are unitary-extractable. In particular, \(E_{B}^{\rm sq}\) can be removed by an appropriate inverse squeezing operation, while the coherent contribution \(E_{B}^{\rm coh}\) can be removed by a displacement.

\suppsubsection{Energies, powers, and efficiencies}
\label{app:energies_powers_gaussian}

In a Gaussian description, all second moments can be expressed in terms of the first moments and the covariance matrix as
\begin{equation}
\frac{1}{2}\langle R_a R_b + R_b R_a \rangle=\mu_a\mu_b+V_{ab},
\label{eq:second_moments_muV}
\end{equation}
where the covariance matrix is partitioned as in Eq.~\eqref{V_block_partition}. Below we summarize the expressions for the main figures of merit used in the main text, including energy, ergotropy, power, and efficiency, in a form that remains valid in the presence of vacuum or thermal noise. The total battery ergotropy for our BKC charging scheme is
\begin{equation}
\mathcal W(t)=\mathcal W_\L(t)+\mathcal W_\R(t).
\label{eq:Wbatt_def}
\end{equation}
Here, \(\mathcal W_{\L}(t)\) and \(\mathcal W_{\R}(t)\) denote the ergotropies of the left and right batteries, respectively, which can be computed from Eq.~\eqref{eq:WB_gaussian} for the corresponding reduced single-mode Gaussian states.

\paragraph{Chain energy above vacuum.}
The energy above vacuum of the BKC is written as
\begin{align}
E_{\rm C}(t)
&\equiv \frac12\omega_{\rm C}\sum_{j=-N}^{N}\Big(\langle x_j^2(t)\rangle+\langle p_j^2(t)\rangle-1\Big)
\nonumber\\
&= \frac12\omega_{\rm C}\sum_{j=-N}^{N}\Big(\mu_{x_j}^2(t)+\mu_{p_j}^2(t)+V_{x_jx_j}(t)+V_{p_jp_j}(t)-1\Big),
\label{eq:Echain_muV}
\end{align}
with $\mu_{x_j}=\langle x_j\rangle$ and $\mu_{p_j}=\langle p_j\rangle$. $\omega_{\rm C}$ is the uniform resonance frequency of each chain site.

\paragraph{Battery energies above vacuum.}
For a battery mode $\alpha\in\{\L,\R\}$ with resonance frequency $\omega$ (we assume $\omega=\omega_{\rm C}$ in the main text), quadratures $(X_\alpha,P_\alpha)$, and reduced $2\times2$ covariance matrix $V_\alpha$ (i.e., the corresponding sub-block of the full covariance matrix), its energy above vacuum is
\begin{equation}
E_\alpha(t)=\frac12\omega\Big(\mu_{X_\alpha}^2(t)+\mu_{P_\alpha}^2(t)\Big)
+\frac12\omega\Big(\mathrm{Tr}\,[V_\alpha(t)]-1\Big).
\label{eq:Ebatt_single_muV}
\end{equation}
The total energy of the two batteries is given by
\begin{equation}
E_{\rm bat}(t)=E_\L(t)+E_\R(t).
\label{eq:E_battery}
\end{equation}


To compute the above quantities, we start from the chain energy $E_{\rm C}(t)$ given in Eq.~(\ref{eq:Echain_muV}).
Differentiating $E_{\rm C}(t)$ yields
\begin{equation}
\dot E_{\rm C}(t)
= \frac12\omega_{\rm C}\sum_{j=-N}^{N}\Big\langle
x_j\dot x_j+\dot x_j x_j
+p_j\dot p_j+\dot p_j p_j
\Big\rangle.
\label{eq:Echain_dot_sym}
\end{equation}
This symmetrized form is useful because it remains valid even if the on-site potential term,
\(\sum_j\omega_{\rm C} a_j^\dagger a_j\), is explicitly retained in $H_{\rm C}$. Such a rotation only mixes \(x_j\) and \(p_j\) locally and therefore does not generate an additional contribution to \(\dot E_{\rm C}\), since it conserves \(x_j^2+p_j^2\). To identify the different energy-flow channels, we now substitute the Langevin equations in Eq.~\eqref{xp_chain}, written in the rotating frame where the on-site frequency is taken as the reference, into Eq.~\eqref{eq:Echain_dot_sym}. This gives
\begin{equation}
\dot E_{\rm C}(t)
=
P_{x\rm -pump}(t) + P_{p\rm -pump}(t) - P_{\rm diss}(t) + P_{\rm pulse}(t) + P_{\rm noise}(t) + P_{\rm bat \to \rm C}(t),
\label{eq:Echain_dot_decomp}
\end{equation}
where
\begin{subequations}\label{eq:P_terms_def}
\begin{align}
P_{x{\rm -pump}}(t)/\omega_{\rm C}
&\equiv
\frac{\tilde h(t)}{2}\sum_{j=-N}^{N}
\Big(
e^{r(t)}\langle x_j x_{j-1}\rangle
-
e^{-r(t)}\langle x_j x_{j+1}\rangle
\Big),
\label{eq:Px_def}
\\[2mm]
P_{p{\rm -pump}}(t)/\omega_{\rm C}
&\equiv
\frac{\tilde h(t)}{2}\sum_{j=-N}^{N}
\Big(
e^{-r(t)}\langle p_j p_{j-1}\rangle
-
e^{r(t)}\langle p_j p_{j+1}\rangle
\Big),
\label{eq:Pp_def}
\\[2mm]
P_{\rm diss}(t)/\omega_{\rm C}
&\equiv
\frac12\sum_{j=-N}^{N}\kappa_j
\Big(
\langle x_j^2\rangle+\langle p_j^2\rangle
\Big),
\label{eq:Pdiss_def_clean}
\\[2mm]
P_{\rm pulse}(t)/\omega_{\rm C}
&\equiv
-\sqrt{\kappa_0}\Big[x_{\rm in}(t)\,\langle x_0\rangle+p_{\rm in}(t)\,\langle p_0\rangle\Big],
\label{eq:Pcenter_def_clean}
\\[2mm]
P_{\rm noise}(t)/\omega_{\rm C}
&\equiv
\sum_{j=-N}^{N}\sqrt{\kappa_j}
\Big(
\langle x_j\,\xi^{(x)}_j(t)\rangle+\langle p_j\,\xi^{(p)}_j(t)\rangle
\Big),
\label{eq:Pnoise_def_clean}
\\[2mm]
P_{\rm bat \to \rm C}(t)/\omega_{\rm C}
&\equiv
g_\R(t)\Big(\langle x_{N} p_\R\rangle-\langle p_{N} x_\R\rangle\Big)
+
g_\L(t)\Big(\langle x_{-N} p_\L\rangle-\langle p_{-N} x_\L\rangle\Big).
\label{eq:Pstor_def_clean}
\end{align}
\end{subequations}

It is convenient to combine the two transport contributions, $P_{x{\rm -pump}}(t)$ and $P_{p{\rm -pump}}(t)$, into a total transport power
\begin{eqnarray}
P_{\rm pump}(t) &\equiv& P_{x{\rm -pump}}(t)+P_{p{\rm -pump}}(t) \nonumber\\
&=& \frac{\omega_{\rm C}\tilde h(t)}{2}\sum_{j=-N}^{N} \Big(e^{r}\langle x_j x_{j-1}\rangle - e^{-r}\langle x_j x_{j+1}\rangle \,+\, e^{-r}\langle p_j p_{j-1}\rangle - e^{r}\langle p_j p_{j+1}\rangle\Big).
\label{eq:transport_split_gain_ref}
\end{eqnarray}
By shifting the summation index $j\mapsto j+1$ in the term containing $\langle x_{j}x_{j-1} \rangle$, and using the open-boundary condition $x_{N+1}=x_{-N-1}=0$, we obtain
\begin{equation}
P_{x{\rm -pump}}(t)
=\frac{\omega_{\rm C}\tilde h(t)}{2}\Big(e^{r}-e^{-r}\Big)\sum_{j=-N}^{N-1} \langle x_j\,x_{j+1}\rangle.
\label{eq:Px_final}
\end{equation}
Similarly, 
\begin{equation}
P_{p{\rm -pump}}(t)
=-\frac{\omega_{\rm C}\tilde h(t)}{2}\Big(e^{r}-e^{-r}\Big)\sum_{j=-N}^{N-1} \langle p_j\,p_{j+1}\rangle.
\label{eq:Pp_final}
\end{equation}
As a result, the total transport power can be simplified to
\begin{eqnarray}
P_{\rm pump}(t)
&=&\frac{\omega_{\rm C}\tilde h(t)}{2}\Big(e^{r}-e^{-r}\Big)\sum_{j=-N}^{N-1}
\Big[\langle x_j(t)\,x_{j+1}(t)-p_j(t)\,p_{j+1}(t)\rangle\Big] \nonumber\\
&=&\omega_{\rm C}\tilde h(t)\,\sinh r(t) \sum_{j=-N}^{N-1}\Big[\langle x_jx_{j+1}\rangle - \langle p_jp_{j+1}\rangle\Big],
\label{eq:Ptransport_final}
\end{eqnarray}
where $\tilde h(t)$ and $r(t)$ are determined by hopping $h(t)$ and pump $\Delta(t)$; see Eq.~(\ref{defhr}). In particular, one has $P_{\rm pump}(t)=0$ in the absence of the parametric pump ($r=0$). We therefore interpret $P_{\rm pump}(t)$ as an instantaneous pump-associated power, and define the corresponding pump energy as
\begin{equation}
E_{\rm pump}(t)=\int_{0}^{t} P_{\rm pump}(t')\,dt'.
\label{eq:Egain_def}
\end{equation}

%


Similarly, the energy cost associated with the coherent input field is
\begin{equation}
E_{\rm pulse}(t)=\int_{0}^{t}P_{\rm pulse}(t')\,dt',
\qquad
\label{eq:E_center}
\end{equation}
and the total (active) energy cost for our charging protocol is given by
\begin{equation}
E_{\rm cost}(t) = E_{\rm pump}(t)+E_{\rm pulse}(t).
\label{eq:Ecost_def_app}
\end{equation}

Within the Gaussian model, all correlators entering the above expressions are evaluated using $\langle R_aR_b\rangle=\mu_a\mu_b+V_{ab}$ [cf. Eq.~\eqref{eq:second_moments_muV}], including the nearest-neighbor site moments $\langle x_j x_{j+1}\rangle$ and $\langle p_j p_{j+1}\rangle$ and the chain-battery cross moments appearing in $P_{\rm bat \to C}(t)$.
In particular,
\begin{equation}
\langle x_jx_{j+1}\rangle=\mu_{x_j}\mu_{x_{j+1}}+V_{x_j,x_{j+1}},
\qquad
\langle p_jp_{j+1}\rangle=\mu_{p_j}\mu_{p_{j+1}}+V_{p_j,p_{j+1}}.
\label{eq:bond_moments_muV}
\end{equation}

The instantaneous power flowing into the battery modes through the chain-battery couplings is given in Eq.~\eqref{eq:Pstor_def_clean}, with the required correlators evaluated as
\begin{align}
\langle x_{N}p_\R\rangle &= \mu_{x_{N}}\mu_{p_\R}+V_{x_{N},p_\R}, &
\langle p_{N}x_\R\rangle &= \mu_{p_{N}}\mu_{x_\R}+V_{p_{N},x_\R},
\nonumber\\
\langle x_{-N}p_\L\rangle &= \mu_{x_{-N}}\mu_{p_\L}+V_{x_{-N},p_\L}, &
\langle p_{-N}x_\L\rangle &= \mu_{p_{-N}}\mu_{x_\L}+V_{p_{-N},x_\L}.
\label{eq:cross_moments_muV}
\end{align}
These terms make explicit how the chain noise (encoded in the full covariance matrix) can feed into the battery modes even if $\gamma_{\L}=\gamma_{\R}=0$. When the batteries are assumed to be lossless and initially in vacuum, the total stored energy can be obtained by defining $P_{\rm C \to \rm bat}=-P_{\rm bat \to \rm C}$ and 
\begin{equation}
E_{\rm bat}(t) = \int_{0}^{t} P_{\rm C \to \rm bat}(t')dt'.
\label{eq:E_battery2}
\end{equation}
This can be understood from the fact that $P_{\rm bat \to \rm C}$ in Eq.~\eqref{eq:Pstor_def_clean} actually describes the energy flow from the batteries to the BKC. The energies stored in the right and left batteries, $E_\R(t)$ and $E_\L(t)$, arise from the first and second terms in Eq.~\eqref{eq:Pstor_def_clean}, respectively.
\color{Black}The average charging power is defined as the energy gained by the batteries per unit charging time,
\begin{equation}
P(t)=\frac{E_{\rm bat}(t)-E_{\rm bat}(0)}{t}
=\frac{E_{\rm bat}(t)}{t},
\label{average_power}
\end{equation}
where the second equality follows from the batteries being initially in vacuum.\color{Black}

The overall charging efficiency of our protocol is instead defined as
\begin{equation}
\eta(t)=\frac{E_{\rm bat}(t)}{E_{\rm cost}(t)},
\label{eq:eta_def_app}
\end{equation}
Equivalently, one may introduce the individual efficiencies for the left and right batteries, $\eta_\L(t)=E_\L(t)/E_{\rm cost}(t)$ and $\eta_\R(t)=E_\R(t)/E_{\rm cost}(t)$, so that $\eta(t)=\eta_\L(t)+\eta_\R(t)$.

As a final rewrite of Eq.~\eqref{eq:Ebatt_single_muV}, it is useful to resolve the stored energies into separate quadrature contributions. For a battery mode \(\alpha\in\{\L,\R\}\), we define
\[
\delta x_\alpha \equiv x_\alpha-\langle x_\alpha \rangle = x_\alpha-\mu_{x_\alpha},\qquad
\delta p_\alpha \equiv p_\alpha-\langle p_\alpha \rangle = p_\alpha-\mu_{p_\alpha}.
\]
The energy stored in this battery can then be decomposed as 
\begin{equation}
\begin{split}
E_\alpha(t)&=E_\alpha^{\rm coh}(t)+E_\alpha^{\rm fluc}(t) \\
&=E_\alpha^{\rm coh}(t)+E_{\alpha,\delta x^2}(t)+E_{\alpha,\delta p^2}(t),
\end{split}
\label{E_QB_decompose}
\end{equation}
where
\begin{subequations}\label{E_QB_components}
\begin{align}
E_\alpha^{\rm coh}(t)&=\frac{1}{2}\omega\left[\mu_{x_\alpha}^2(t)+\mu_{p_\alpha}^2(t)\right], \label{eq:Ecoh_SM}\\
E_{\alpha,\delta x^2}(t)&=\frac{1}{2}\omega\left(\langle\delta x_\alpha^2(t)\rangle-\frac{1}{2}\right)
=\frac{1}{2}\omega\left(V_{x_\alpha, x_\alpha}(t)-\frac{1}{2}\right), \label{eq:Edeltax2_SM}\\
E_{\alpha,\delta p^2}(t)&=\frac{1}{2}\omega\left(\langle\delta p_\alpha^2(t)\rangle-\frac{1}{2}\right)
=\frac{1}{2}\omega\left(V_{p_\alpha, p_\alpha}(t)-\frac{1}{2}\right). \label{eq:Edeltap2_SM}
\end{align}
\end{subequations}
The subtraction of \(1/2\) removes the vacuum contribution from each quadrature, such that \(E_{\alpha,\delta x^2}\) and \(E_{\alpha,\delta p^2}\) vanish for the vacuum state, while one of them may become negative for a squeezed state. No mixed-covariance term appears in this decomposition because the local stored energy (of each battery) depends on \(x_\alpha^2+p_\alpha^2\), rather than on the symmetrized cross term \(x_\alpha p_\alpha+p_\alpha x_\alpha\). These quadrature-resolved quantities are used in the main text to show that the amplified fluctuations are directed into orthogonal phase-space directions at opposite ends of the BKC.
\begin{figure}
\centering
\includegraphics[width=0.7\linewidth]{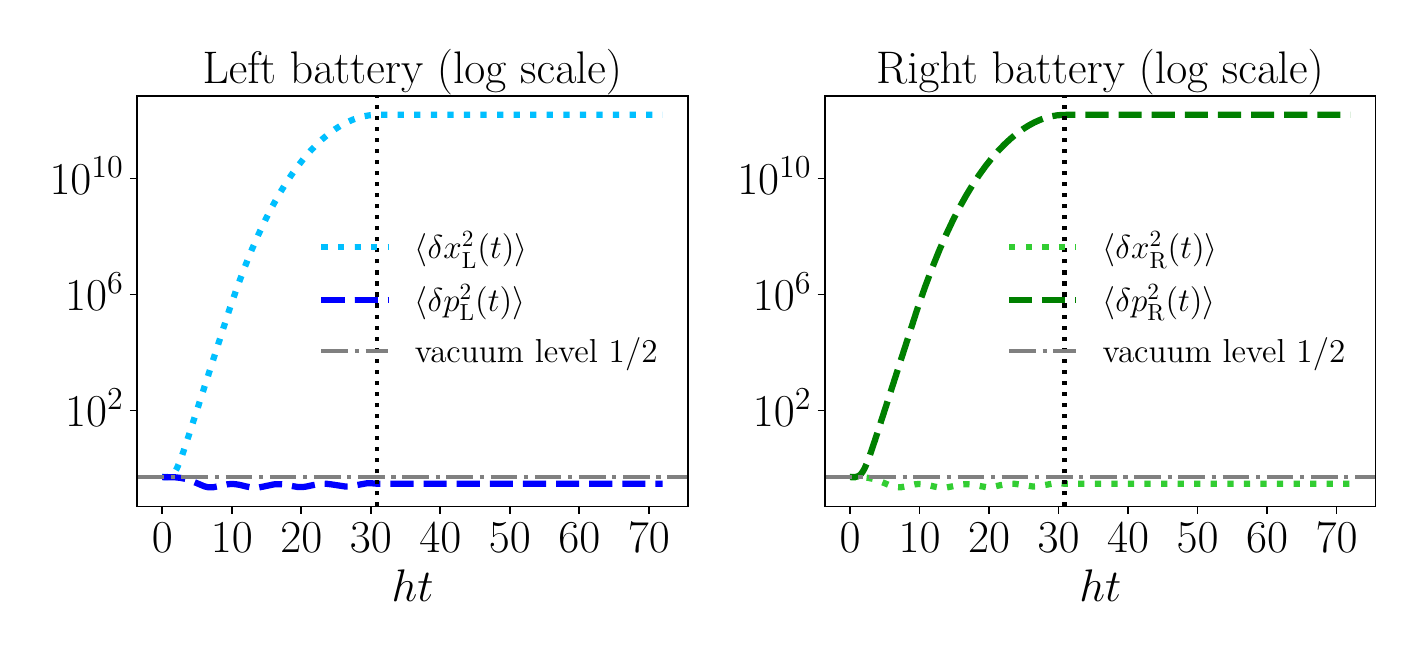}
\caption{
Quadrature variances of the left and right battery modes on a logarithmic scale. 
Left panel: \(\langle \delta x_\L^2(t)\rangle\) and \(\langle \delta p_\L^2(t)\rangle\). 
Right panel: \(\langle \delta x_\R^2(t)\rangle\) and \(\langle \delta p_\R^2(t)\rangle\). 
The vertical dotted lines indicate the switch-off time \(t^\star\), while the horizontal dash-dotted line marks the vacuum level \(1/2\). 
The amplified fluctuations are directed into different quadratures at the two ends of the chain: the left battery is dominated by \(\langle \delta x_\L^2(t)\rangle\), whereas the right battery is dominated by \(\langle \delta p_\R^2(t)\rangle\). 
At the same time, the complementary quadratures are squeezed below the vacuum level at \(t=t^\star\), with \(\langle \delta p_\L^2(t^\star)\rangle<1/2\) and \(\langle \delta x_\R^2(t^\star)\rangle<1/2\). 
Parameters: $\Delta/h=11/12$, $g/h=5/12$, $\kappa_j/h=8.3\times 10^{-4}$ ($j=-N,\cdots,N$), \(n_{\rm th}=0\), \(\gamma_{\rm L}=\gamma_{\rm R}=0\), $\mathcal{A}/\sqrt{h}=2.887$, $\theta=\pi/2$, $h\sigma=1.2$, and $ht_0=1.2$. 
\justifying}
    \label{fig:storage_quadrature_variances}
\end{figure}

The decomposition in Eq.~\eqref{E_QB_decompose} can be carried over directly to the other
figures of merit. Defining
\[
E_{\rm bat}^{\chi}(t)
=E_{\rm L}^{\chi}(t)+E_{\rm R}^{\chi}(t),
\qquad
\chi\in\{\mathrm{coh},\mathrm{fluc}\},
\]
we have
$E_{\rm bat}=E_{\rm bat}^{\rm coh}+E_{\rm bat}^{\rm fluc}$.
The corresponding contributions to the average charging power and efficiency
are
\[
P^{\chi}(t)
=\frac{E_{\rm bat}^{\chi}(t)-E_{\rm bat}^{\chi}(0)}{t},
\qquad
\eta^{\chi}(t)
=\frac{E_{\rm bat}^{\chi}(t)}{E_{\rm cost}(t)},
\]
so that
$P=P^{\rm coh}+P^{\rm fluc}$ and
$\eta=\eta^{\rm coh}+\eta^{\rm fluc}$. For the ergotropy, Eq.~\eqref{eq:WB_gaussian} gives
$\mathcal{W}_{\alpha}
=E_{\alpha}^{\rm coh}+E_{\alpha}^{\rm sq}$.
We therefore identify
$\mathcal{W}_{\alpha}^{\rm coh}=E_{\alpha}^{\rm coh}$ and
$\mathcal{W}_{\alpha}^{\rm fluc}=E_{\alpha}^{\rm sq}
=E_{\alpha}^{\rm fluc}-E_{\alpha}^{\rm th}$.
Thus, only the ordered, squeezing-generated part of the fluctuation energy
contributes to the fluctuation ergotropy, while the thermal-like part is
passive.

Figure~\ref{fig:storage_quadrature_variances} directly confirms the phase-sensitive and spatially asymmetric squeezing of the battery fluctuations. For the left storage mode, the amplified fluctuation energy is carried almost entirely by the \(x\)-quadrature, while the conjugate \(p\)-quadrature remains squeezed below the vacuum value. Conversely, for the right storage mode, the amplified fluctuations are concentrated in the \(p\)-quadrature, whereas the \(x\)-quadrature is the one squeezed below vacuum. In particular, at the protocol-defined switch-off time \(t=t^\star\), we find
\[
\langle \delta p_\L^2(t^\star)\rangle < \frac12,
\qquad
\langle \delta x_\R^2(t^\star)\rangle < \frac12.
\]
Thus the two batteries receive fluctuations that are squeezed into orthogonal phase-space directions when propagating toward opposite ends of the bosonic Kitaev chain. This provides a direct covariance-level signature of the chiral squeezing mechanism discussed in the main text. It also explains why the stored energy is almost entirely non-passive at late times: one quadrature is strongly amplified, while the conjugate one remains suppressed, so the battery covariance becomes highly anisotropic rather than thermal-like. In the figure, the squeezed quadratures and the vacuum reference line are visually compressed near zero because the vertical scale is set by the very large amplified variance in the orthogonal quadrature.

\color{Black}
\suppsubsection{Charging power dynamics and scaling}

\begin{figure}
\centering
\includegraphics[width=0.8\linewidth]{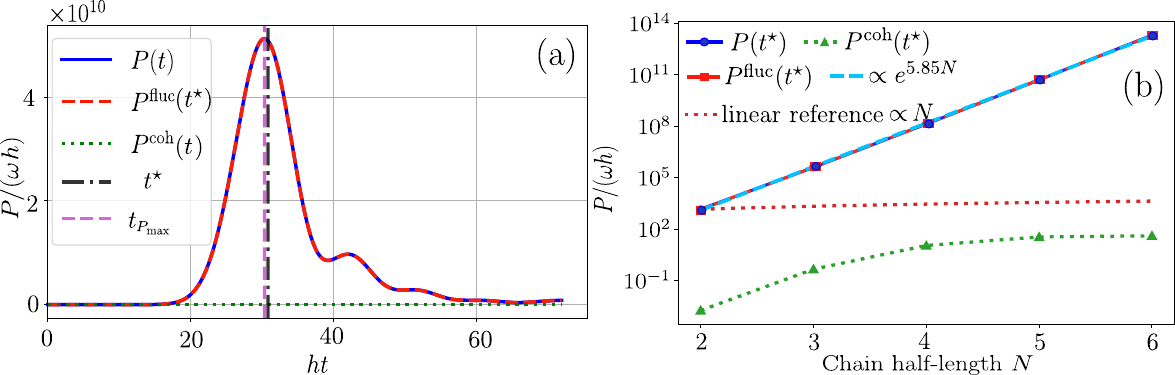}
\caption{Average charging power and its size scaling.
(a) Time evolution of the average charging power $P(t)$ and its fluctuation and coherent contributions when the setup is kept on continuously. The vertical lines mark the optimal times $t^\star$ and $t_{P_{\max}}$ for the charging efficiency and power, respectively.
(b) $P(t^\star)$ and its fluctuation and coherent contributions versus the chain half-length $N$. The cyan dashed line shows the fit $P(t^\star)\propto e^{5.85N}$, while the crimson dotted line denotes the extensive reference $P(t^\star)\propto N$. All other parameters are the same as in Fig.~2 of the main text. \justifying}
\label{fig:PowerScaling}
\end{figure}

To complement the finite-time dynamics shown in Fig.~2(f) of the main text, we examine the average charging power $P(t)$ when the setup is kept on continuously (i.e., without switching off the setup at $t=t^\star$). As shown in Fig.~\figpanel{fig:PowerScaling}{a}, $P(t)$ initially increases, reaches a maximum at $t=t_{P_{\max}}$, and subsequently decreases. The power is almost entirely dominated by the fluctuation contribution, while the coherent contribution remains negligible.

The optimal time $t_{P_{\max}}$ at which $P(t)$ reaches the maximum is close to, but does not exactly coincide with, the efficiency-optimal time $t^\star$. This is expected because the two quantities optimize different ratios, $\eta=E_{\rm bat}/E_{\rm cost}$ and $P=E_{\rm bat}/t$. Accordingly, their maxima satisfy
$\dot E_{\rm bat}/E_{\rm bat}=\dot E_{\rm cost}/E_{\rm cost}$
and
$\dot E_{\rm bat}=E_{\rm bat}/t$,
respectively. In the main text, we nevertheless choose to switch off the setup at $t=t^\star$.

Figure~\figpanel{fig:PowerScaling}{b} further shows that $P(t^\star)$ exhibits a strongly \emph{superextensive scaling}, approximately
$P(t^\star)\propto e^{5.85N}$
over the investigated range $N=2$--$6$, substantially exceeding the extensive scaling $P\propto N$. This enhancement is again dominated by the fluctuation contribution.
\color{Black}

\suppsubsection{Short- and long-time behaviors}

Let us revisit the general single-bosonic-mode case. A convenient way to understand the difference between the stored energy \(E_B(t)\) and the ergotropy \(\mathcal W_B(t)\) at short times [see Fig. 2(d) in the main text] is to examine its Gaussian properties. For a bosonic mode with frequency $\omega_{B}$, first moments \(\mu_B(t)=(\langle X(t)\rangle,\langle P(t)\rangle)^{\mathsf T}\), and covariance matrix \(V_B(t)\), the total energy and the corresponding ergotropy are given in Eqs.~\eqref{E_coh_fluc} and \eqref{eq:WB_gaussian}, respectively.
Their difference is therefore
\begin{equation}
E_B(t)-\mathcal W_B(t)=\omega_{B}\left(\nu_B(t)-\frac12\right),
\label{eq:EW_difference_shorttime}
\end{equation}
which is exactly the passive, thermal-like contribution to the stored energy. Equation~\eqref{eq:EW_difference_shorttime} shows that exact equality \(E_B(t) = \mathcal W_B(t)\) requires
\(
\nu_B(t) = \frac12
\). More generally, near-unity extractability requires the passive contribution \(\omega_B(\nu_B-1/2)\) to be small compared with the total stored energy.

At short times, this condition is generally not satisfied. Indeed, the battery mode first receives covariance contributions that are not yet strongly ordered, so \(\mathrm{Tr}[V_B]\) and therefore the stored energy start to grow. At the same time, however, the determinant \(\det[V_B]=\nu_{B}^{2}\) [cf. Eq.~\eqref{nu_det}] also increases, which implies that part of the stored energy is still passive and cannot be extracted by a unitary operation. In this sense, the battery acquires finite energy before the stored fluctuations become sufficiently ordered.

This behavior can be made explicit by expressing the storage covariance in the form
\begin{equation}
V_B(t)=
\begin{pmatrix}
\frac12+n(t)+s(t) & z(t)\\[1mm]
z(t) & \frac12+n(t)-s(t)
\end{pmatrix},
\end{equation}
where \(n(t)\) measures the isotropic broadening of both quadratures, while \(s(t)\) and \(z(t)\) quantify, respectively, the anisotropy between the two quadratures (due to the chiral squeezing) and their correlations. These quantities satisfy 
\begin{equation}\label{SM:nt}
\frac12\,\mathrm{Tr}[V_B(t)]-\frac12 = n(t),
\end{equation}
and
\begin{equation}\label{SM:sc}
\nu_B(t)=\sqrt{\mathrm{det}[V_{B}(t)]}=\sqrt{\left(\frac12+n(t)\right)^2-s^2(t)-z^2(t)}.
\end{equation}
Equation~\eqref{SM:nt} shows that $n(t)$ gives the total fluctuation occupation above vacuum, whereas Eq.~\eqref{SM:sc} shows how the passive part depends on the balance between the isotropic broadening and anisotropic squeezing. If \(n(t)\) grows before sizable anisotropy \(s(t)\) and correlation \(z(t)\) have developed, then \(\nu_B(t)-1/2\) is non-negligible and one has \(E_B(t)>\mathcal W_B(t)\). By contrast, once the fluctuations become strongly ordered due to the chiral squeezing, \(s^2(t)+z^2(t)\) reduces the determinant contribution and drives \(\nu_B(t)\) back toward \(1/2\). The passive contribution then becomes small, and the stored energy becomes almost fully extractable.

The dynamical origin of this crossover is that the battery covariance does not become strongly squeezed instantaneously. Rather, the diffusion term $D(t)$ continuously injects fluctuations into the chain, while the drift matrix $A(t)$ transports and chirally reshapes them before they are loaded into the batteries. At early times, the covariance transferred to the batteries still has a mixed, partly isotropic character, so the passive term given in Eq.~\eqref{eq:EW_difference_shorttime} remains visible. After a certain propagation-and-loading time, the BKC converts the incoming fluctuations into a strongly anisotropic squeezed state of the battery mode. The squeezing contribution then dominates over the passive thermal-like contribution, and the ergotropy catches up with the total energy. Therefore, the timescale on which $E_{B}(t)$ and $\mathcal{W}_{B}(t)$ become nearly equal is set by the time required for the ordered squeezing sector to build up and reach the battery. This timescale is controlled by the same transport and loading mechanisms that govern the overall charging dynamics: the drift set by the BKC hopping \(h\), the parametric pump \(\Delta\), the edge coupling \(g\), and the half-chain length \(N\).


\color{Black}
\suppsubsection{Effect of finite battery dissipation}

\begin{figure}
\centering
\includegraphics[width=0.75\linewidth]{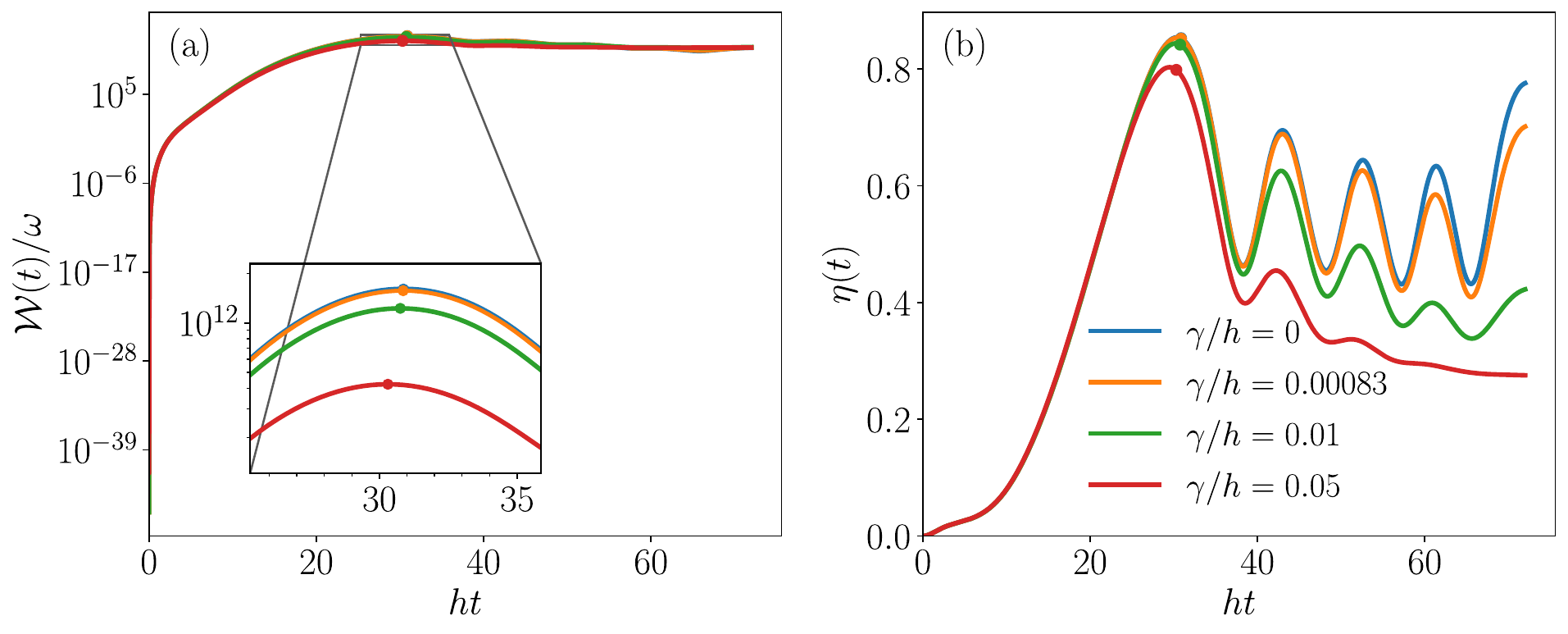}
\caption{Effect of intrinsic battery loss. Time evolution of (a) the total battery ergotropy $\mathcal{W}(t)$ and (b) the charging efficiency $\eta(t)$ for $\gamma_{\rm L}=\gamma_{\rm R}\equiv\gamma$ with $\gamma/h=0,\,8.3\times10^{-4},\,0.01,\,0.05$. The inset in panel (a) shows a zoom-in view around the optimal switch-off time $t^\star$. The battery reservoirs are taken to be at zero temperature, \(n_{\rm L}=n_{\rm R}=0\). All other parameters are the same as in Fig.~2 of the main text.  
\justifying}
\label{fig:intrinsicLoss}
\end{figure}

In the main calculations, we assume the battery modes to be lossless, $\gamma_{\rm L}=\gamma_{\rm R}=0$, which is relevant when the battery lifetimes are much longer than the charging time. Here we examine the effect of finite intrinsic battery loss by setting $\gamma_{\rm L}=\gamma_{\rm R}=\gamma>0$. Figure~\ref{fig:intrinsicLoss} shows the dynamics of the total battery ergotropy and charging efficiency for several values of $\gamma$. For weak battery loss ($\gamma/h\lesssim10^{-2}$), the peak values of $\mathcal{W}$ and $\eta$ remain close to the lossless case. As $\gamma$ increases, both quantities are gradually reduced because part of the stored energy is dissipated. Nevertheless, the qualitative charging behavior remains unchanged. These results show that the fluctuation-driven charging mechanism remains effective in the presence of finite battery dissipation and does not rely on perfectly lossless storage modes.

\suppsubsection{Work-like signal-to-noise ratio and its scaling}

\begin{figure}
\centering
\includegraphics[width=0.95\linewidth]{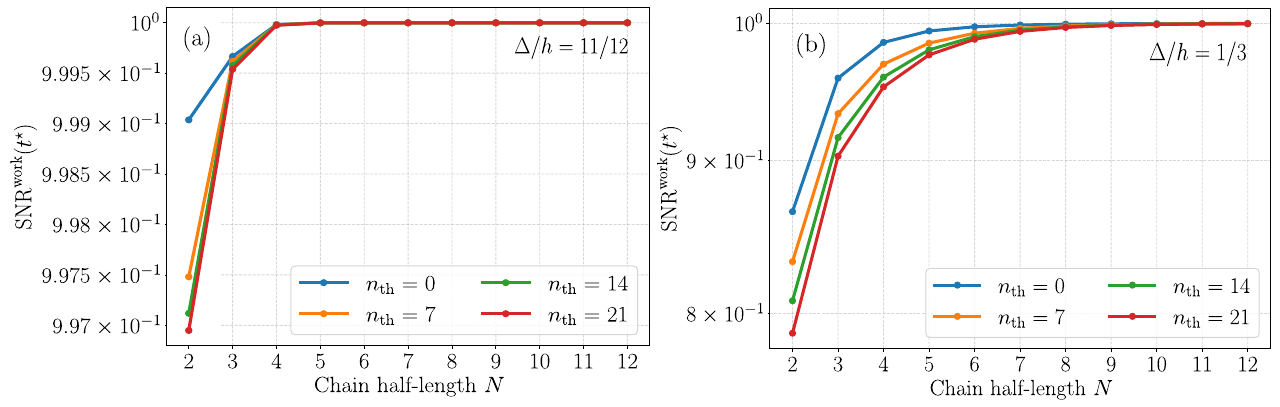}
\caption{Size dependence of the work-like SNR for different parametric pump strengths. $\mathrm{SNR}^{\rm work}(t^\star)$ versus chain half-length $N$ for (a) $\Delta/h=1/3$ and (b) $\Delta/h=11/12$, for several thermal occupancies $n_{\rm th}$. The latter corresponds to the pump strength used in the main text. All other parameters are the same as in Fig.~3 of the main text. 
\justifying}
\label{fig:SNR_Scaling}
\end{figure}

In addition to the standard figures of merit for quantum-battery charging, such as stored energy, ergotropy, efficiency, and power, we also characterize our protocol using a work-like signal-to-noise ratio (SNR) 
\begin{equation}
\mathrm{SNR}^{\mathrm{work}}(t)=\frac{\mathcal{W}(t)}{\sigma_E(t)},
\label{SM_SNR_work}
\end{equation}
where $\mathcal{W}(t)$ is the total battery ergotropy given in Eq.~\eqref{eq:Wbatt_def} and $\sigma_E(t)=\sqrt{\mathrm{Var}[\hat{E}_{\rm bat}](t)}$ is the energy fluctuation of the corresponding battery state. In the main text, we show that $\mathrm{SNR}^{\mathrm{work}}(t^\star)$ quickly approaches unity for our BKC-based charging setup as the half-chain length $N$ increases over the considered range. 

To examine whether the near-unity work-like SNR is specific to the strong
parametric pump used in the main text, we compare its size dependence for two
substantially different pump strengths. Figure~\ref{fig:SNR_Scaling} shows
$\mathrm{SNR}^{\mathrm{work}}(t^\star)$ up to $N=12$ for
$\Delta/h=1/3$ and $11/12$ and for several thermal occupancies
$n_{\rm th}$.

For the main-text value $\Delta/h=11/12$, the SNR rapidly approaches unity
and remains essentially unchanged as $N$ is further increased. Thus, although
a longer chain contains more dissipative channels, the associated energy
fluctuations do not outgrow the extractable work over the investigated range.
For the ideal harmonic model, increasing $N$ leads to progressively
stronger amplification and correspondingly larger battery occupations while
preserving the near-unity SNR. In realistic implementations, however,
arbitrarily large occupations will eventually be limited by practical constraints such as nonlinearities, pump depletion, and other saturation mechanisms. It is therefore useful to examine whether the same favorable SNR
behavior persists in a substantially weaker-amplification regime.

For the weaker pump, $\Delta/h=1/3$, the growth of the work-like SNR is more gradual because the chiral amplification and squeezing are reduced. Nevertheless,
$\mathrm{SNR}^{\mathrm{work}}(t^\star)$ again approaches unity as $N$
increases, for all thermal occupancies considered here. This demonstrates that the near-unity work-like SNR is not tied to the particular strong-pump regime used
in the main text, but persists for a
substantially weaker parametric pump over the extended range of system sizes.

\color{Black}

\suppnote{Estimation of the optimal switch-off time $t^\star$}\label{tstar-proof}

\suppsubsection{General formula}

In this Note, we explain how to estimate the optimal switch-off time $t^\star$ for a given chain half-length $N$. As emphasized in the main text, the total energy stored in the batteries is dominated by amplified ordered fluctuations. However, these ordered fluctuations are predominantly non-passive and squeezing-generated, rather than thermal-like. Therefore the relevant timescale is not set by passive bath-induced heating, but by the drift-controlled propagation and amplification dynamics of the BKC, through which this squeezing-dominated energy is transported to the batteries and subsequently loaded into them. Since both the coherent first moments and the ordered fluctuation sector propagate according to the same BKC drift matrix, the characteristic charging time can still be estimated from the same ballistic transport mechanism that underlies the propagation of coherent excitations.

We thus describe the charging process as a cascade of three stages: (\textit{i}) an effective seeding stage in which the dominant energy contribution becomes appreciable in the chain, (\textit{ii}) a transport stage in which this contribution propagates chirally toward the ends of the chain, and (\textit{iii}) a final loading stage in which it is exchanged into the battery mode at the edge.


\paragraph*{Capture-time estimate.}
Let $t^\star$ denote the optimal switch-off time at which the charging efficiency $\eta$ reaches its peak value. This optimal time can be estimated directly from the propagation picture discussed in~\ref{app:gaussian_formalism}. After an initial onset time \(t_{\rm on}\), the relevant amplified component propagates from the chain bulk to the edge and is subsequently transferred to the battery by the edge coupling. We therefore write the optimal switch-off time as the sum of three causal delays,
\begin{equation}
t^\star(N)\simeq t_{\rm on}+t_{\rm prop}(N)+t_{\rm load}.
\label{tstar_decomp}
\end{equation}
For the estimation below, we use \(h t_{\rm on}=1.2\) (i.e., the pulse-center time; see the main text and~\ref{app:hle}) as an \(N\)-independent effective onset time. Its precise value affects the intercept but not the predicted linear scaling with \(N\).

The propagation time can be inferred from the chirally squeezed dynamics of the BKC. In this picture, the chain is mapped to a nearest-neighbor tight-binding problem with dispersion
\(
\omega_k=\tilde h\sin k
\)
and group velocity
\(
v_{k}=\partial_k\omega_k=\tilde h\cos k 
\)~\cite{AshcroftMermin1976,Du2025prr}.
For a narrow-band component centered around a wave number $k_0$, the propagation delay is the travelled distance divided by the group velocity, i.e.,
\begin{equation}
t_{\rm prop}(N)\simeq \frac{L_{\rm eff}(N)}{\tilde h\cos k_0}.
\label{tprop_k0}
\end{equation}
Here \(L_{\rm eff}\) denotes the \emph{effective propagation length} of this component before it is loaded into the batteries, which has a simple geometric meaning. For a coherent pulse injected at the center, one simply has \(L_{\rm eff}=N\). For the fluctuation sector, the Markovian noise is injected at every chain site and there is no unique injection point. Nevertheless, the covariance reaching the batteries is propagated by the same drift matrix \(A\) that gives the above BKC dispersion. We therefore use the ballistic scaling form
\begin{equation}
t_{\rm prop}(N)\simeq
\frac{N}{\tilde h\cos k_{\rm eff}}+\mathcal{O}(h^{-1}),
\label{tprop_keff}
\end{equation}
where \(k_{\rm eff}\) denotes the \emph{effective wave number} of the dominant covariance component. The \(\mathcal{O}(h^{-1})\) term accounts for the fact that the source is spatially distributed and that the finite chain has boundaries.

Finally, the loading time is set by the edge beam-splitter coupling. For a resonant two-mode exchange interaction,
\[
H_{\rm bs}=g(a^\dagger b+ab^\dagger),
\]
the dynamics exhibits Rabi oscillation, with \(b(t)=b(0)\cos(gt)-i a(0)\sin(gt)\), so the transferred energy is proportional to \(\sin^2(gt)\) and reaches its first maximum at \(gt=\pi/2\)~\cite{gardiner2004quantum,WallsMilburn2008}. We therefore write
\begin{equation}
t_{\rm load}=\beta/g,\qquad \beta=\pi/2
\end{equation}
in the ideal two-mode limit, with finite bandwidth, weak loss, and boundary reflections producing only \(\mathcal{O}(1)\) corrections to \(\beta\). Combining the above estimates gives
\begin{equation}\label{tstar_expression}
t^\star(N)\simeq t_{\rm on} + \frac{\pi}{2g} + \frac{N}{\tilde h \cos k_{\rm eff}}+\mathcal{O}(h^{-1}).
\end{equation}

Equation~\eqref{tstar_expression} should be read as a ballistic scaling estimate. It predicts that, after subtracting the onset time and loading delay, the remaining \(N\)-dependence of \(t^\star\) is controlled by the BKC group velocity. In the numerical analysis below, we check this ballistic scaling directly by extracting \(t^\star(N)\) and verifying that a nearly \(N\)-independent \(k_{\rm eff}\) captures the results.

\suppsubsection{Numerical analysis}

\begin{figure}[t]
    \centering
    \begin{subfigure}[b]{0.34\textwidth}
        \includegraphics[width=\textwidth]{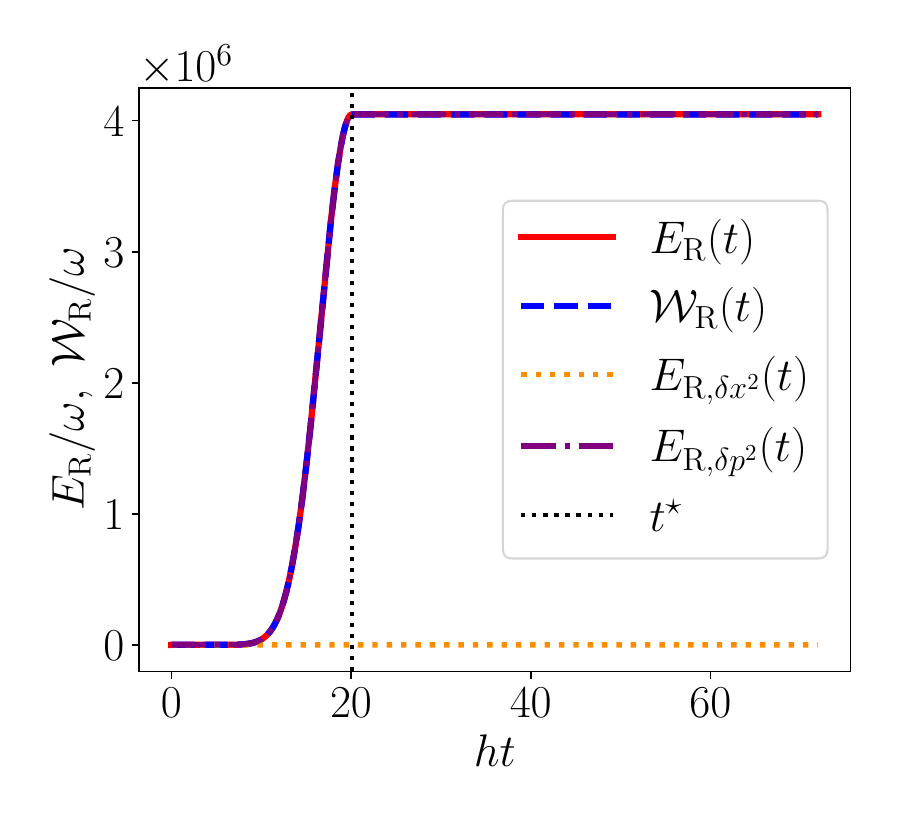}\caption{$N=3$}
        \label{3_RightEnergy}
    \end{subfigure}
    \begin{subfigure}[b]{0.34\textwidth}
\includegraphics[width=\textwidth]{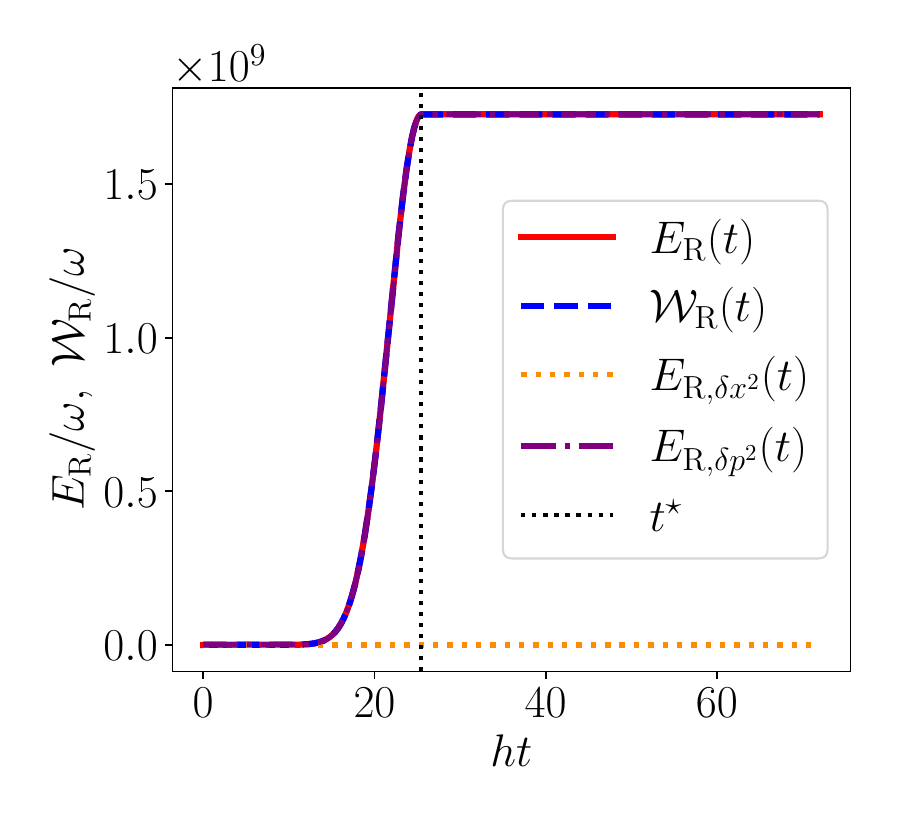}
        \caption{$N=4$}\label{4_RightEnergy}
    \end{subfigure}
    \begin{subfigure}[b]{0.34\textwidth}
\includegraphics[width=\textwidth]               {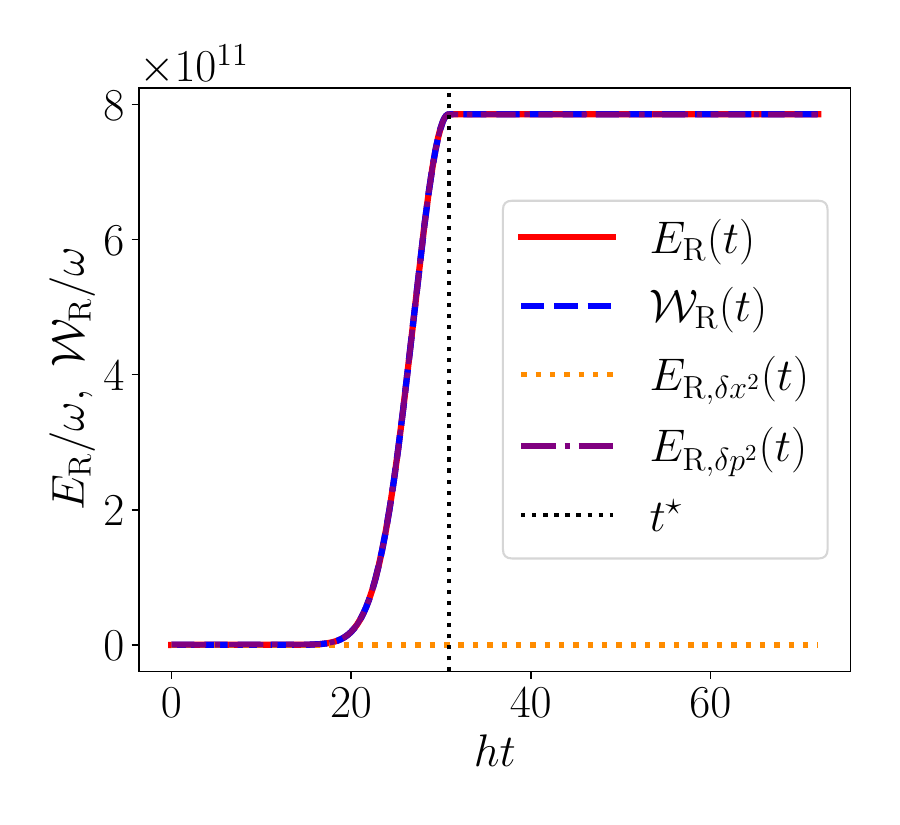}
        \caption{$N=5$}\label{5_RightEnergy}
    \end{subfigure}
    \begin{subfigure}[b]{0.34\textwidth}
\includegraphics[width=\textwidth]{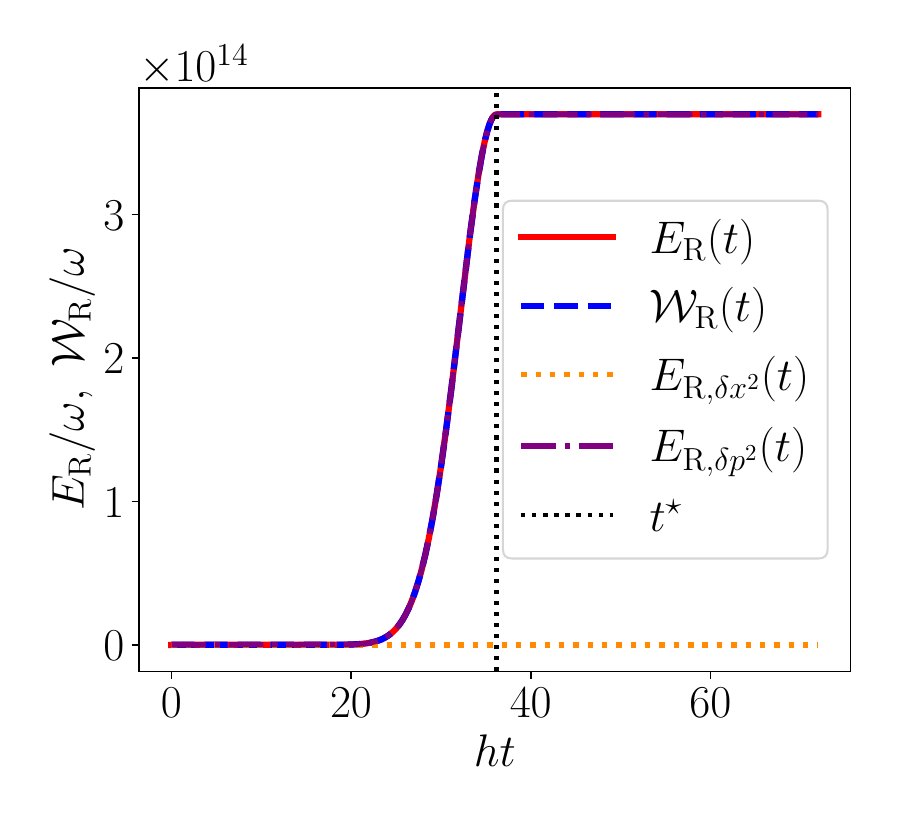}
        \caption{$N=6$}\label{6_RightEnergy}
    \end{subfigure}
   \captionsetup{justification=justified}
    \caption{Time evolution of the right-battery stored energy $E_\R(t)$ and ergotropy $\mathcal{W}_\R(t)$, together with the quadrature-resolved fluctuation contributions $E_{\R,\delta x^2}(t)$ and $E_{\R,\delta p^2}(t)$, for different chain half-length $N$: (a) $N=3$, (b) $N=4$, (c) $N=5$, and (d) $N=6$. The vertical dotted lines mark the optimal switch-off time $t=t^\star$, at which the setup is turned off by setting $h(t^\star)=\Delta(t^\star)=g(t^\star)=0$. Parameters: $\Delta/h=11/12$, $g/h=5/12$, $\kappa_j/h=8.3\times 10^{-4}$ ($j=-N,\cdots,N$), \(n_{\rm th}=0\), \(\gamma_{\rm L}=\gamma_{\rm R}=0\), $\mathcal{A}/\sqrt{h}=2.887$, $\theta=\pi/2$, $h\sigma=1.2$, and $ht_0=1.2$. \justifying}
\label{EnergyRightSupllementary}
\end{figure}

For the data analysis below, we adopt a simplified effective description and use $L_{\rm eff}(N)\approx N$ as a phenomenological transport length. This choice is exact for the coherent pulse injected at the center site. For the fluctuation-dominated dynamics, it effectively absorbs the difference between the actual fluctuation propagation distance and the half-chain length into an effective wave number $k_{\rm eff}$ extracted from the numerical data. This fitted wave number should therefore be understood as an \emph{effective transport parameter} that captures the optimal switch-off time $t^\star$, rather than as the exact carrier wave number of a narrowband coherent packet.

Since the plots in Fig.~\ref{EnergyRightSupllementary} use the dimensionless time variable \(ht\), it is convenient to rewrite the estimate in the same units. Defining
\[
\tilde t \equiv ht,\qquad
\tilde t^\star \equiv h t^\star,\qquad
\tilde t_{\rm on} \equiv h t_{\rm on},\qquad
\tilde t_{\rm prop}\equiv h t_{\rm prop},
\]
one has
\begin{equation}
    \tilde t^\star(N)\approx \tilde t_{\rm on}+\frac{\beta}{g/h}+\tilde t_{\rm prop}(N).
\label{tstar_decomp_dimless}
\end{equation}
For the parameters used in Fig.~\ref{EnergyRightSupllementary}, the loading delay is
\begin{equation}
    \frac{\beta}{g/h}=\frac{\pi}{2(g/h)}\approx 3.77.
\end{equation}
With the observed peak times,
\begin{equation}\label{tstar_data_dimless}
\tilde t^\star(2)=14.64,\qquad
\tilde t^\star(3)=20.04,\qquad
\tilde t^\star(4)=25.44,\qquad
\tilde t^\star(5)=30.84,\qquad
\tilde t^\star(6)=36.20,\qquad
\tilde t^\star(7)=41.50,
\end{equation}
and choosing \(\tilde t_{\rm on} = 1.2\), the propagation delays
\(
\tilde t_{\rm prop}(N)=\tilde t^\star(N)-\tilde t_{\rm on}-\frac{\pi}{2(g/h)}
\)
are inferred as
\begin{equation}
\begin{aligned}
\tilde t_{\rm prop}(2)&\approx 9.67,\qquad\qquad
\tilde t_{\rm prop}(3)\approx 15.07,\qquad\qquad
\tilde t_{\rm prop}(4)\approx 20.47,\\
\tilde t_{\rm prop}(5)&\approx 25.87,\qquad\qquad
\tilde t_{\rm prop}(6)\approx 31.23,\qquad\qquad
\tilde t_{\rm prop}(7)\approx 36.53.
\end{aligned}
\end{equation}
Then the effective wave number can be estimated through
\begin{equation}\label{k0_fit_dimless}
\cos k_{\rm eff}(N)\approx \frac{N}{(\tilde h/h)\,\tilde t_{\rm prop}(N)}.
\end{equation}
For \(\Delta/h=11/12\), one has
\(
\tilde h/h=\sqrt{1-(\Delta/h)^2}\approx 0.40
\). This yields
\begin{equation}
\cos k_{\rm eff}\simeq 0.52,\ 0.50,\ 0.49,\ 0.48,\ 0.48,\ 0.48
\qquad
\textrm{for} \quad N=2,3,4,5,6,7,
\end{equation}
corresponding to
\begin{equation}\label{k0_value_dimless}
k_{\rm eff} \simeq 1.03,\ 1.05,\ 1.06,\ 1.07,\ 1.07,\ 1.07.
\end{equation}
We therefore identify an approximately \(N\)-independent effective wave number \(k_{\rm eff}\simeq 1.07\) for longer chains. Substituting this value into Eq.~\eqref{tstar_decomp_dimless} yields an essentially \emph{linear scaling} with \(N\) over this range,
\[
\tilde t^\star(N)\approx \tilde t_{\rm on}+\frac{\beta}{g/h}+\frac{N}{(\tilde h/h)\cos k_{\rm eff}}.
\]
The convergence of the extracted $k_{\rm eff}$ with increasing \(N\) is physically expected. The quantities \(t_{\rm on}\), \(\beta/g\), and the boundary-induced reflection corrections are $N$-independent local delays, whereas the bulk propagation time grows linearly with the chain length. Thus, for longer chains the peak time is increasingly dominated by the bulk group delay,
\[
\tilde t^\star(N)=\tilde t_{\rm on}+\frac{\beta}{g/h}
+\frac{N}{(\tilde h/h)\cos k_{\rm eff}}+\mathcal{O}(1),
\]
or equivalently
\[
\frac{d\tilde t^\star}{dN}
\longrightarrow
\frac{1}{(\tilde h/h)\cos k_{\rm eff}}.
\]
The effective wave number \(k_{\rm eff}\) is selected by the propagating components of the wavepacket that reach the edges and are loaded into the batteries. For finite \(N\), the inferred value is affected by the finite pulse width, boundary loading, and residual reflections. As \(N\) increases, these finite-size offsets become negligible such that the extracted value converges to an \(N\)-independent bulk value. In this sense, the average value \(k_{\rm eff}\simeq 1.07\) provides an increasingly accurate description of the dominant propagation delay for larger $N$. This convergence means that the optimal switch-off time is governed primarily by the chiral transport encoded in the drift matrix \(A\), rather than being an outcome of an uncontrollable numerical search.

\color{Black}

\begin{figure}[t]
    \centering
    \includegraphics[width=0.45\linewidth]{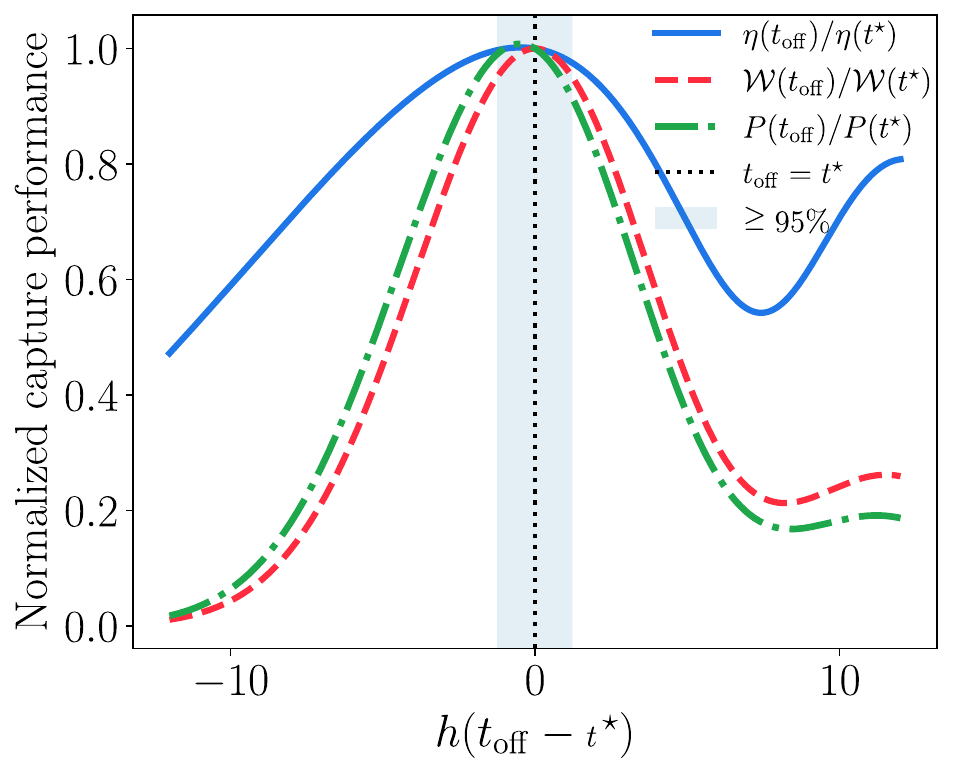}
    \caption{Robustness against deviations from the optimal switch-off time.
    Normalized charging efficiency $\eta(t_{\rm off})/\eta(t^\star)$ (blue solid), ergotropy $\mathcal{W}(t_{\rm off})/\mathcal{W}(t^\star)$ (red dashed), and average charging power $P(t_{\rm off})/P(t^\star)$ (green dot-dashed) as functions of the switch-off time deviation $t_{\rm off}-t^\star$.
    The vertical dotted line marks $t_{\rm off}=t^\star$, while the shaded region indicates the interval in which all the normalized quantities remain $\geq 95\%$ of their optimal values. All other parameters are the same as in Fig.~2 of the main text.\justifying
    }
\label{fig:switch_off_sensitivity}
\end{figure}

Having estimated the optimal switch-off time $t^\star$, we further examine the sensitivity of the charging performance to deviations from this optimal timing. Figure~\ref{fig:switch_off_sensitivity} shows three normalized figures of merit, i.e., charging efficiency $\eta(t_{\rm off})/\eta(t^\star)$, ergotropy $\mathcal{W}(t_{\rm off})/\mathcal{W}(t^\star)$, and average charging power $P(t_{\rm off})/P(t^\star)$, as functions of the switch-off time deviation $t_{\rm off}-t^\star$. All these normalized quantities remain close to their optimal values (staying above $95\%$) within a finite window around $t_{\rm off}=t^\star$. This demonstrates that the finite-time charging protocol is robust against moderate deviations from the optimal switch-off time. Note that, for simplicity, we model the switch-off as instantaneous in the numerical simulations. In practice, a smooth ramp is expected to yield comparable charging performance, provided that the ramp duration is sufficiently short compared with the relevant dynamical timescales.

\color{Black}

\suppnote{Conventional phase-preserving amplification: Added noise and signal-to-noise ratio}
\label{app:local_gain_snr}

\begin{figure}[t]
    \centering
    \includegraphics[width=0.8\linewidth]{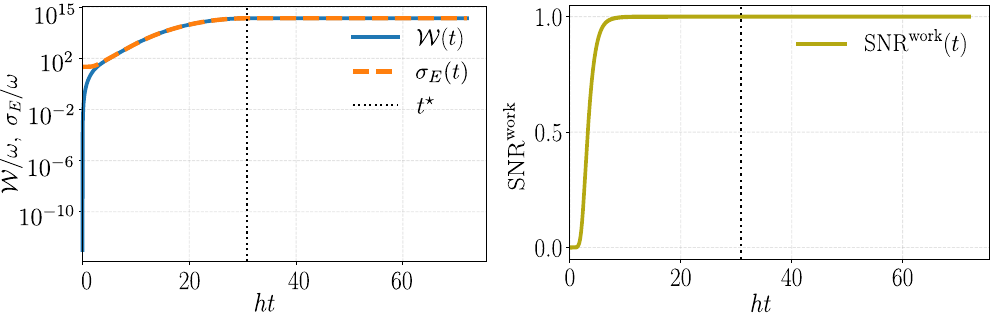}
    \caption{
    Time evolution of the ergotropy, energy fluctuation, and work-like SNR for the BKC charging protocol.
    Left: total battery ergotropy $\mathcal{W}(t)$ (blue solid) and energy fluctuation $\sigma_E(t)$ (orange dashed).
    Right: the corresponding work-like SNR, $\mathrm{SNR}^{\rm work}(t)=\mathcal{W}(t)/\sigma_E(t)$.
    The vertical dotted lines mark the optimal switch-off time $t^\star$.
    The work-like SNR rapidly approaches unity at an early stage of the charging process and remains close to unity up to $t^\star$.
    All other parameters are the same as in Fig.~2 of the main text. \justifying
    }
    \label{fig:time_evolution_snr}
\end{figure}

\color{Black}
In this Note, we provide a representative benchmark to illustrate why a conventional \emph{phase-preserving} amplifier (e.g., a distributed gain medium) does not generically yield an improved work-like SNR, even though it amplifies the total energy. As a reference for this comparison, we first briefly examine the behavior of the work-like SNR in the proposed BKC scheme. Figure~\ref{fig:time_evolution_snr} shows the time evolution of the total battery ergotropy $\mathcal{W}(t)$, the corresponding energy fluctuation $\sigma_E(t)$, and their ratio [i.e., the work-like SNR defined in Eq.~\eqref{SM_SNR_work}].
As the charging proceeds, $\mathcal{W}(t)$ and $\sigma_E(t)$ grow in a closely correlated manner, which is consistent with their correlated scaling shown in Fig.~5 of the main text. Correspondingly, $\mathrm{SNR}^{\rm work}(t)$ rapidly approaches unity at an early stage of the charging process and subsequently remains close to unity up to the optimal switch-off time $t^\star$. This shows that the near-unity work-like SNR is established well before the charging dynamics is switched off. The subsequent charging dynamics primarily increases the extractable work while preserving this favorable work-to-fluctuation ratio.

\color{Black}

 
We now turn to a conventional phase-preserving chain amplifier and examine whether such a favorable work-to-fluctuation ratio can also arise from standard phase-insensitive amplification. The key point is that a phase-preserving amplifier necessarily introduces additional fluctuations together with the amplification. As shown below, these added fluctuations are predominantly isotropic and passive, and therefore contribute strongly to the battery-energy fluctuation $\sigma_{E}$ while providing little extractable work $\mathcal{W}$. As a result, the work-like SNR remains small when the incoherent background dominates over the coherent contribution. Moreover, increasing the gain amplifies both the coherent contribution and the added fluctuations with the same leading amplification factor, so that this common gain factor cancels in their ratio. Consequently, increasing the gain does not provide a parametric enhancement of the work-like SNR. This sharply contrasts with the mechanism studied in the main text, where the \emph{phase-sensitive} amplification arises from the Hamiltonian drift engineering (encoded in the drift matrix $A$) rather than from coupling each chain site to an independent active reservoir.

Specifically, we consider a one-dimensional bosonic tight-binding chain with chain-site annihilation (creation) operators \(a_j\) ($a_j^\dag$), nearest-neighbor hopping \(h\), and local gain/loss on every site. Importantly, a completely positive Markovian description of an \emph{active reservoir} is not obtained by simply inserting a negative damping rate into a passive Lindblad term. Instead, the minimal physical model is the gain-loss master equation
\begin{equation}
\dot \rho = \kappa_\downarrow \,\mathcal D[a]\rho
+ \kappa_\uparrow \,\mathcal D[a^\dagger]\rho,
\qquad
\mathcal D[L]\rho = L\rho L^\dagger - \frac{1}{2}\{L^\dagger L,\rho\},
\label{eq:app_gain_loss_master_single}
\end{equation}
which yields net gain when \(\kappa_\uparrow>\kappa_\downarrow\). For comparison, a passive mode coupled to a thermal bath is commonly described by  $\dot{\rho}=\kappa(n_{\rm th}+1)\mathcal{D}[a]\rho+\kappa n_{\rm th}\mathcal{D}[a^\dag]\rho$. The corresponding Heisenberg equation for the mode amplitude reads
\begin{equation}
\frac{d}{dt}\langle a\rangle = \frac{\kappa_\uparrow-\kappa_\downarrow}{2}\,\langle a\rangle,
\label{eq:app_single_mode_mean}
\end{equation}
whereas the mean occupation obeys
\begin{equation}
\frac{d}{dt}\langle a^\dagger a\rangle =
(\kappa_\uparrow-\kappa_\downarrow)\langle a^\dagger a\rangle
+\kappa_\uparrow.
\label{eq:app_single_mode_number_ode}
\end{equation}
The second term in Eq.~\eqref{eq:app_single_mode_number_ode} is the spontaneous-emission contribution generated by the active medium itself. Solving Eq.~\eqref{eq:app_single_mode_number_ode} gives
\begin{equation}
\langle a^\dagger a\rangle_t = G_t\,\langle a^\dagger a\rangle_0
+ \frac{\kappa_\uparrow}{\kappa_\uparrow-\kappa_\downarrow}\,(G_t-1),
\qquad
G_t = e^{(\kappa_\uparrow-\kappa_\downarrow)t}.
\label{eq:app_single_mode_number_solution}
\end{equation}
Thus, even for vacuum input, the active source populates the mode. This spontaneous-emission-induced population is the basic origin of the noise penalty.

For a direct comparison with the BKC charging protocol, we inject the
finite-duration coherent Gaussian pulse (as defined in the main text and in~\ref{app:hle}) through the central site of the conventional amplifier chain. We take $M_{\rm conv}=2N+1$ as the total site number of the conventional amplifier chain and denote the
central site by $j_c=N+1$. 

In the rotating frame with respect to the site frequency $\omega_{\rm C}$, the Heisenberg-Langevin equation of each chain site reads
\begin{equation}
\dot a_j
= -ih(a_{j+1}+a_{j-1})
+\frac{\kappa_\uparrow-\kappa_\downarrow}{2}\,a_j + d_j(t)
+\sqrt{\kappa_\uparrow}\,\xi_j^\dagger(t)
+\sqrt{\kappa_\downarrow}\,\epsilon_j(t),
\label{eq:app_chain_langevin_local_gain}
\end{equation}
where $d_j(t)=-\sqrt{\kappa_{\downarrow}}\alpha_{\rm in}(t)\delta_{j,j_c}$ is the coherent input term, with $\alpha_{\rm in}(t)=[x_{\rm in}(t)+ip_{\rm in}(t)]/\sqrt{2}=\mathcal{A}e^{i\theta}f(t)/\sqrt{2}$. \(\xi_j\) and \(\epsilon_j\) are independent vacuum input fields associated with amplification and loss, respectively. We assume the standard Markovian correlations
\begin{equation}
\begin{split}
&\langle\xi_{j}(t)\rangle = \langle\xi_j^\dag(t)\rangle = \langle\epsilon_j(t)\rangle = \langle\epsilon_j^\dag(t)\rangle = 0, \\
&\langle \xi_j(t)\xi_\ell^\dagger(t')\rangle = \delta_{j\ell}\delta(t-t'),
\quad
\langle \epsilon_j(t)\epsilon_\ell^\dagger(t')\rangle = \delta_{j\ell}\delta(t-t'),
\end{split}
\label{eq:app_chain_noise_corr}
\end{equation}
with all other normally ordered correlations vanishing. Defining the column vectors
\begin{equation}
\bm{a} =
\begin{pmatrix}
a_1\\
a_2\\
\vdots\\
a_{M_{\rm conv}}
\end{pmatrix},
\qquad
\bm{d} =
\begin{pmatrix}
d_1\\
d_2\\
\vdots\\
d_{M_{\rm conv}}
\end{pmatrix},
\qquad
\bm{\xi}^{\dagger} =
\begin{pmatrix}
\xi_1^{\dagger}\\
\xi_2^{\dagger}\\
\vdots\\
\xi_{M_{\rm conv}}^{\dagger}
\end{pmatrix},
\qquad
\bm{\epsilon} =
\begin{pmatrix}
\epsilon_1\\
\epsilon_2\\
\vdots\\
\epsilon_{M_{\rm conv}}
\end{pmatrix},
\end{equation}
and the tight-binding Hamiltonian matrix \(H_{\rm TB}\in
\mathbb{C}^{M_{\rm conv}\times M_{\rm conv}}\) via
\begin{equation}
(H_{\rm TB})_{j\ell}
=
h\left(\delta_{j,\ell+1}+\delta_{j,\ell-1}\right),
\qquad
j,\ell=1,\ldots,M_{\rm conv},
\end{equation}
we have, for bulk sites,
\begin{equation}
(H_{\rm TB}\bm{a})_j
=
h(a_{j+1}+a_{j-1}),
\qquad
(-iH_{\rm TB}\bm{a})_j
=
-ih(a_{j+1}+a_{j-1}),
\end{equation}
with open-boundary conditions \(a_0=a_{M_{\rm conv}+1}=0\).
Equation~\eqref{eq:app_chain_langevin_local_gain} can thus be written in vector form as
\begin{equation}
\dot{\bm a}
=
\left(\mathcal{G} \mathbb I_{M_{\rm conv}}-iH_{\rm TB}\right)\bm a + \bm{d}(t)
+\sqrt{\kappa_{\uparrow}}\,\bm{\xi}^{\dagger}
+\sqrt{\kappa_{\downarrow}}\,\bm{\epsilon},
\label{eq:app_chain_vector_langevin}
\end{equation}
where \[\mathcal{G}=\frac{\kappa_{\uparrow}-\kappa_{\downarrow}}{2}.\] Since the gain term is proportional to the identity, the propagator factorizes exactly as
\begin{equation}
e^{(\mathcal{G}\mathbb I_{M_{\rm conv}}-iH_{\rm TB})t}
=
e^{\mathcal{G}t}U(t),
\qquad
U(t)=e^{-iH_{\rm TB}t}.
\label{eq:app_factorized_propagator}
\end{equation}
This therefore separates the uniform amplification factor \(e^{\mathcal{G}t}\) from the passive unitary transport generated by the lattice.

Consider first the coherent part and define \(\alpha_j(t)=\langle a_j(t)\rangle\). Taking the expectation value of Eq.~\eqref{eq:app_chain_vector_langevin} and using Eq.~\eqref{eq:app_chain_noise_corr} yields
\begin{equation}
\dot{\bm \alpha}
=
(\mathcal{G}\mathbb I-iH_{\rm TB})\bm \alpha + \bm d,
\label{eq:app_alpha_ode}
\end{equation}
whose solution is
\begin{equation}
\bm \alpha(t)
=
e^{\mathcal{G}t}U(t)\bm \alpha(0) + \int_{0}^{t}ds e^{\mathcal{G}(t-s)}U(t-s)\bm d(s).
\label{eq:app_alpha_solution}
\end{equation}
For an initially empty chain, $\boldsymbol{\alpha}(0)=0$, the coherent occupation at site \(j\) is therefore
\begin{equation}
n^{\rm coh}_j(t)
=
|\alpha_j(t)|^2
=
\left|
\int_0^t ds\,
e^{\mathcal{G}(t-s)}
\left[U(t-s)\bm d(s)\right]_j
\right|^2.
\label{eq:app_local_coherent_energy}
\end{equation}

We next characterize the incoherent fluctuations generated by the active reservoirs. Writing the field operator as \(a_j(t)=\alpha_j(t)+\delta a_j(t)\), where \(\delta a_j(t)\) denotes the corresponding fluctuation operator, we define the equal-time normal covariance matrix
\begin{equation}
\mathbb{N}_{j\ell}(t)
=
\bigl\langle \delta a_j^\dagger(t)\,\delta a_\ell(t)\bigr\rangle .
\label{eq:app_equal_time_noise_covariance_def}
\end{equation}
In particular, the diagonal entries \(\mathbb{N}_{jj}(t)\) give the added incoherent occupation on site \(j\), while the off-diagonal entries quantify spatial correlations of the incoherent fluctuations.

Using Eq.~\eqref{eq:app_chain_vector_langevin}, the fluctuation operator \(\delta \bm a(t)=\bm a(t)-\bm\alpha(t)\) obeys
\begin{equation}
\frac{d}{dt}\,\delta \bm a(t)
=
\bigl(\mathcal{G}\mathbb I-iH_{\rm TB}\bigr)\delta \bm a(t)
+\sqrt{\kappa_\uparrow}\,\bm \xi^\dagger(t)
+\sqrt{\kappa_\downarrow}\,\bm \epsilon(t).
\label{eq:app_fluctuation_vector_eq}
\end{equation}
Its formal solution is
\begin{equation}
\delta \bm a(t)
=
e^{(\mathcal{G}\mathbb I-iH_{\rm TB})t}\delta \bm a(0)
+\int_0^t ds\;
e^{(\mathcal{G}\mathbb I-iH_{\rm TB})(t-s)}
\Bigl[
\sqrt{\kappa_\uparrow}\,\bm \xi^\dagger(s)
+\sqrt{\kappa_\downarrow}\,\bm \epsilon(s)
\Bigr].
\label{eq:app_fluctuation_formal_solution}
\end{equation}
Retaining only the reservoir-driven contribution,
\begin{equation}
\delta\bm{a}_{\rm noise}(t)
=
\int_0^t ds\;
e^{(\mathcal{G}\mathbb{I}-iH_{\rm TB})(t-s)}
\Bigl[
\sqrt{\kappa_\uparrow}\,\bm{\xi}^{\dagger}(s)
+\sqrt{\kappa_\downarrow}\,\bm{\epsilon}(s)
\Bigr],
\label{eq:deltaa_noise_part}
\end{equation}
we obtain
\begin{eqnarray}
\mathbb{N}_{j\ell}(t)
&=&
\int_0^t ds\int_0^t ds'\;\sum_{m,n}\left[e^{(\mathcal{G}\mathbb{I}-iH_{\rm TB})(t-s')}\right]^*_{jm}\left[
e^{(\mathcal{G}\mathbb{I}-iH_{\rm TB})(t-s)}\right]_{\ell n} \nonumber\\
&&\times \left\langle
\left[
\sqrt{\kappa_\uparrow}\,\xi_m(s')
+\sqrt{\kappa_\downarrow}\,\epsilon_m^{\dagger}(s')
\right]
\left[
\sqrt{\kappa_\uparrow}\,\xi_n^{\dagger}(s)
+\sqrt{\kappa_\downarrow}\,\epsilon_n(s)
\right]
\right\rangle.
\label{eq:N_double_integral_start}
\end{eqnarray}
For vacuum input fields, the Markovian correlations read
\begin{equation}
\left\langle\xi_m(s')\,\xi_n^{\dagger}(s)\right\rangle
=\delta_{mn}
\delta(s-s'), \quad \left\langle\epsilon_m^{\dagger}(s')\,\epsilon_n(s)\right\rangle=0,
\label{eq:xi_corr_matrix}
\end{equation}
with all mixed \(\xi\)-\(\epsilon\) correlations vanishing. Consequently, the normally ordered covariance \(\mathbb{N}(t)\) receives contributions solely from the gain bath, and the double integral reduces to
\begin{eqnarray}
\mathbb{N}_{j\ell}(t)
&=&\kappa_\uparrow
\int_0^t ds \sum_{m}
\left[ e^{(\mathcal{G}\mathbb{I}-iH_{\rm TB})(t-s)}\right]^*_{jm} 
\left[e^{(\mathcal{G}\mathbb{I}-iH_{\rm TB})(t-s)}\right]_{\ell m} \nonumber\\
&=& \kappa_\uparrow
\int_0^t ds \; e^{2\mathcal{G}(t-s)} \sum_{m}
U^*_{jm}(t-s) 
U_{\ell m}(t-s).
\label{eq:app_noise_cov_integral_derived}
\end{eqnarray}
Since \(U(t)\) is unitary and commutes with the identity diffusion matrix, the integral can be evaluated directly:
\begin{equation}
\mathbb{N}(t)
=
\frac{\kappa_\uparrow}{2\mathcal{G}}\left(e^{2\mathcal{G}t}-1\right)\mathbb I
=
\frac{\kappa_\uparrow}{\kappa_\uparrow-\kappa_\downarrow}\left(G_t-1\right)\mathbb I.
\label{eq:app_noise_cov_solution}
\end{equation}
Equation~\eqref{eq:app_noise_cov_solution} shows that each lattice site acquires the same added incoherent occupation,
\begin{equation}
n^{\rm inc}_j(t)
=
\frac{\kappa_\uparrow}{\kappa_\uparrow-\kappa_\downarrow}(G_t-1).
\label{eq:app_spontaneous_local}
\end{equation}
For the special case of pure gain, i.e., \(\kappa_\downarrow=0\), this reduces to
\begin{equation}
n^{\rm inc}_j(t)=G_t-1=e^{2\mathcal{G}t}-1.
\label{eq:app_quantum_limited_gain_noise}
\end{equation}

We now define a \emph{local} work-like SNR for each lattice site. For site \(a_j\), the local energy operator is
\begin{equation}
\hat{E}_j=\omega_{\rm C}\hat{a}_j^\dagger \hat{a}_j
=\frac{1}{2}\omega_{\rm C}\left(\hat{x}_j^2+\hat{p}_j^2-1\right),
\label{eq:app_local_energy_operator}
\end{equation}
and we accordingly define
\begin{equation}
{\rm SNR}^{\rm work}_j(t)
=
\frac{\mathcal W_j(t)}{\sigma_{E,j}(t)},
\qquad
\sigma_{E,j}(t)=\sqrt{{\rm Var}[\hat{E}_j](t)}.
\label{eq:app_local_work_snr_def}
\end{equation}

Note that the present benchmark describes a \emph{phase-insensitive} (phase-preserving) amplifier: the drift dynamics contains no anomalous \(a^\dagger\)-mixing terms; see Eq.~\eqref{eq:app_chain_vector_langevin}. Consequently, no anomalous correlators are generated. More precisely, we define
\begin{equation}
\mathbb{M}_{j\ell}(t)=\langle \delta a_j(t)\,\delta a_\ell(t)\rangle.
\label{eq:app_anomalous_cov_def}
\end{equation}
For an initially vacuum chain, $\mathbb{M}(0)=0$. Since the coherent drive affects
only the first moments and Eq.~\eqref{eq:app_fluctuation_vector_eq} contains no anomalous noise terms,
one has $\mathbb{M}(t)=0$ for all $t$. Hence, each single-site reduced state remains a \emph{displaced thermal Gaussian state} with an isotropic quadrature covariance matrix,
\begin{equation}
V_j(t)=\left[n_j^{\rm inc}(t)+\tfrac12\right]\mathbb I_2.
\label{eq:app_local_isotropic_cov}
\end{equation}
Since the ergotropy of a displaced thermal state is entirely due to its coherent displacement, we have
\begin{equation}
\mathcal W_j(t)=E^{\rm coh}_j(t)=\omega_{\rm C}n^{\rm coh}_j(t).
\label{eq:app_local_ergotropy_equals_coh}
\end{equation}

For a displaced thermal state with coherent occupation \(n^{\rm coh}_j(t)\) and incoherent occupation \(n^{\rm inc}_j(t)\), the energy variance reads
\begin{equation}
{\rm Var}[E_j](t)
=
\omega_{\rm C}^2n^{\rm inc}_j(t)\left[n^{\rm inc}_j(t)+1\right]
+\omega_{\rm C}^2\left[2n^{\rm inc}_j(t)+1\right]n^{\rm coh}_j(t).
\label{eq:app_local_energy_variance_displaced_thermal}
\end{equation}
Combining Eqs.~\eqref{eq:app_local_ergotropy_equals_coh} and \eqref{eq:app_local_energy_variance_displaced_thermal}, the local work-like SNR becomes
\begin{equation}
{\rm SNR}^{\rm work}_j(t)
=
\frac{n^{\rm coh}_j(t)}
{\sqrt{
n^{\rm inc}_j(t)\left[n^{\rm inc}_j(t)+1\right]
+\left[2n^{\rm inc}_j(t)+1\right]n^{\rm coh}_j(t)
}}.
\label{eq:app_local_work_snr_exact}
\end{equation}
Using Eqs.~\eqref{eq:app_local_coherent_energy} and \eqref{eq:app_spontaneous_local}, and defining
\begin{equation}
s_j(t):=
\left|
\int_0^t ds\,
e^{-\mathcal{G}s}
\left[U(t-s)\bm d(s)\right]_j
\right|^2,
\qquad
\chi:=\frac{\kappa_\uparrow}{\kappa_\uparrow-\kappa_\downarrow},
\label{eq:app_sj_chi_def}
\end{equation}
we obtain
\begin{equation}
n_j^{\rm coh}(t)=G_t\,s_j(t),
\qquad
n_j^{\rm inc}(t)=\chi\,(G_t-1),
\label{eq:app_local_coh_inc_compact}
\end{equation}
and therefore
\begin{equation}
{\rm SNR}^{\rm work}_j(t)
=
\frac{G_t\,s_j(t)}
{\sqrt{
\chi(G_t-1)\bigl[\chi(G_t-1)+1\bigr]
+
\bigl[2\chi(G_t-1)+1\bigr]G_t\,s_j(t)
}}.
\label{eq:app_local_work_snr_exact_compact}
\end{equation}
In the strong-gain limit \(G_t\gg 1\), both the numerator and the denominator scale linearly with \(G_t\), so the common amplification factor cancels:
\begin{equation}
{\rm SNR}^{\rm work}_j(t)
\longrightarrow
\frac{s_j(t)}{\sqrt{\chi\left[\chi+2s_j(t)\right]}}.
\label{eq:app_local_work_snr_saturated}
\end{equation}
Thus, the explicit exponential amplification factor cancels from the
work-like SNR in the strong-gain regime. Increasing the phase-preserving gain
therefore does not provide a parametric enhancement of the SNR, although the
pulse-dependent prefactor $s_j(t)$ can retain a residual dependence on the
gain rate. The difference of ${\rm SNR}^{\rm work}_j$ from the simpler ratio \(n_j^{\rm coh}/n_j^{\rm inc}\) is that here the denominator is the standard deviation of the corresponding energy observable.
Moreover, when $s_j(t)\ll\chi$, Eq.~\eqref{eq:app_local_work_snr_saturated} further reduces to
\begin{equation}
\mathrm{SNR}^{\rm work}_j(t)
\simeq
\frac{s_j(t)}{\chi}
=
\frac{\kappa_\uparrow-\kappa_\downarrow}
{\kappa_\uparrow}s_j(t),
\label{eq:app_local_work_snr_weak_seed}
\end{equation}
which coincides with the strong-gain limit of the ratio \(n_j^{\rm coh}/n_j^{\rm inc}\). Hence, when the coherent seed arriving at the site is weak, the corrected work-like SNR is numerically very close to the simpler benchmark, even though the underlying definition is conceptually different.
\begin{figure}
    \centering
    \includegraphics[width=0.7\linewidth]{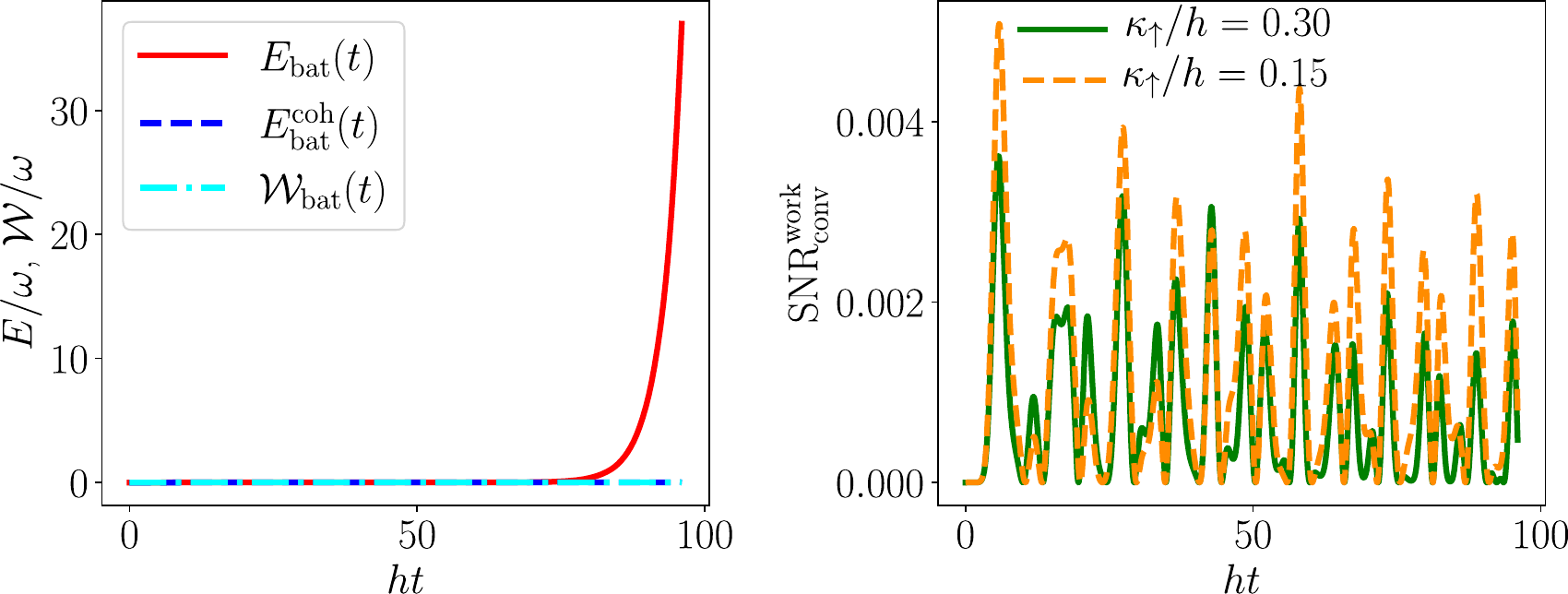}
    \caption{Conventional phase-preserving chain-amplifier benchmark. Left panel: dynamics of the total stored energy \(E_{\rm bat}(t)\), its coherent contribution \(E_{\rm bat}^{\rm coh}(t)\), and the ergotropy \(\mathcal{W}_{\rm bat}(t)\) for a one-dimensional tight-binding chain with a uniform on-site gain rate \(\kappa_{\uparrow}/h=0.3\). Right panel: dynamics of the work-like SNR for \(\kappa_{\uparrow}/h=0.15\) and \(0.3\). The conventional chain has \(M_{\rm conv}=2N+1\) sites and is driven at its central site by the same coherent Gaussian pulse as in the BKC scheme. Parameters: \(M_{\rm conv}=11\) (corresponding to \(N=5\)), \(\kappa_{\downarrow}/h=8.3\times10^{-4}\), \(g_{\rm L}/h=g_{\rm R}/h=5/12\), \(n_{\rm th}=0\), \(\gamma_{\rm L}=\gamma_{\rm R}=0\), \(\mathcal{A}/\sqrt{h}=2.887\),
\(\theta=\pi/2\), \(ht_0=1.2\), and \(h\sigma=1.2\). 
\justifying}\label{LocalGain_SummaryRow}
\end{figure}

For the full charging setup, which includes the chain and the coupled batteries, we verify the same behavior numerically in Fig.~\ref{LocalGain_SummaryRow}. In this case, we again consider two batteries coupled to the two ends of the lattice with rates $g_\L$ and $g_\R$, respectively, and evaluate the battery SNR  
\begin{equation}
{\rm SNR}^{\rm work}_{\rm conv}(t)
=
\frac{\mathcal W_{\rm bat}(t)}{\sigma_{E,{\rm bat}}(t)},
\label{eq:app_battery_work_snr_def}
\end{equation}
where \(\mathcal W_{\rm bat}(t)=\mathcal W_\L(t)+\mathcal W_\R(t)\) and \(\sigma_{E,{\rm bat}}(t)=\sqrt{\mathrm{Var}[\hat{E}_{\rm bat}](t)}\), with $\hat{E}_{\rm bat}=\omega \left(\hat{b}_{\L}^\dag \hat{b}_{\L}+\hat{b}_{\R}^\dag \hat{b}_{\R}\right)$. Numerically, the battery modes remain very close to displaced thermal states, so that \(\mathcal W_{\L,\R}(t)\simeq E^{\rm coh}_{\L,\R}(t)\). Meanwhile, the coherent seed transferred to the batteries is small compared with the gain-induced incoherent background. Consequently, although the total stored energy grows rapidly, the work-like SNR in Eq.~\eqref{eq:app_battery_work_snr_def} remains very small (\(\sim 10^{-3}\)) throughout the evolution.

The physical message is therefore clear. In a conventional phase-preserving gain medium, the amplified coherent contribution is accompanied by added incoherent fluctuations. Since these fluctuations are essentially isotropic, they contribute predominantly as passive, thermal-like energy: they strongly increase the total stored energy and its fluctuation, but provide little additional ergotropy. This is why the work-like SNR remains very small when the gain-induced incoherent background dominates over the coherent seed. Increasing the gain does not overcome this limitation, because both the coherent contribution and the added fluctuations acquire the same leading amplification factor, which cancels in the work-like SNR.

By contrast, in the phase-sensitive BKC scheme considered in the main text, the parametric dynamics not only amplifies the fluctuations but also reshapes them into highly anisotropic, squeezed states. These amplified fluctuations are therefore non-passive and contribute directly to the ergotropy. This allows the work-like SNR to remain close to unity, without the same passive-noise penalty of the conventional phase-preserving amplification.

\suppnote{On-site detuning disorder and its impact on chirality and stability}\label{app:onsite-perturb}

\suppsubsection{Robustness and Stability}

The phase-sensitive chirality of the BKC relies on the complete decoupling between the $x$- and $p$-quadrature sectors in the ideal model; see, e.g., the block structure in Eq.~\eqref{app:A_block}, where the $x$-$p$ drift blocks vanish. In this Note, we consider a more realistic situation, where on-site detuning perturbations break this decoupling and induce local quadrature rotations $x_j \leftrightarrow p_j$. These perturbations play the role of the ``on-site disorder''~\cite{PhysRevX.8.041031} and constitute the most detrimental imperfection for chiral amplification.

To study the impact of these imperfections, we add to the chain Hamiltonian a site-dependent detuning term
\begin{equation}
H_{\rm disorder}=\sum_{j=-N}^{N}\delta\omega_j\, a_j^\dagger a_j,
\label{eq:Hpert}
\end{equation}
where $\delta\omega_j$ is a random on-site frequency shift of site $j$. In the quadrature basis, this term generates a local phase-space rotation,
\begin{equation}
\dot x_j \big|_{\mathrm{disorder}}=+\delta\omega_j\, p_j,\qquad
\dot p_j \big|_{\mathrm{disorder}}=-\delta\omega_j\, x_j,
\label{eq:onsite-rotation}
\end{equation}
which explicitly couples the two orthogonal quadratures at each site. Incorporating Eq.~\eqref{eq:onsite-rotation} into the chain dynamics [cf. Eq.~\eqref{xp_chain}], we obtain
\begin{align}
\dot x_j &=
\frac{\tilde h(t)}{2}\!\left[e^+(t)\, x_{j-1}-e^-(t)\, x_{j+1}\right]
-\frac{\kappa_j}{2}x_j
-\sqrt{\kappa_0}\,x_{\mathrm{in}}(t)\,\delta_{j,0}
+\sqrt{\kappa_j}\,\xi^{(x)}_j(t)
+ g_\R(t)\,p_\R\,\delta_{j,+N}+g_\L(t)\,p_\L\,\delta_{j,-N}
+\delta\omega_j\, p_j,
\label{eq:HLx-pert}
\\[2mm]
\dot p_j &=
\frac{\tilde h(t)}{2}\!\left[e^-(t)\, p_{j-1}-e^+(t)\, p_{j+1}\right]
-\frac{\kappa_j}{2}p_j
-\sqrt{\kappa_0}\,p_{\mathrm{in}}(t)\,\delta_{j,0}
+\sqrt{\kappa_j}\,\xi^{(p)}_j(t)
- g_\R(t)\,x_\R\,\delta_{j,+N}-g_\L(t)\,x_\L\,\delta_{j,-N}
-\delta\omega_j\, x_j.
\label{eq:HLp-pert}
\end{align}
The equations for the batteries' dynamics, e.g., Eq.~\eqref{storLR}, and the Markovian noise statistics given in Eqs.~\eqref{noisexi} and \eqref{app:markov_noise_corr} remain unchanged. In the compact form in Eq.~\eqref{app:linear_R}, on-site detunings modify the drift matrix by introducing nonzero $x$-$p$ blocks. Specifically, Eq.~\eqref{app:A_block} generalizes to
\begin{equation}
A(t)=
\begin{pmatrix}
A_{x_{\rm C}, x_{\rm C}}(t) & A_{x_{\rm C}, p_{\rm C}} & A_{x_{\rm C}, \R}(t) & A_{x_{\rm C}, \L}(t) \\
A_{p_{\rm C}, x_{\rm C}} & A_{p_{\rm C}, p_{\rm C}}(t) & A_{p_{\rm C}, \R}(t) & A_{p_{\rm C}, \L}(t) \\
A_{\R, x_{\rm C}}(t) & A_{\R, p_{\rm C}}(t) & A_{\R,\R}(t) & 0 \\
A_{\L, x_{\rm C}}(t) & A_{\L, p_{\rm C}}(t) & 0 & A_{\L,\L}(t)
\end{pmatrix}.
\label{app:A_block2}
\end{equation}
The only additional nonzero blocks are
\begin{equation}
    A_{x_{\rm C},p_{\rm C}} = \mathrm{diag}(\delta\omega_{-N},\ldots,\delta\omega_{+N}),\qquad
    A_{p_{\rm C},x_{\rm C}} = -A_{x_{\rm C},p_{\rm C}},
\label{eq:AxpApx}
\end{equation}
which arise from the on-site detuning terms that mix the $x$- and $p$-quadrature sectors locally. All other blocks remain unchanged from Eq.~\eqref{app:A_block}.
\begin{figure}
    \centering
    \includegraphics[width=0.5\linewidth]{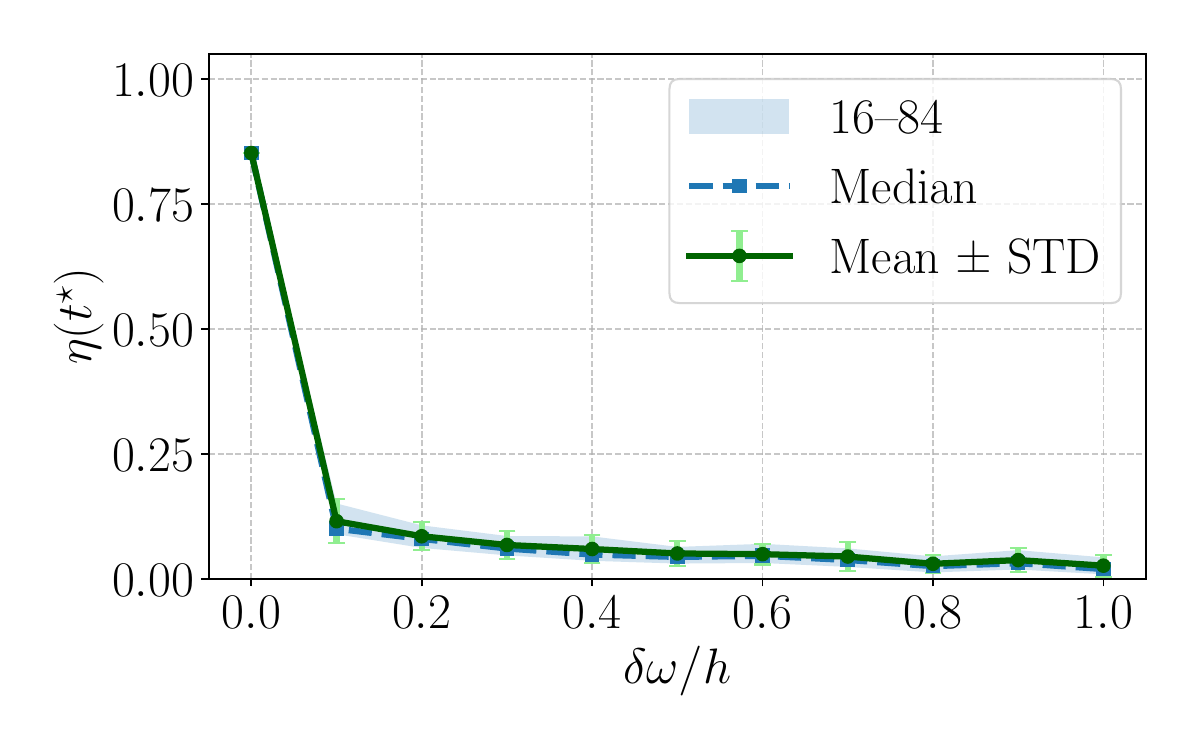}
    \caption{Overall charging efficiency $\eta(t^\star)$ versus on-site detuning amplitude $\delta \omega$. For each $\delta\omega$ we draw independent detunings
    $\delta\omega_j\sim\mathcal{U}[-\delta\omega, +\delta\omega]$ and run the two-pass protocol, recording $\eta(t^\star)$. The blue dashed line shows the median, the shaded region indicates the $16$th--$84$th percentile interval, and the markers with error bars give the mean $\pm$ standard deviation. Parameters: $N=5$, $\Delta/h=11/12$, $g/h=5/12$, $\kappa_j/h=8.3\times 10^{-4}$ ($j=-N,\cdots,N$), \(n_{\rm th}=0\), \(\gamma_{\rm L}=\gamma_{\rm R}=0\), $\mathcal{A}/\sqrt{h}=2.887$, $\theta=\pi/2$, $h\sigma=1.2$, and $ht_0=1.2$. \justifying}
    \label{EfficiencydOmega}
\end{figure}

Equations \eqref{app:A_block2} and \eqref{eq:AxpApx} show the key qualitative difference from the ideal BKC: for $\delta\omega_j\neq 0$, the $x-$ and $p-$quadrature sectors are no longer completely decoupled. Since the perturbation $H_{\rm disorder}$ does not introduce additional dissipative channels, the diffusion matrix $D(t)$ remains unchanged and is still fully determined by the Markovian baths; see Eqs.~\eqref{app:V_eq}--\eqref{DRL}.
\begin{figure}
    \centering
    \includegraphics[width=0.7\linewidth]{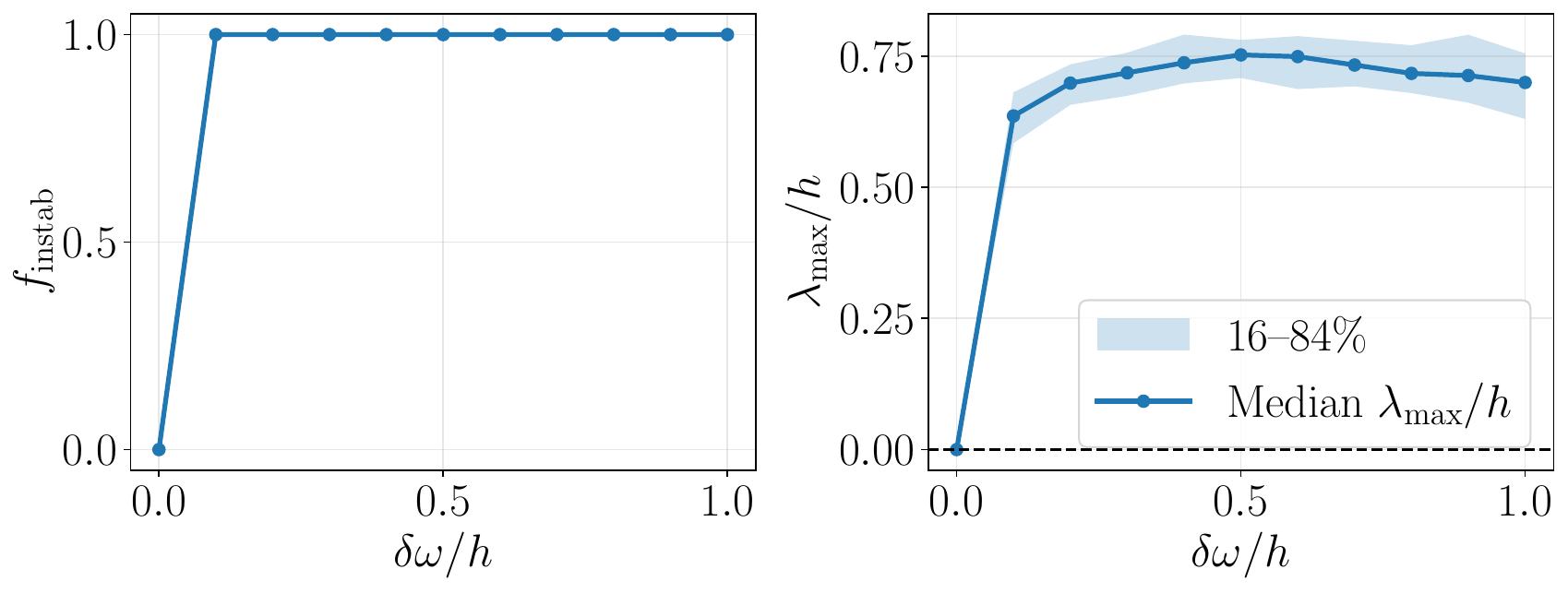}
    \caption{Stability analysis for the same on-site detuning-disorder sweep (50 realizations) as in Fig.~\ref{EfficiencydOmega}. Left: fraction of unstable realizations, $f_{\rm instab} = \Pr[\lambda_{\max}>0]$. Right: median spectral abscissa \(\lambda_{\max}=\max_k {\rm Re}[\lambda_k(A)]\), with the shaded region showing the $16-84\%$ interval across the disorder realizations. While the ideal system is stable, for the strong-pump parameters considered here, the finite on-site detuning disorder rapidly drives the dynamics unstable by mixing the \(x\)- and \(p\)-quadrature sectors. All other parameters are the same as in Fig.~\ref{EfficiencydOmega}. \justifying}
    \label{InstabilityFraction_LambdaMax}
\end{figure}
\begin{figure}
    \centering
    \includegraphics[width=0.45\linewidth]{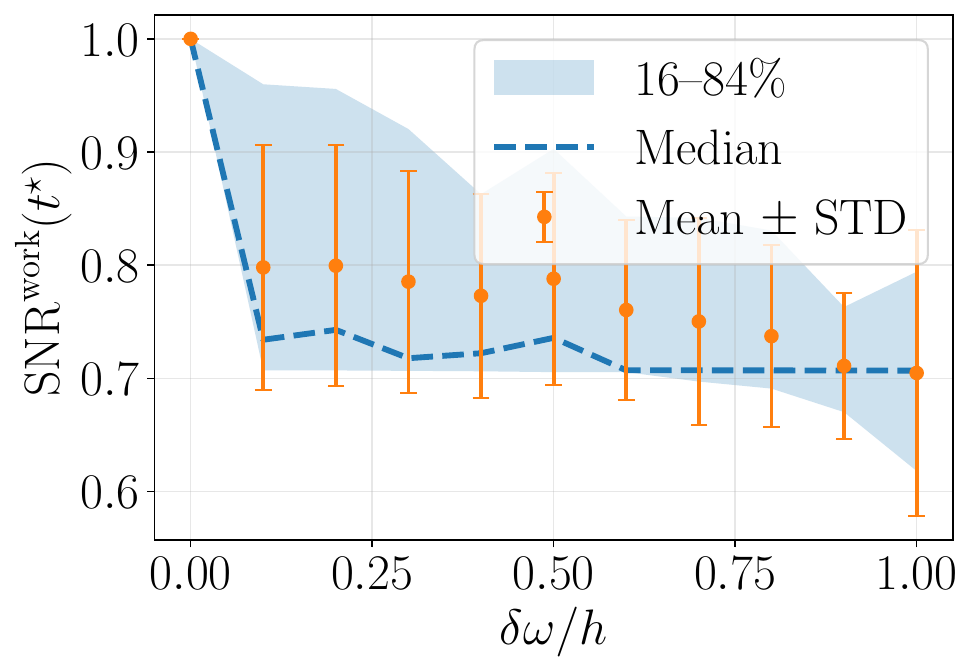}
    \caption{Work-like signal-to-noise ratio at the optimal switch-off time, ${\rm SNR}^{\rm work}(t^\star)$, versus the on-site detuning-disorder strength \(\delta\omega\) (50 disorder realizations with \(\delta\omega_j\in[-\delta\omega,\delta\omega]\)). The blue dashed line shows the median, the shaded region indicates the $16\%-84\%$ interval, and the markers with error bars give the mean $\pm$ standard deviation. All other parameters are the same as in Fig.~\ref{EfficiencydOmega}. \justifying}\label{SNRwork_vs_OnSiteDetuning}
\end{figure}

To quantitatively assess the impact of the on-site disorder, we draw independent detunings $\delta\omega_j\sim\mathcal{U}[-\delta\omega,+\delta\omega]$ and, for each disorder realization, run the same two-pass protocol as in the main text: we first determine $t^\star$ according to the analytical expression Eq.~\eqref{tstar_decomp} and the numerical simulations, and then switch off the setup ($h=\Delta=g=0$) at $t=t^\star$. In Fig.~\ref{EfficiencydOmega} (and similar figures hereafter), the median is the middle value of $\eta(t^\star)$ across realizations at a given disorder level ($50$th percentile). Half of the runs yield $\eta(t^\star)$ above it and half below. The $16$th--$84$th percentile band spans the range between the $16$th and $84$th percentiles of $\eta(t^\star)$ across realizations, where the $16$th ($84$th) percentile is the value below which $16\%$ ($84\%$) of the runs fall. This band summarizes the empirical spread across realizations (i.e., sample-to-sample variability), and should not be interpreted as an uncertainty of the mean. For comparison, we also report the mean and standard deviation; see the curve with error bars. Specifically, the mean is given by $\bar{\eta} = \frac{1}{M'}\sum_{j=1}^{M'} \eta_j$, and the standard deviation is
\[
\text{STD} = \sqrt{\frac{1}{M'-1}\sum_{j=1}^{M'} (\eta_j - \bar{\eta})^2}.
\]
While the mean $\pm$ STD summarizes the distribution through its moments, the median and $16$th--$84$th band are more robust to outliers and can better reveal asymmetry or skewness. 
As shown in Fig.~\ref{EfficiencydOmega}, the charging efficiency is highly sensitive to this type of disorder. It drops from \(\eta(t^\star)\simeq 0.852\) at \(\delta\omega/h=0\) to \(\eta(t^\star)=0.1155\pm0.0437\) already at \(\delta\omega/h=0.1\) (over \(50\) realizations). For stronger disorder the efficiency decreases more gradually, reaching \(\eta(t^\star)=0.0267\pm0.0221\) at \(\delta\omega/h=1\). Therefore, for the strong-pump parameters used in Fig.~\ref{EfficiencydOmega}, the protocol rapidly enters a low-efficiency regime as the detuning disorder is increased, with \(\eta(t^\star)\lesssim 0.1\) over most of the investigated range, \(\delta\omega/h\gtrsim0.2\).

To understand the origin of this degradation, we monitor the spectral abscissa of the drift matrix \(A(t)\), defined as the largest real part among its eigenvalues,
\begin{equation}
    \lambda_{\max}\equiv \max_k \Re\!\left[\lambda_k(A)\right].
\label{eq:lmax-def}
\end{equation}
This quantity provides a direct diagnostic of the stability of the linear Langevin dynamics: \(\lambda_{\max}<0\) means that, in the absence of drive and noise, small perturbations decay in time, whereas \(\lambda_{\max}>0\) indicates the onset of exponentially growing modes. In the ideal system without on-site detuning disorder, we find \(\lambda_{\max}/h\simeq -2.7\times10^{-4}\), indicating stable dynamics; see Fig.~\ref{InstabilityFraction_LambdaMax}. By contrast, on-site detuning disorder induces local quadrature rotations [cf. Eq.~\eqref{eq:onsite-rotation}], which mix the counter-propagating quadrature sectors and can drive the system into an unstable regime. This is precisely the mechanism identified in Ref.~\cite{PhysRevX.8.041031}: impurity scattering converts one quadrature into the other, and because the two quadratures experience opposite directional amplification, repeated scattering can generate an effective amplified feedback. For the strong-pump parameters used in Fig.~\ref{InstabilityFraction_LambdaMax}, all disorder realizations become unstable over the investigated range, $\delta\omega/h \in[0.1,1]$, with $f_{\rm instab}=\Pr[\lambda_{\max}/h>0]=1$ and a typical median $\lambda_{\max}/h$ in the range $0.63$--$0.75$.

This instability mechanism also impacts the work-like SNR [see Eq.~\eqref{SM_SNR_work}] in a related way. Figure~\ref{SNRwork_vs_OnSiteDetuning} shows ${\rm SNR}^{\rm work}(t^\star)$ as a function of the disorder strength $\delta\omega$. The ideal protocol starts near ${\rm SNR}^{\rm work}(t^\star)\simeq 1$, while the detuning disorder reduces it to a lower value, which remains of order unity. Importantly, this modest reduction should not be interpreted as robustness of the charging dynamics itself. As discussed above, for the strong-pump parameters considered here, the detuning-induced quadrature mixing that degrades the charging efficiency also drives the system into an unstable regime (see Fig.~\ref{InstabilityFraction_LambdaMax}). Here we report ${\rm SNR}^{\rm work}$ at the optimal switch-off time $t^\star$, before the late-time runaway growth becomes dominant. The reason ${\rm SNR}^{\rm work}(t^\star)$ does not collapse as dramatically as $\eta(t^\star)$ is that, once unstable modes are present, the active Gaussian dynamics amplifies both the useful non-passive contribution $\mathcal{W}(t^\star)$ and the battery-energy fluctuation $\sigma_E(t^\star)$, so the numerator and denominator co-vary. Accordingly, the values of ${\rm SNR}^{\rm work}$ shown in Fig.~\ref{SNRwork_vs_OnSiteDetuning} should be understood as \emph{finite-time performance metrics} evaluated at $t^\star$, rather than as indicators of asymptotically stable operation. In a physical implementation, the late-time runaway would ultimately be regularized by nonlinearities, pump depletion, or other saturation mechanisms that are absent from the linear Gaussian model.

\begin{figure}
    \centering
    \includegraphics[width=0.7\linewidth]{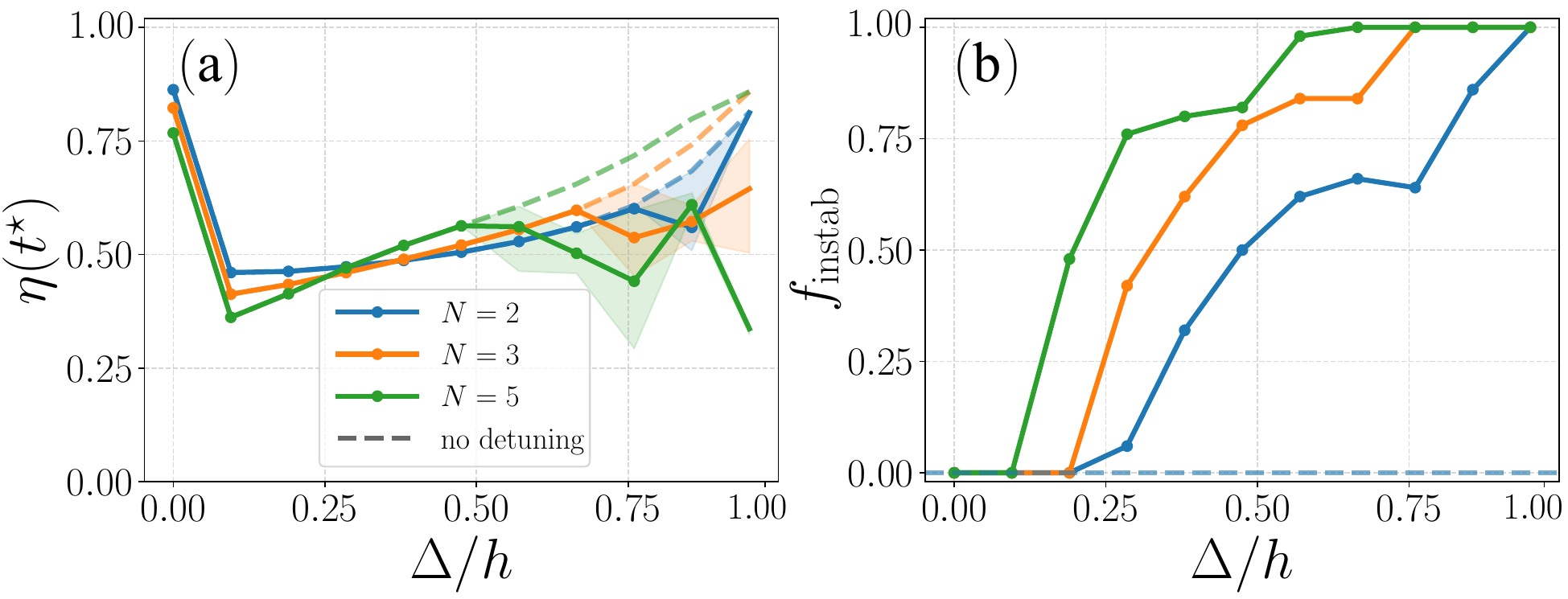}
    \caption{Mitigating sensitivity to on-site detuning disorder by reducing the pump strength and/or chain length. We fix weak on-site disorder, $\delta\omega_j\sim\mathcal{U}[-10^{-3}h,+10^{-3}h]$, and sweep the pump strength $\Delta$ (in units of $h$) for several chain half-lengths $N$. (a) Charging efficiency $\eta(t^\star)$ in the presence of on-site detuning disorder (solid lines; shaded regions indicate the $16-84\%$ interval over disorder realizations) compared with the no-detuning baseline (dashed lines). (b) Fraction of unstable realizations $f_{\rm instab}=\Pr[\lambda_{\max}/h>0]$ as a function of $\Delta$. All other parameters are the same as in Fig.~\ref{EfficiencydOmega}. \justifying}
    \label{KnobStudy}
\end{figure}

\suppsubsection{Mitigation of the disorder-induced performance degradation}

The strong sensitivity found above is closely related to the strong parametric pump ($\Delta/h=11/12$) used in the main text. It does not imply that arbitrarily
weak on-site detuning disorder necessarily destabilizes the BKC. The
susceptibility to detuning-induced quadrature mixing depends strongly on the
parametric pump strength and the chain length.

A practical way to mitigate the detuning-induced performance degradation is therefore to use a shorter BKC (i.e., with smaller $N$) and/or reduce the parametric pump strength $\Delta$. To quantify this trade-off, we consider a weak but experimentally reasonable disorder level with $\delta\omega=10^{-3}h$~\cite{PhysRevX.8.041031} and sweep the pump strength $\Delta$ for several values of $N$. Figure~\ref{KnobStudy} shows that for small $\Delta$ the protocol remains insensitive to the detuning disorder. The curves for the disordered cases coincide with the ``no-detuning'' baseline, and the drift dynamics remains stable, with $f_{\rm instab}\simeq 0$ and $\lambda_{\max}$ slightly negative. As $\Delta$ increases, the system becomes gradually more susceptible to the detuning disorder, and the onset of instability shifts to smaller $\Delta$ for longer chains, consistent with the cumulative effect of quadrature mixing along the transport path. In the strong-amplification regime, $\Delta/h\to 1$, the ideal (no-detuning) protocol can still yield high efficiencies, but this same regime becomes quite sensitive to on-site disorder, especially for larger $N$. Overall, these results identify a clear experimental knob: reducing $\Delta$ and/or using a shorter chain strongly suppresses the detuning-induced degradation, at the cost of weaker directional amplification.


\begin{figure}
    \centering
    \includegraphics[width=0.55\linewidth]{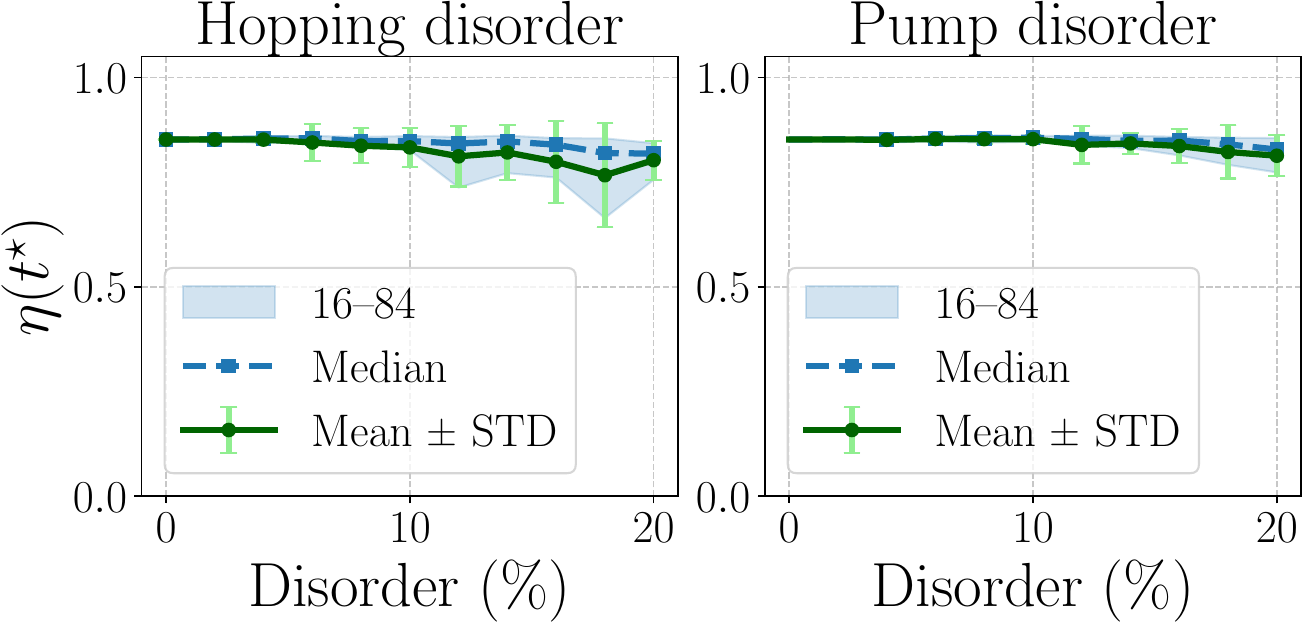}
     \caption{Optimal charging efficiency $\eta(t^\star)$  versus (left) hopping disorder and (right) pump disorder. For each disorder level, $\eta(t^\star)$ is evaluated from 50 independent disorder realizations. The dashed lines show the medians, the shaded regions indicate the $16-84\%$ interval, and the markers with error bars show the mean $\pm$ standard deviation. All other parameters are the same as in Fig.~\ref{EfficiencydOmega}. \justifying}
    \label{Disorder}
\end{figure}

\suppsubsection{Disorder in hopping $h$ and parametric pump $\Delta$}

In the main text, we show that the work-like SNR is robust against symmetry-preserving disorder, such as random perturbations in the hopping $h$ and parametric pump $\Delta$ of the BKC. Here we further demonstrate that the charging efficiency exhibits a similar robustness against these two disorder types.   Figure~\ref{Disorder} plots the optimal efficiency $\eta(t^\star)$ for static imperfections in both $h$ and $\Delta$. As discussed in the main text, moderate hopping and pump disorder mainly renormalize local parameters rather than opening strong back-scattering pathways, so $\eta(t^\star)$ remains close to its ideal value over a broad disorder range. Similar to the work-like SNR, the degradation of $\eta(t^\star)$ is more pronounced for hopping disorder (see the left panel of Fig.~\ref{Disorder}) because randomness in the hopping strengths $h_{j,j+1}$ directly distorts the transport backbone and makes the effective hopping scale $\tilde h=\sqrt{h^2-\Delta^2}$ position dependent, creating impedance mismatches that enhance reflections and delay the energy delivery before it reaches the chain edges. By contrast, disorder in the parametric pump (see the right panel of Fig.~\ref{Disorder}) primarily modulates the local gain/squeezing strength [cf. Eq.~\eqref{defhr}] while leaving the underlying connectivity intact. As a result, its effect tends to self-average along the chain, and the efficiency remains nearly flat until the disorder becomes sufficiently strong.

\color{Black}
\suppnote{Experimental implementation of battery--BKC interactions}

In the main text, we have mentioned that the BKC has been experimentally realized in both superconducting quantum circuits~\cite{BKCrealizationCircuit2024} and optomechanical arrays~\cite{BKCrealizationOM2024}. In this Note, we briefly discuss feasible implementations of the additional battery modes and the corresponding BKC--battery interactions. 

In the superconducting synthetic-frequency implementation reported in Ref.~\cite{BKCrealizationCircuit2024}, the BKC sites correspond to distinct modes of a multimode cavity. The hopping and pairing interactions are independently activated by parametric drives at the \emph{difference} and \emph{sum} frequencies, respectively, of the two relevant modes. Thus, an additional cavity mode can naturally serve as a battery mode and be connected to a BKC edge mode through a hopping-only link. Specifically, the beam-splitter battery--chain interaction $g\left(a_{\pm N}^{\dag}b_{\alpha}+\textrm{H.c.}\right)$ ($\alpha=\rm L, R$) can be realized by applying only a difference-frequency drive at $\Omega$, corresponding to the frequency detuning between the battery modes $b_{\rm L,R}$ and the BKC edge sites $a_{\pm N}$.

Likewise, in the optomechanical realization reported in Ref.~\cite{BKCrealizationOM2024}, parametrically controlled optomechanical interactions already generate beam-splitter and two-mode-squeezing couplings between mechanical modes. An additional mechanical mode can be used as a battery and coupled to an edge mode by activating only the beam-splitter interaction. In both cases, the battery modes remain outside the parametrically paired BKC network, while their beam-splitter coupling strengths $g_{\rm L}$ and $g_{\rm R}$ can be tuned through the corresponding drive amplitudes.

\color{Black}

\end{document}